\documentclass[prb,superscriptaddress, reprint]{revtex4-1}
\pdfoutput=1
\usepackage{comment}
\usepackage{amssymb}
\usepackage{amsmath}
\usepackage{braket}
\usepackage{graphicx}
\usepackage{enumerate}
\usepackage[usenames,dvipsnames]{color}
\usepackage[colorlinks,bookmarks=false,citecolor=NavyBlue,linkcolor=OliveGreen,urlcolor=blue]{hyperref}
\usepackage{mathrsfs}

\newcommand{\be}{\begin{equation}}
\newcommand{\ee}{\end{equation}}
\newcommand{\ba}{\begin{aligned}}
\newcommand{\ea}{\end{aligned}}

\newcommand{\1}{\mathrm{I}}

\newcommand{\ratioF}{0.40}
\newcommand{\ratioFthree}{0.323607}
\newcommand{\Hspace}{0.8cm}

\DeclareMathOperator{\arctanh}{arctanh}

\newenvironment{remark}[1][Remark]{\begin{trivlist}
\item[\hskip \labelsep {\bfseries #1}]}{\end{trivlist}}

\begin{document}
\title{Universal prethermalization dynamics of entanglement entropies after a global quench}
\author{Maurizio Fagotti}
\affiliation{D\'epartement de Physique, \'Ecole normale sup\'erieure, CNRS, 24 rue Lhomond,75005 Paris, France}
\author{Mario Collura}
\affiliation{SISSA \& INFN, via Bonomea 265, 34136 Trieste, Italy}
\begin{abstract}
We consider the quantum XY model and study the effects of interacting perturbations on the time evolution  
of the von Neumann and R\'enyi entropies of spin blocks after global quenches.
We show that the entropies are sensitive to perturbations that break hidden symmetries behind the integrability of the model. At times much larger than the characteristic time of the well-known linear increase of the entropies, we identify a time window characterized by a novel linear growth followed by saturation. The typical time of the phenomenon is inversely proportional to the perturbation strength and the behavior is trigger off by the extinction of an infinite number of local conservation laws following a non-abelian algebra. 
The universality of the crossover is revealed by a semi-classical picture that captures the leading behavior of the entropies.
We check our theoretical predictions against iTEBD simulations.
The good agreement between theory and numerics 
substantiates the method developed in [Bertini and Fagotti, J. Stat. Mech. (2015) \href{http://dx.doi.org/10.1088/1742-5468/2015/07/P07012}{P07012}] for investigating a pre-relaxation limit in weakly interacting models.
\end{abstract}
\maketitle


\section{Introduction}  %

A global Hamiltonian parameter is suddenly changed and the state, originally in equilibrium, evolves under a different Hamiltonian. This simple  protocol of non-equilibrium time evolution is generally called a ``global quench''. Since the first studies on global quenches, many exciting theoretical aspects have been uncovered (emergence of statistical descriptions, universality of dynamics, etc.)\cite{CC:05,CCcorr:05,C-06,rigol,BS-08,C-08,S:08,CEF, CEF1,EEF:dyn,FE:13,HPK:13,CE-13,M-13,CSC-13,SC:cluster,E:preT,QAXXZ-14,PMWKZT-14,B:sG,D-14,F:super,CBSF:glassy, BEL:14} and observed in numerical simulations\cite{RSMS:08,ban-11,CK,KRSM:12,FCEC-14,CCS-P:15,rigol:num, BEL:14} and real experiments\cite{gm-02, kww-06,Getal1,Getal2,Tetal-12, chetal-12, schetal-12, Metal-13, Fateal-13, FSetal-13, Metal-14}.  One of the first theoretical insights was that  the entanglement entropies (von Neumann and R\'enyi entropies) of subsystems follow a universal behavior\cite{CC:05}.  In the framework of conformal field theory (CFT), Ref.~[\onlinecite{CC:05}] showed that the entropies of an interval grow linearly in time until a particular time proportional to the subsystem length $\ell$. 
Then, they saturate to stationary values. 
It was soon realized that the linear behavior is almost independent of the system details:
despite CFTs describing only a class of critical models, a similar time evolution appears in the (space-time) scaling limit $\ell\sim t\gg 1$ also in noncritical spin chains. The semiclassical interpretation proposed in  Ref.~[\onlinecite{CC:05}] is indeed expected to hold rather generally\cite{lk2008,dmcf2008,FC:08}.
Furthermore, the extent of the time window of linearity is inversely proportional to the (Lieb-Robinson\cite{LR:velocity}) maximal velocity $v_M$ at which information propagates, so the analysis of the entropies represents a simple way to extract $v_M$.

After Ref.~[\onlinecite{CC:05}] there has been an increasing interest in the non-equilibrium time evolution of the entanglement entropies. A large part of works concerned the dynamics induced by short-range Hamiltonians \cite{lk2008,dmcf2008,isl2012,ttd2014,KBC:exc,CTC:14}, but recently some attention moved also to global quenches with long-range interactions \cite{SLRD:13, gr2014}.
In addition, the entanglement entropy has been at the center of intensive research in other non-equilibrium problems, like local and geometric quenches\cite{sd2011,ep2012, CCo:13, ah2014}. It is also worth mentioning that there are a few proposals, based on local quenches,  to measure the entanglement entropies in real experiments \cite{cardy2011,ad2012}.

In spite of the numerous studies on global quenches, not much attention was paid to the relaxation process of the entropies:
the lack of a simple picture explaining the approach to the stationary values has  probably been a major deterrent to the research on that topic. Indeed, to the best of our knowledge, the long time behavior of the von Neumann entropy after a global quench has been analyzed analytically only in the quantum XY model\cite{FC:08}, where, depending on the system details, it was found either a simple power-law relaxation $\sim \ell^2/t$ or a more elaborate decay $\sim \ell^4 \log (t/\ell)/t^3$. 

The goal of this paper is to identify some universal aspects that emerge in a long time limit. We choose a backward point of view, asking firstly what are the features that dramatically affect the values of the entropies at infinite time after the quench. 

It is widely believed that at late times after global quenches in generic models local degrees of freedom can be described by a Gibbs ensemble (GE) with an effective temperature depending on the initial state\cite{deu-91,sred-94,rignat-08,bir-10,sir-14,A:ETH}. In integrable models, instead, infinite information about the initial state is retained and the stationary properties are 
described by a generalized Gibbs ensemble (GGE) constrained by infinitely many \mbox{(quasi-)local} conservation laws\cite{rigol,FE:13}. 
Among the integrable models it is also convenient to distinguish the ones with an infinite number of local conservation laws that satisfy a non-abelian algebra \cite{F:super}, which will be refereed to as \emph{non-abelian integrable models}\cite{SUPER}. Also in the latter case a GGE is expected to describe the stationary properties of local observables, but infinitely more information than usual is retained about the initial state at late times. 
Clearly, the entanglement entropies are very sensitive to constraints on the dynamics, and indeed relaxation to a GE, a GGE, or a ``non-abelian'' GGE result in different asymptotic values.

Our strategy is to perturb the Hamiltonian in such a way that the limit of infinite time does not commute with the limit of zero perturbation. In other words, we consider  perturbations that break hidden symmetries behind the integrability of the model.
It is then reasonable to expect that, as long as the time is sufficiently
large but much smaller than some characteristic time determined by the
perturbation, the entropies settle close to the stationary values
associated with the unperturbed integrable model (prethermalization
plateau): the GGE of the unperturbed model, possibly including also
non-commuting charges, is a good local approximation of the state.
Whenever the unperturbed set of local charges is abelian and the
perturbation breaks integrability, at larger times the expectation values
move from the pre-thermalization plateau and reach the values predicted by
the GE.
When instead the perturbation breaks non-abelian integrability into integrability, 
a subset of charges is destroyed and the remaining ones are deformed; as a consequence, the
earliest plateau (pre-relaxation plateau) is generally followed by a less
constrained GGE (although persistent oscillatory behavior could remain).
If the perturbation breaks also integrability, in even longer times all charges but the Hamiltonian become extinct, inexorably leading to a GE.
Therefore, we recognize the strength of a perturbation that breaks integrability or non-abelian integrability as a relevant parameter to the late time behavior of local observables and entanglement entropies.

Since infinitesimal perturbations are sufficient to trigger off a crossover between macroscopically different ensembles and the entropy per unit length is  robust under a large class of transformations, we  can expect most of the details of the system and of the perturbations to be qualitatively irrelevant: 
the time evolution of the entanglement entropies in the time window of the crossover could display some form of universality.   

\paragraph{The model.} %

We consider the Hamiltonian of the XYZ model 
\be\label{eq:HD}
H_\Delta=\sum_\ell\frac{1+\gamma}{4}\sigma_\ell^x\sigma_{\ell+1}^x+\frac{1-\gamma}{4}\sigma_\ell^y\sigma_{\ell+1}^y+\frac{\Delta}{4}\sigma_\ell^z\sigma_{\ell+1}^z\, ,
\ee
where $\sigma_\ell^\alpha$ act like Pauli matrices on site $\ell$ and like the identity elsewhere. 
We have not explicitly written an overall multiplicative constant with the dimensions of an energy, which can be viewed as implicitly attached to the time. 
The XYZ spin-$\frac{1}{2}$ chain is an interacting integrable model that can be solved by algebraic Bethe ansatz~\cite{Baxter}. For $\Delta= 0$ it is reduced to the quantum XY model and can be mapped to a noninteracting chain of spinless fermions~\cite{LSM:XY}. The XY model has  an infinite non-abelian set of local conservation laws~\cite{F:super}, so, in the limit $\Delta\rightarrow 0$, \eqref{eq:HD} describes a non-abelian integrable model. 
We also consider two perturbations that break integrability (for generic $\gamma$ and $\Delta$):
\be
S^{z\cdot z}=\frac{1}{4} \sum_\ell \sigma_\ell^z\sigma_{\ell+2}^z\quad\text{and}\quad S^z=\frac{1}{2} \sum_\ell \sigma_\ell^z\, .
\ee
Thus, in its most general form, the Hamiltonian reads as
\be\label{eq:H}
H=H_\Delta+\Delta (U S^{z\cdot z}+ h S^z)\, .
\ee
We multiplied the terms that (are supposed to) break non-abelian integrability by $\Delta$, so $\Delta$ is expected to characterize 
the timescale at which the hidden symmetry breaking becomes manifest. On the other hand,  for $h$ or $U$ different from zero, relaxation to a thermal ensemble generally occurs on much longer timescales that depend on $\Delta U$ and $\Delta h$ in a different way (the results of Ref.~[\onlinecite{BEGR:preT}] suggest timescales $\sim (\Delta U)^{-2},(\Delta h)^{-2}$). 

In Refs~[\onlinecite{F:super}][\onlinecite{BF:mf}] some effects of hidden symmetry breaking have been identified in the time evolution of local observables, provided that the initial state breaks one-site shift invariance, like the ground state of $H_\Delta$ in the antiferromagnetic phase. 
\emph{A posteriori}, this explains why the original investigations~\cite{FC:08} on the time evolution of the entropy after quantum quenches in the XY model did not show any peculiar behavior beyond the scaling limit of Ref.~[\onlinecite{CC:05}]. 
We therefore consider initial states in which the symmetry with respect to translations by one site is (spontaneously) broken.

\paragraph{Results.}  %

Let us denote by
\be
\rho_\ell=\mathrm{Tr}_{\bar A}[\ket{\Psi(t)}\bra{\Psi(t)}]
\ee
the reduced density matrix of a subsystem $A$ consisting of $\ell$ adjacent spins ($\bar A$ is the complement of $A$) and by
\be
S_\ell^{(\alpha)}=\frac{\log\mathrm{Tr}\rho_\ell^\alpha}{1-\alpha}
\ee
the corresponding R\'enyi entropies (the von Neumann entropy being $S_\ell=\lim_{\alpha\rightarrow 1^+} S^{(\alpha)}_\ell$). 
The main result of this paper is that, up to corrections that approach zero in the limit of small perturbation, the entropies per unit length asymptotically behave as follows
\be\label{eq:scaling}
\frac{S_\ell^{(\alpha)}(t)}{\ell}\rightarrow \begin{cases}
\kappa_0^{(\alpha)} \frac{t}{\ell}&1\ll t<\frac{\ell}{2 v_M}(\ll \Delta^{-1})\\
\sigma_0^{(\alpha)}+\kappa_1^{(\alpha)} \frac{\Delta t}{\ell}& 1\ll \Delta t<\frac{\ell}{2 \tilde v_M}\\
\sigma_1^{(\alpha)}&\frac{\ell}{2 \tilde v_M}\ll \Delta t\ll \dots\, .
\end{cases}
\ee
Here $v_M$ is the maximal velocity at which information propagates and $\tilde v_M$ is another velocity emerging at intermediate times; $\sigma_0^{(\alpha)} $ is the entropy per unit length in the earliest plateau while $\sigma_1^{(\alpha)} $ is the value reached at larger times. 
The last undefined inequality signifies that the final regime is not expected to extend over infinite times.
In particular, $\sigma_1^{(\alpha)} $ can be still different from the value approached at infinite times after the quench. 

The first time window in \eqref{eq:scaling} corresponds to a well-known limit \cite{CC:05} and will not be discussed here. On the other hand, times $t\sim\Delta^{-1}$ have not yet been explored, so we will focus on the limit $\Delta\ll 1 $ with $\Delta t\sim O(\Delta^0)$. 
We will give a semi-classical interpretation to the scaling law~\eqref{eq:scaling} and  demonstrate the formula both numerically and, in some cases, analytically, by deriving an asymptotic analytic expression for $S_\ell/\ell$ (\emph{cf.} \eqref{eq:Sellstar}).

\paragraph{Organization of the paper.}%

Section~\ref{s:non-abelian} is a concise review of what is meant by non-abelian integrability and what are the expected effects on quench dynamics. In Section~\ref{s:pre-relax} we summarize the results of Refs~[\onlinecite{F:super}][\onlinecite{BF:mf}], which are focussed on the dynamics in an intermediate time window called pre-relaxation limit. Section~\ref{s:picture} provides a physical interpretation of \eqref{eq:scaling}, based on very general arguments. The subsequent sections are instead devoted to the explicit solution of the dynamics in the model with Hamiltonian \eqref{eq:H}. Specifically, Section~\ref{s:state0} describes the initial state. 
Section~\ref{s:Neel} is devoted to the analytical and numerical analysis of the pre-relaxation limit after quenches from the N\'eel state. In Section~\ref{s:oscillations} dynamics characterized by persistent oscillations are considered. We summarize our results in Section~\ref{s:conclusions}. Secondary aspects and some details of the iTEBD simulations are reported in the supplemental material~[\onlinecite{SM}].

\section{Non-abelian integrability}\label{s:non-abelian} %

The notion of integrability for a quantum model like \eqref{eq:HD} is deeply connected with  its conservation laws. Using algebraic Bethe ansatz~\cite{KIB:ABA}, one can indeed construct an infinite family of local\cite{f:0} conservation laws $Q_j$ in involution with one another
\be\label{eq:abelian}
[Q_j,H_\Delta]=0,\qquad [Q_i,Q_j]=0\, .
\ee
In principle, there could be other independent \mbox{(quasi-)local} charges\cite{P:XXZbc,PPSA:XXZ,IMP:XXX}, however, 
in the absence of manifest symmetries (like $SU(2)$ for $\gamma=0$ and $\Delta=1$), the set of local conservation laws is generally assumed to be abelian \eqref{eq:abelian}. 

As a matter of fact, there are exceptions in which hidden symmetries generate an infinite number of additional local conservation laws, the simplest example being the quantum XY model, with Hamiltonian $H_0$ \cite{F:super}. The XY model can be mapped to free fermions and the dispersion relation satisfies $\varepsilon_k=\varepsilon_{k+\pi}$. This in turn produces an infinite number of local conservation laws $Q_j$ that break one-site shift invariance and do not commute with one another
\be\label{eq:nonabelian}
[Q_j,H_0]=0,\qquad i [Q_i,Q_j]=f_{i j k}Q_k\, .
\ee 
We stress that, relaxing the requirement of locality, the existence of noncommuting conservation laws is not exceptional. For example, any reflection symmetric spin model that can be mapped to noninteracting fermions has a dispersion relation with the symmetry  $\varepsilon_k=\varepsilon_{-k}$. 
This generates an infinite number of noncommuting charges, which however are nonlocal. 
On the other hand, the non-abelian set of charges of the XY model is \emph{local} and \emph{infinitely} larger than the abelian set found in the absence of the symmetry $\varepsilon_k=\varepsilon_{k+\pi}$, justifying the appellation of ``non-abelian integrable model''. 

The immediate consequence of \eqref{eq:nonabelian} is that  in the thermodynamic limit not all local charges are needed to classify the states.
Any maximal abelian subset of $\{Q_j\}$ can be used to that aim, and indeed the classification is not unique but depends on the subset.

\subsection{Local charges in noninteracting spin chains}\label{ss:Q} %

Spin-$\frac{1}{2}$ models that can be mapped to noninteracting fermionic chains can be investigated with specific techniques that, to some extent, go beyond the standard Bethe ansatz solution of the model. Here we briefly remind the reader of the free-fermion formalism to construct the local conservation laws. We focus on one-site shift invariant models that are mapped to noninteracting fermions by the Jordan-Wigner transformation (here written in terms of Majorana fermions $\{a_\ell,a_n\}=2\delta_{\ell n}$)
\be\label{eq:J-W}
a_{2\ell-1}=\prod_{j<\ell}\sigma_j^z \sigma_\ell^{x}\qquad a_{2\ell}=\prod_{j<\ell}\sigma_j^z \sigma_\ell^{y}\, .
\ee
This means that, apart from boundary terms (which are irrelevant, if neglected both in the mapping from spins to fermions and in the reverse mapping to the spins) the noninteracting Hamiltonian $\tilde{H}$ can be written as
\be\label{eq:Hsymbol}
\tilde H\sim \frac{1}{4}\sum_{i, j=1}^{2L}a_i \mathcal {\tilde H}_{i j}a_j\, ,
\ee
where $\mathcal {\tilde H}$ is a block circulant matrix with 2-by-2 blocks
\begin{multline}
\mathcal {\tilde H}^{(1)}_{\ell n}=\begin{pmatrix}
\mathcal {\tilde H}_{2\ell-1, 2n-1}&\mathcal {\tilde H}_{2\ell, 2n-1}\\
\mathcal {\tilde H}_{2\ell-1, 2n}&\mathcal {\tilde H}_{2\ell, 2n}
\end{pmatrix}\\
=\frac{1}{L}\sum_{k|e^{i k L}=1}\mathcal {\tilde H}^{(1)}(k) e^{i(\ell-n) k}
\end{multline}
with $\ell,n\in \{1,\hdots, L\}$. The sum is over inequivalent momenta of the form $\frac{2\pi n}{L}$, with $n$ integer (in the thermodynamic limit the quantization rule is not important). 
The 2-by-2 matrix $\mathcal {\tilde H}^{(1)}(k)$ is usually called \emph{symbol}. If $s$ is a divisor of $L$ we can also define the $s$-site representation $\mathcal {\tilde H}^{(s)}(k)$ of the symbol, which is obtained by gathering together blocks of $s$ sites (reducing thus the effective size of the chain $L\rightarrow \frac{L}{s}$ at the price of increasing the local space):
\be\label{eq:ssymb}
\mathcal {\tilde H}^{(s)}_{\ell n}=\frac{s}{L}\sum_{k|e^{i k L/s}=1}\mathcal {\tilde H}^{(s)}(k) e^{i(\ell-n) k}\, .
\ee
Here $\ell, n\in \{1,\hdots, L/s\}$ and $\mathcal {\tilde H}^{(s)}(k)$ is a $(2s)$-by-$(2s)$ matrix with the following properties:
\be
 {{\mathcal{\tilde H}}^{(s)\dag}}(k)=\mathcal {\tilde H}^{(s)}(k)\qquad  {\mathcal {\tilde H}^{(s)t}}(k)=-\mathcal {\tilde H}^{(s)}(-k)\, .
\ee
To be explicit, $\mathcal{\tilde H}_{i+2s(l-1),j+2s(n-1)} = (\mathcal {\tilde H}^{(s)}_{\ell n} )_{i,j}$.

A conservation law $Q$ that can be mapped to noninteracting fermions by \eqref{eq:J-W} and that is translation invariant by $s$ sites has a symbol $\mathcal Q^{(s)}(k)$ that commutes with $\mathcal {\tilde H}^{(s)}(k)$. The conservation law is local if $\mathcal Q^{(s)}(k)$ has a finite number of nonzero Fourier coefficients. 
Remarkably, if the spectrum of $\mathcal {\tilde H}^{(s)}(k)$ is degenerate, one can generally find more than $2s$ linearly independent symbols $\mathcal Q^{(s)}(k)$ commuting with $\mathcal {\tilde H}^{(s)}(k)$. Any time that this happens independently of $k$ the Hamiltonian has an infinite non-abelian set of local conservation laws, infinitely larger than the set that arises when $\mathcal {\tilde H}^{(s)}(k)$ is non-degenerate. We refer the reader to Ref.~[\onlinecite{F:super}] for a more extensive discussion. 

To make the notation simpler, since we are going to consider the time evolution of two-site shift invariant states, in the rest of the paper the superscript identifying the representation ${}^{(s)}$ will be omitted in the two-site representation ($s=2$) of the symbols: $\mathcal Q^{(2)}(k)\rightarrow \mathcal Q(k)$. 

\subsection{Stationary properties after a global quench} %

The additional conservation laws can play a crucial role in non-equilibrium time evolution. The initial state somehow selects the maximal abelian subset of $\{Q_j\}$ that will be relevant \cite{F:super}.
The mechanism is essentially the following.  Noncommuting conservation laws generate degeneracy in the spectrum.  
There are ``privileged'' bases in which the Hamiltonian is diagonal and in each degenerate energy level a single state has a nonzero overlap with the initial state: the dynamics are essentially the same as in the non-degenerate case, provided that the Hilbert space is properly reduced.
From this point of view, we can identify the relevant conservation laws as those that are diagonal in a ``privileged'' basis.

Using the standard arguments behind the emergence of stationary behavior\cite{C-08,FE:13},  
it is reasonable to expect that at late times local degrees of freedom will relax to stationary values that can be described by a GGE
\be
\rho_{\rm GGE}=\frac{1}{\tilde Z}e^{-\sum_j\tilde \lambda_ j\tilde Q_j}\, .
\ee
Here $\tilde \lambda_j$ are Lagrange multipliers fixed by the integrals of motion and $\tilde Q_j$ are relevant charges, which belong to an abelian subset of $\{Q_j\}$.
\be\label{eq:subset}
[\tilde Q_i,\tilde Q_j]=0\qquad \tilde Q_j=\sum_\ell A_{j\ell} Q_\ell\, .
\ee
The  real rectangular infinite matrix $A$  depends on the initial state, which indeed has a twofold effect, determining both the Lagrange multipliers $\tilde\lambda_j$ and the charges $\tilde Q_j$. 
Alternatively, one could express the ensemble  in terms of the whole set of linearly independent local conservation laws
\be\label{eq:GGE*}
\rho_{\rm GGE}=\frac{1}{Z}e^{-\sum_\ell \lambda_\ell Q_\ell}\, .
\ee 
In this way one removes the subtle dependence of the operators on the initial state with the drawback of having a representation in terms of noncommuting operators.

So far, non-abelian integrability has simply resulted in a (complicated) reduction of the Hilbert space. There are however nontrivial effects due to the noncommutativity of local charges.
We focus on the interesting situation in which a small perturbation breaks the hidden symmetries behind the additional conservation laws.
Roughly speaking, two phenomena compete with each other: the perturbation tends to approximately preserve a subset of the conservation laws, which is however different from the abelian set selected (in the sense of \eqref{eq:subset}) by the initial state in the absence of perturbations.  This gives rise to non-trivial time evolution that is disclosed in long timescales. 

This phenomenon can be partially understood in the framework of stationary perturbation theory in quantum mechanics~\cite{CT:QM} as follows.
Let us choose a ``privilaged basis'' of eigenstates $\ket{\bar \Psi^{(0)}_{n,i}}$ of the unperturbed Hamiltonian, where $n$ identifies the energy level and $i$ the possible degenerate states. 
The perturbation introduces $O(\Delta)$ corrections to energies and eigenfunctions. In non-degenerate energy levels, up to phases, the eigenfunctions remain ``close'' to their original values also at times $t\sim O(\Delta^{-1})$. For degenerate states the situation is instead different: the perturbation splits each degenerate energy level (like in the Stark/Zeeman effect) in such a way that the energies are approximately equal to the unperturbed ones ($E_{n,i}\sim E_n^{(0)}+\Delta W_{n,i}$) but the eigenfunctions mix within the subspace. At times $t\sim O(\Delta^{-1})$, this results in $O(\Delta^0)$  discrepancies between $ \ket{\bar \Psi^{(0)}_{n,i}}$  and the corresponding time evolving state:
\be
\ba
&\ket{\Psi_{n,i}(t)}\sim e^{-i \Delta W_{n,i}t}\ket{\Psi_{n,i}}\\
&\ket{\bar \Psi^{(0)}_{n,i}(t)}\sim \sum_{j,j'} e^{-i \Delta W_{n,j}t}(R^{(n)}_{j i})^*R^{(n)}_{j j'}\ket{\bar \Psi^{(0)}_{n,j'}}.
\ea
\ee
Here we indicated with $ \ket{\Psi_{n,i}}\sim\sum_j R^{(n)}_{i j}\ket{\bar \Psi^{(0)}_{n,j}}+O(\Delta)$ the eigenfunctions that diagonalize the full Hamiltonian. 

This analogy fails however in explaining the importance of locality and infiniteness of non-commuting charges in the case of a quantum \emph{many-body} system. 

\section{A pre-relaxation limit}\label{s:pre-relax}  %

If the quantum XY model is perturbed so as to break non-abelian integrability, \emph{e.g.} with a term proportional to $S^z$, the expectation values of local observables experience a crossover\cite{F:super} in a time window that scales as $\Delta^{-1} $.  In generic models, the appearance of plateaux long before the typical relaxation times is commonly referred to as \emph{prethermalization}\cite{Getal1,Getal2,QHubbard1,QHubbard2,metaOpt,Kollar,IsingNI,E:preT,BEGR:preT,NI:15,BDK:15}. In Refs~[\onlinecite{F:super}][\onlinecite{BF:mf}] the crossover at intermediate times was instead called \emph{pre-relaxation limit}, in order to emphasize  that it is not driven by the breaking of integrability, but rather by the breaking of the hidden symmetries behind \emph{non-abelian} integrability; as a result, plateaux at intermediate times can be observed also in integrable models.

For noninteracting perturbations one can rigorously show \cite{F:super} that, in the limit $\Delta\ll 1$ with $\Delta t\sim O(\Delta^0)$, local degrees of freedom  are described by a time-dependent GGE of the form \eqref{eq:GGE*}, with time-dependent Lagrange multipliers $\lambda_\ell(\Delta t)$:
\be\label{eq:GGEt}
e^{-i H t}\ket{\Psi_0}\bra{\Psi_0}e^{i H t}\sim  \frac{e^{-\sum_\ell \lambda_\ell(\Delta t) Q_\ell}}{Z}+O(\Delta)\, .
\ee 

In Ref.~[\onlinecite{BF:mf}] this picture was generalized to interacting perturbations and, under a few assumptions, the same result has been obtained. 
The crucial difference from the noninteracting case is related to the effective description of the dynamics, which is now generated by a time-dependent (noninteracting) Hamiltonian.

\subsection{Mean-field solution} %

The effective theory proposed in Ref.~[\onlinecite{BF:mf}] is like a mean-field approximation that takes into account the presence of conservation laws breaking one-site shift invariance. However, it is not a real approximation, being instead an emergent description valid in the pre-relaxation limit. 

Specifically, it was shown that in the limit of small $\Delta$ and finite $\Delta t$, 
the time evolution under \eqref{eq:H} of local observables, starting from a two-site shift invariant state with \emph{cluster decomposition properties}, can be written as follows 
\be\label{eq:O1}
\braket{\mathcal O}(t)\xrightarrow{1\ll t\sim O(\Delta^{-1})} \braket{\mathcal O}_{{\rm GGE}_{0}}^{\rm MF}(\Delta t)+o(\Delta^0)\, ,
\ee
where the range of the operator must be small in comparison with $t$ and $1/\Delta$. 
In \eqref{eq:O1} $\braket{\mathcal O}^{\rm MF}_{{\rm GGE}_0}(\Delta t)$ is the expectation value of the observable $\mathcal O$ in the mixed state $\rho^{\rm MF}(\Delta t)$, time evolving under a quasi-local time-dependent noninteracting Hamiltonian 
$H^{\rm MF}(\Delta t)$ and equivalent, at the initial time, to the stationary state that emerges  for $\Delta=0$:
\be\label{eq:rho0}
\ba
& \braket{\mathcal O}_{{\rm GGE}_{0}}^{\rm MF}(T)=\mathrm{Tr}[\rho^{\rm MF}(T) \mathcal O]\\
&i \partial_{T} \rho^{\rm MF}(T)= [H^{\rm MF}(T),\rho^{\rm MF}(T)]\\
&\rho^{\rm MF}(0)=\lim_{|A|\rightarrow\infty}\lim_{t\rightarrow\infty}\mathrm{Tr}_{\bar A}[e^{i H_0 t}\ket{\Psi_0}\bra{\Psi_0}e^{-i H_0 t}]\, .
\ea
\ee
We introduced the notation $T=\Delta t$ to help one identify the slowly evolving quantities. 
The mean-field Hamiltonian can be formally written as follows~\cite{f:m1}
\be\label{eq:HMFform}
\ba
H^{\rm MF}(T)=\sum_j \mu_j(T) Q_j,&&\mu_j(T)=\Braket{\frac{\delta \bar V}{\delta Q_j}}_{\!\!\rm GGE_0}^{\!\!\rm MF}\!\!\!\!\!\!\!\!\!(T)\, ,
\ea
\ee
where the sum is over linearly independent local conservation laws of the unperturbed Hamiltonian and $\bar V$ is the time-averaged perturbation in the interacting picture
\be
\bar V=\lim_{t\rightarrow\infty}\frac{1}{t}\int_0^t\mathrm d \tau\ e^{i H_0\tau}\ \frac{\partial H}{\partial \Delta}\Bigr|_{\Delta=0}\ e^{-i H_0 \tau}\, .
\ee
The compact notation of \eqref{eq:HMFform} means that we take only the part of the time average $\bar V$ that depends on the local conservation laws and then work out the various terms in a mean-field fashion.
There are indeed also purely nonlocal terms in $\bar V$, which however are not expected to contribute at these time scales\cite{BF:mf}. The remaining terms are polynomials of the local charges, so \eqref{eq:HMFform} simply means that, in each product of charges, all but one charge, in turn, contribute as simple c-numbers. 

We would like to place some emphasis upon the efficacy of the mean-field mapping: quench dynamics in an interacting model are reduced to noninteracting (though complicated) dynamics!

\subsection{Generalities}\label{ss:pre-relaxB} %

We collect here some useful observations.
\begin{description}

\item[\small \emph{Ensemble representation}]\hfill\\
Since everything is written in terms of the local conservation laws of $H_0$, the mean-field description is consistent with \eqref{eq:GGEt}. Indeed the time evolution of a local charge (of the unperturbed model) under \eqref{eq:HMFform} commutes with the unperturbed Hamiltonian (\emph{cf.}~\eqref{eq:nonabelian}) and hence the mean-field dynamics are reduced to time evolving Lagrange multipliers~\eqref{eq:GGEt}.

\item[\small \emph{Paradox of relaxation in closed systems}]\hfill\\
In the limit $1\ll \ell\ll t$  the entanglement entropy per unit length  of a subsystem is generally equivalent to the (thermodynamic) entropy density of the GGE describing the late time dynamics\cite{CKC:2014}. Since mean-field time evolution is unitary, one could expect the entropy of the \mbox{(quasi-)stationary} state to be conserved. 
In fact, this is the same paradox arising in the non-equilibrium time evolution of pure states: despite the system being closed, at long times reduced density matrices can be described by more entangled mixed states (\emph{e.g.} Gibbs ensembles in generic models).
Rather counterintuitively, this means that if any local observable relaxes in the pre-relaxation limit, the  mixed state that describes the stationary properties has the same form as \eqref{eq:GGEt}, but with a different partition function $Z$. 

\item[\small \emph{One-site shift invariance}]\hfill\\
It is important to note that the pre-relaxation limit for \eqref{eq:H} can be nontrivial only if the initial state breaks one-site shift invariance (1-ssi).  
This can be seen as follows. 
The unperturbed GGE corresponding to a \mbox{1-ssi} initial state is \mbox{1-ssi}, being also the Hamiltonian \eqref{eq:H} \mbox{1-ssi}.
Analogously, the time averaged perturbation is  \mbox{1-ssi}.
On the other hand, the conservation laws that break \mbox{1-ssi} are odd under translation by one site \cite{BF:mf} $Q_i\rightarrow -Q_i$. 
From this it follows that a generic term of the polynomial part of $\bar V$ is generally written as $Q_1 Q_2\cdots Q_n$, where $Q_{i}$ and $\prod_{j\neq i}Q_j$ take the same sign under a shift by one site. By symmetry, the expectation value of any odd operator is zero, therefore in \eqref{eq:HMFform} only the terms proportional to \mbox{1-ssi} charges give a non zero contribution: \emph{the mean-field Hamiltonian is one-site shift invariant}.  
We now remind the reader that the \mbox{1-ssi} charges of the quantum XY model commute with one another\cite{F:super}. Consequently \mbox{$[H^{\rm MF}(T),\rho^{\rm MF}(T)]=0$} and  $\rho^{\rm MF}(T)=\rho^{\rm MF}(0)$. In other words, the expectation values do not move from the values corresponding to the first plateau. 

\item[\small \emph{On the number of non-commuting charges}] \hfill\\
One could wonder whether a \emph{finite} dimensional non-abelian subalgebra of local conservation laws could be sufficient to trigger a crossover in the pre-relaxation limit of the entanglement entropy.
In that case mean-field time evolution would result in oscillations in a finite number of Lagrange multipliers of the time-dependent GGE \eqref{eq:GGE*} (the ones corresponding to the non-abelian subset of charges). 
This is like a quantum problem with a finite number of degrees of freedom, so time evolution is expected to be quasi-periodic, with recurrence times dependent on the subalgebra dimension and not significantly affected by the subsystem length. The oscillations should not modify the extensive part of the entanglement entropies, which are in turn expected not to move from the first plateau (see also \ref{p:osc}).
As a result, any nontrivial crossover in the entanglement entropies can be seen as an effect of the unperturbed model having an infinite number of non-commuting local charges. 

\item[\small \emph{Problem setting}]\hfill\label{d:ps}\\
For Hamiltonians of the form \eqref{eq:H}, in which the interaction terms consist of four fermion operators, 
the (mean-field) dynamics of local observables are reduced to the solution of an infinite system of quadratic first order differential equations, with an extra linear term originated by the noninteracting perturbation $S^z$.
The system of differential equations can be derived as follows. 
First of all one has to compute the time average of the perturbation. Then, the purely nonlocal contributions are removed and the mean-field differential equations for the time evolution of the local integrals of motion of the unperturbed model are set up. This  determines the mean-field Hamiltonian in a self-consistent way.  
Being $\rho^{\rm MF}(T)$ gaussian, the expectation value of any local observable can be obtained from the two-point fermionic correlation function 
($\Gamma_{i j}=\delta_{i j}-\braket{a_i a_j}$) using the Wick's theorem. Indeed
one can show that $\Gamma$ is written in terms of the integrals of motion (of the unperturbed model); therefore, it can be extracted from the solution of the system of differential equations above. We point out that, in order to deal with the infinite system, the introduction of a cut-off is unavoidable; this can be done for example confining the system in a fictitious finite box. If the cut-off has been chosen in a sensible way the expectation values of local observables will be stable under its variation. 

We refer the reader to Ref.~[\onlinecite{BF:mf}] for the explicit derivation of the differential equations and for a preliminary analysis of their solutions. In this paper we will only exhibit the equations in a specific example (Section \ref{s:Neel}). 

\item[\small \emph{Assumptions and checks}]\hfill\\
 For interacting perturbations, \eqref{eq:O1} was derived under two assumptions: the first is similar to the hypothesis behind perturbation theory in quantum mechanics, in particular the non-diagonal elements of the perturbation must approach zero as the corresponding unperturbed energy differences vanish. The second is that the purely nonlocal part of the time averaged perturbation is not relevant for the time evolution of local observables in the pre-relaxation limit (Ref.~[\onlinecite{BF:mf}] established that this approximation is self-consistent). 

We will give grounds for \eqref{eq:O1}  by checking the mean-field predictions against iTEBD simulations.
This will be the most delicate point of this work, being difficult to reliably simulate long times.

\end{description}

\section{Physical interpretation}\label{s:picture}  %

\setcounter{paragraph}{0}
\begin{figure}[tbp]
\begin{center}
\includegraphics[width=0.48\textwidth]{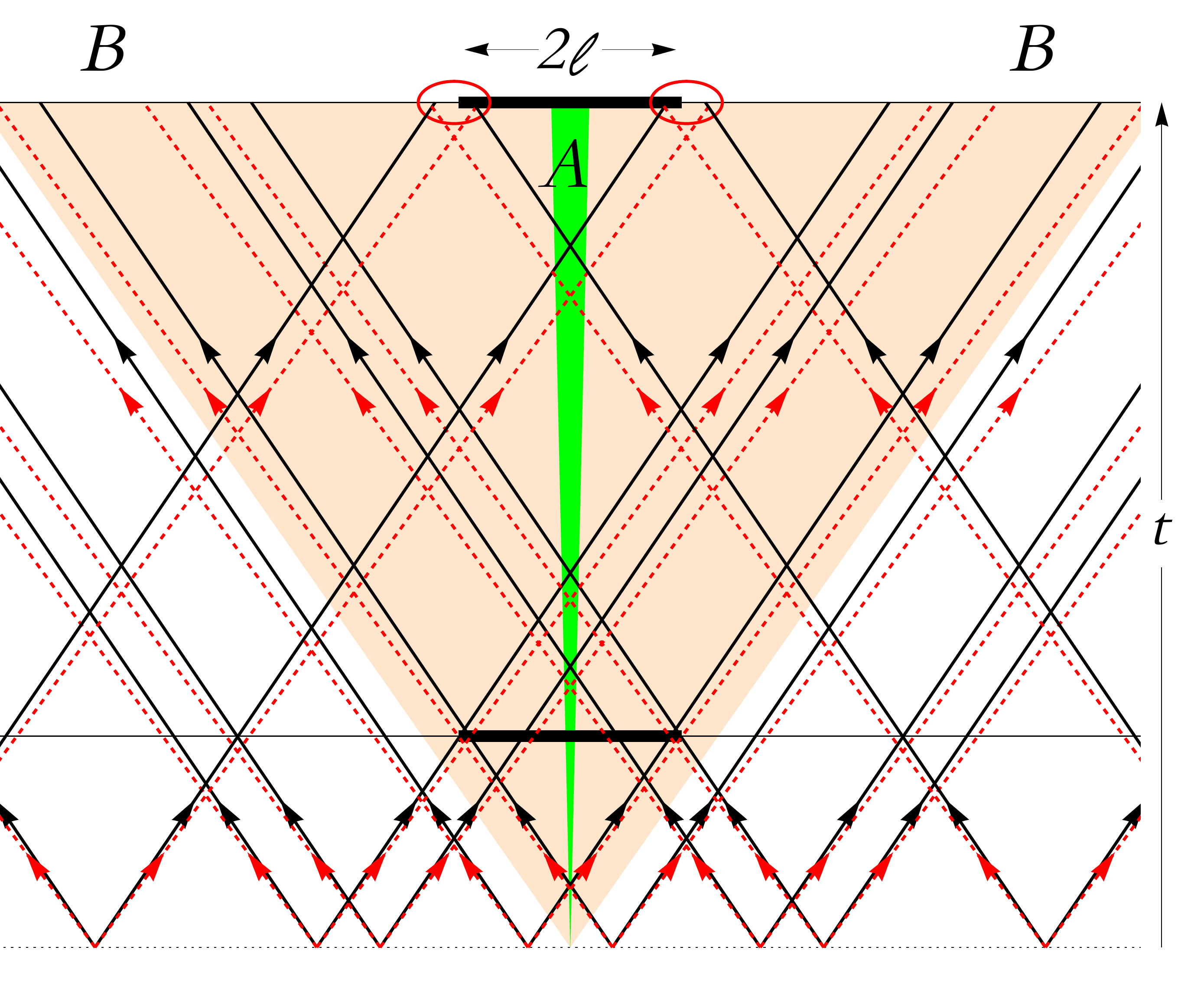}
\end{center}\caption{\label{f:1}Simplified space-time picture illustrating the scaling behavior \eqref{eq:scaling} of the entanglement entropy between an interval $A$ and the rest of the system $B$. 
The non-abelian integrability of the unperturbed model is manifested by the fact that at the initial time \emph{two} pairs of  quasiparticles (black and red arrows), \emph{correlated} with one another, with almost identical velocity and traveling in both directions are emitted from each point (which represents two sites). The light orange region (biggest background triangle) shows the subsystems (symmetric about the centre) that can be approximately described by the stationary state emerging in the absence of perturbations (for which the velocities of the black and red quasiparticles would have been exactly equal). In that regime only the correlations between the pairs of particles traveling in opposite directions are important (see \emph{e.g.} the subsystem, horizontal thick line, at earlier times). 
However, at larger times the contributions from the cases (highlighted by red circles at the top of the diagram) in which a single particle (of the four) crosses the subsystem are no more negligible. 
Such contributions increase linearly in time up to the time at which no \emph{pair} of quasiparticles originated at the same point can cross the subsystem together.  After that a new equilibrium is reached and the subsystems can be approximately described by a stationary state that in some cases is the GGE of the full Hamiltonian (the green region, namely the smallest background triangle). 
The pre-relaxation limit describes the crossover between the light orange region and the green one.
}
\end{figure}

Let us indicate with $r_H$ the typical range (generally independent of time) of $H^{\rm MF}(\Delta t)$ \eqref{eq:HMFform}.

 Ref.~[\onlinecite{BF:mf}] pointed out that, in the pre-relaxation limit, the system can experience either local relaxation or persistent oscillatory behavior. 
 \subsection{Local relaxation}
In the hypothesis of local relaxation, at late (rescaled) times $\Delta t$, $H^{\rm MF}$ becomes independent of time. Concomitantly, the expectation values of operators with larger ranges start relaxing to stationary values over a timescale $\Delta t$ that is expected to be proportional to their range $\ell$. As a matter of fact, this is analogue of what happens in the space-time scaling limit \mbox{$1\ll \ell\sim t$} considered in Ref.~[\onlinecite{CC:05}]. 
There, the qualitative behavior of the entropy has been understood  through a semiclassical reasoning that we briefly repeat here.

Being the energy extensively higher than the ground state one, the initial state acts like a source of pairs of
quasiparticle excitations traveling in opposite directions.
By translation invariance they are originated everywhere. 
Because the initial state is low entangled, one can assume quasiparticles originated from different points to be incoherent. 
There is also another assumption, namely distinct pairs of quasiparticles originated from the same point are not entangled. 
In noninteracting quenches with one-site shift invariance one can convince oneself of this property by simply writing the initial state in a basis that diagonalizes the Hamiltonian. 
More generally, the incoherence of pairs does not seem to be heuristically justifiable, but it is in fact appropriate to a large class of quenches.
Thus, taking that assumption for granted, the qualitative behavior of the entanglement entropy can be understood in terms of the motion of the quasiparticle excitations.  Since the entropy measures the entanglement between a subsystem and the rest, it should be proportional to the number of entangled quasiparticles that, at a given time $t$, are both inside and outside the subsystem. This simplified picture explains the linear growth followed by saturation observed in the space-time scaling limit $1\ll \ell\sim t$, at least in the integrable case. 

We borrow this physical interpretation  
and propose some modifications that allow us to understand the behavior claimed in \eqref{eq:scaling}.
Since now translation invariance is realized by a two-site shift, each point of the effective description corresponds to two sites and, in turn, the number of quasi-particles originated from the same point is doubled. 
Non-abelian integrability (of the unperturbed model) is manifested by the presence of entanglement between pairs of pairs of quasiparticles.  This is a consequence of the freedom in the choice of the basis diagonalizing the Hamiltonian and identifying different abelian subsets of charges. The change of basis is indeed equivalent to mixing  pairs of quasiparticles. 

As depicted in Fig. \ref{f:1} and explained in its caption, this results in linear behavior followed by saturation also in the pre-relaxation limit. There are however some differences with respect to the space-time scaling limit:

\begin{enumerate}

\item \label{i:1}Being $\Delta\ll 1$, at times $t\sim\Delta^{-1}$ the initial state can be replaced by the stationary state that locally emerges when time evolution is generated by the unperturbed Hamiltonian $H_0$, \emph{i.e.} $\rho^{\rm MF}(0)$ in \eqref{eq:rho0} (in Fig.~\ref{f:1},  
quasiparticles that contribute to the crossover in the entropy pass through the light orange region).

\item After the substitution of Point \ref{i:1}, the relevant evolution occurs in a rescaled time $\Delta t$ (in Fig. \ref{f:1}, the angle between the black and red arrows is $\mathcal O(\Delta)$).

\item The extent of the linear regime is not related to the maximal velocity at which information propagates but a new velocity emerges (in Fig. \ref{f:1}, the angle between red and black arrows per unit of $\Delta$ is not in relation to the direction of the arrows). 

\end{enumerate}

\subsection{Oscillations}\label{p:osc}  %

While for noninteracting perturbations like $S^z$ relaxation is the norm, it is very common to find persistent oscillatory behavior when the perturbations consist of interacting terms. If we insist on the interpretation sketched in Figure \ref{f:1}, we realize that oscillations should not affect the behavior of the extensive part of the entropy. Indeed we can imagine to deform the quasiparticle trajectories
including some oscillatory behavior. The amplitude of the oscillations would depend on the quench parameters but not on the subsystem length. As a consequence, the effects of oscillations would be concentrated on the boundaries of the subsystem, modifying just the leading correction $O(\ell^0)$ to the entanglement entropies.  
This conjecture will be corroborated by a numerical analysis of the mean-field solution of the problem reported in Section \ref{s:oscillations}.

This is the last section of the first part of the paper, designed to give the reader the opportunity to gain a physical intuition about the problem under investigation.
In the following sections the specific non-equilibrium time evolution under \eqref{eq:H} will be considered and the mean-field solutions worked out in order to derive an analytic expression for the time evolution of the entanglement entropies. 

\section{The initial state}\label{s:state0} %

Mean-field dynamics are trivial if the initial state is invariant under translations by one site (\emph{cf}. \emph{One-site shift invariance} in \ref{ss:pre-relaxB}), so we must consider states that break such symmetry. We note however that the pre-quench Hamiltonian can be still one-site shift invariant: the symmetry turns out to be spontaneously broken in antiferromagnetic or dimerized phases. 
For the sake of simplicity,  we choose initial states with factorized form
\be\label{eq:state}
\ket{\Psi_0}=\bigotimes
\Bigl[\cos\frac{\beta}{2}\ket{\uparrow\downarrow}+e^{i\phi}\sin\frac{\beta}{2} \ket{\downarrow\uparrow}\Bigr]\, ,
\ee
where $\sigma^z\ket{\uparrow}=\ket{\uparrow}$ and $\sigma^z\ket{\downarrow}=-\ket{\downarrow}$.
It is not difficult to show that, for any value of $\beta$ and $\phi$, $\ket{\Psi_0}$ is a Slater determinant for the Jordan-Wigner fermions \eqref{eq:J-W}. 
Despite this property being unnecessary for \eqref{eq:O1}\eqref{eq:rho0}, it will be convenient when, in Section \ref{ss:intMF}, we will introduce an intermediate mean-field description. 

There are two relevant states in the class \eqref{eq:state}:
\begin{itemize}
\item $\beta=0\vee \pi$: N\'eel state (the ground state of $H_\infty$);
\item $(\beta,\phi)=(\frac{\pi}{2},\pi)$: product state of singlets (the ground state of the Majumdar-Gosh Hamiltonian, \emph{i.e.} \eqref{eq:H} with $\gamma=0$, $h=0$, $\Delta=1$ and $U=\frac{1}{2}$).
\end{itemize}

The correlation matrix 
\be
\Gamma_{\ell n}=\delta_{\ell n}-\braket{\Psi_0|a_{\ell} a_n|\Psi_0}
\ee 
of \eqref{eq:state} is a 4-by-4 block circulant matrix. As any block-circulant matrix, it is completely characterized by its symbol, \emph{i.e.} the Fourier transform of a block-row (\emph{cf.} \eqref{eq:ssymb}). This is given by
\be\label{eq:Gamma0}
\Gamma(k,0)=\cos\beta\, \sigma^z \otimes \sigma^y-\sin\beta\bigl(\cos\phi\, \sigma^x \otimes \sigma^y+\sin\phi\, \sigma^y\otimes\mathrm I\bigr)\, .
\ee 
In the pre-relaxation limit the part of the dynamics with a typical relaxation time much smaller than $1/\Delta$ has already reached (quasi-)stationarity. 
As a result, the correlation matrix can be replaced by the one corresponding to the stationary state for $\Delta=0$. This can be done by projecting \eqref{eq:Gamma0} onto the symbols of the local conservation laws of the quantum XY model $H_0$ \eqref{eq:HD}, which results in
\begin{multline}\label{eq:Gamma0MF}
\Gamma^{\rm MF}(k,0)=\\
\frac{\gamma\cos\beta \mathcal Q_8(k)-\sin\beta\cos\phi\bigl(\cos^2\frac{k}{2} \mathcal Q_1(k)-\gamma  \sin^2\frac{k}{2} \mathcal Q_5(k)\bigr)}{\cos^2\frac{k}{2}+\gamma^2\sin^2\frac{k}{2}}\\
-\sin\beta\sin\phi \mathcal Q_4(k)\, .
\end{multline}
Here $\mathcal Q_j(k)$ identify the eight families of local conservation laws of $H_0$:
\be
\ba
 \mathcal{Q}_1(k)&=\varepsilon_k\,\bigl[\sigma^xe^{i\frac{k}{2}\sigma^z}\bigr]\otimes\bigr[\sigma^y e^{i\theta(k)\sigma^z}\bigr]\\
\mathcal{Q}_2(k)&=\cos(k/2) \varepsilon_k \,\1\otimes\bigl[\sigma^y e^{i\theta(k)\sigma^z}\bigr]\\
\mathcal{Q}_3(k)&=\sin( k)\, \1\otimes\1 \\
\mathcal{Q}_4(k)&=\sin(k/2)\,\bigl[\sigma^xe^{i\frac{k}{2}\sigma^z}\bigr]\otimes\1\\
\mathcal{Q}_5(k)&=\varepsilon_k\, \bigl[\sigma^ye^{i\frac{k}{2}\sigma^z}\bigr]\otimes\bigl[\sigma^x e^{i\theta(k)\sigma^z}\bigl]\\
\mathcal{Q}_6(k)&=\cos(k/2)\,\bigl[\sigma^ye^{i\frac{k}{2}\sigma^z}\bigr]\otimes\sigma^z\\
\mathcal{Q}_7(k)&=\sin(k)\,\sigma^z\otimes\sigma^z \\
\mathcal{Q}_8(k)&=\sin(k/2)\varepsilon_k\,\sigma^z\otimes\bigr[\sigma^x e^{i\theta(k)\sigma^z}\bigl]\label{eq:Q}\, ,
\ea
\ee
where $\tan\theta_k=\gamma \tan\frac{k}{2}$ and $\varepsilon_k=(\cos^2\frac{k}{2}+\gamma^2\sin^2\frac{k}{2})^{1/2}$ is the dispersion relation of the two species of quasiparticles that diagonalize the unperturbed model in the representation in which the local space consists of two sites\cite{f:3}. 
The symbols of a generic local charge can indeed be written as $\cos(nk)\mathcal Q_j(k)$, with integer $n$. We point out that for  $j\leq 4$ the associated conservation laws are one-site shift invariant (they correspond to the two families of local charges identified in Ref.~[\onlinecite{GM:charges}]). 

Replacing $\Gamma(k,0)$ by $\Gamma^{\rm MF}(k,0)$ reflects the third equation of \eqref{eq:rho0}.
 
The mean-field description then establishes that  the noninteracting structure is preserved at $O(\Delta^0)$ up to times $t\sim O(\Delta^{-1})$ and 
\emph{the symbol of the correlation matrix remains a linear combination of the symbols of $H_0$'s local charges}.

\section{Quenches from the N\'eel state}\label{s:Neel} %

Let us now consider in detail the case $\beta=0\vee \pi$ and $U=0$, namely the non-equilibrium time evolution of the N\'eel state under the Hamiltonian of the XYZ model with an integrability breaking perturbation proportional to $S^z$. 

The mean-field equations of motion have been worked out in Ref.~[\onlinecite{BF:mf}] following the \emph{problem setting} described in \ref{d:ps}. 
We do not repeat here the (manageable but tedious) calculations and simply report the results. 

For states that are reflection symmetric about a site, it turns out that the symbol of the correlation matrix of $\rho^{\rm MF}(T)$ \eqref{eq:rho0} can be written as follows
\be\label{eq:Gamma}
\Gamma^{\rm MF}(k,T)=\sum_{j\in\{2,7,8\}} \frac{8 y_j(k,T)}{\mathrm{Tr}[\mathcal Q_j^2(k)]}\mathcal Q_j(k)\, ,
\ee 
where $\mathcal Q_j(k)$ were defined in \eqref{eq:Q}. 

The functions $y_j(k,T)$ are solutions of the \mbox{(integro-)}differential equations (for $U=0$!) 
\begin{multline}\label{eq:sysNeel}
\partial_T y_j(k,T)=\sum_{\ell,n\in\{2,7,8\}}A_{j}^{\ell n} (k)\!\!\int\!\frac{\mathrm d p}{2\pi}\frac{y_\ell (p,T)}{\varepsilon_p^2} y_n(k,T)\\
+h \sum_{n\in\{2,7,8\}}\epsilon^{j 2 n}g_j(k) y_n(k,T)\, ,
\end{multline}
where $\epsilon^{i j k}$ is the (3d) totally antisymmetric Levi-Civita symbol with $\epsilon^{278}=1$.
The only nonzero elements of tensor $A$ are given by 
\be
\ba
&A_{8}^{2 7}(k)=2 &&A_{7}^{28}(k)=-\frac{8}{1+\gamma^2 \tan^2\frac{k}{2}}\\
&A_{2}^{8 7}(k)=2\gamma^2&&A_{7}^{8 2}(k)=-\frac{8\gamma^2}{\cot^2\frac{k}{2}+\gamma^2 }
\ea
\ee
while $g_j$ read as
\be
g_2(k)=0\, ,\quad g_7(k)=\frac{4}{1+\gamma^2\tan^2\frac{k}{2}}\, ,\quad g_8(k)=1\, .
\ee
The initial conditions are as follows (\emph{cf.} \eqref{eq:Gamma0MF} and \eqref{eq:Gamma})
\be\label{eq:initialcond}
\ba
&y_2(k,0)=y_7(k,0)=0\\
&y_8(k,0)=\pm \gamma\frac{1-\cos k}{4}\, .
\ea
\ee
where $+$ and $-$ correspond to $\beta=0$ and $\beta=\pi$, respectively, in \eqref{eq:Gamma0}. 

The magnetic field appears as a multiplicative factor of the linear term in \eqref{eq:sysNeel}. This is because $h$ is the coupling constant of $S^z$, which is a quadratic operator in the J-W fermions \eqref{eq:J-W}. Because only $y_{8}(k,0)$ is different from zero, the pre-relaxation limit is nontrivial only if $h\neq 0$ (at the initial time the quadratic part of \eqref{eq:sysNeel} is zero).  

We identify two non-interacting limits: $\Delta=0$ and $\Delta\rightarrow 0$ at fixed $\Delta h$. The former is trivial (without perturbation the expectation values can not move from the first plateau), while the latter corresponds to neglecting the quadratic part in \eqref{eq:sysNeel}.
In the second case, \eqref{eq:sysNeel} can be exactly solved and the solution is such that the Fourier coefficients of $y_j(k,t)$ relax to stationary values. This is the simplest nontrivial case of local relaxation in the pre-relaxation limit and was considered in Ref.~[\onlinecite{F:super}] (see also the supplemental material~[\onlinecite{SM}]). 
Any displacement from the two noninteracting limits above is an effect of the interaction. 
 
It is important to note that this formalism can not be applied to the shift symmetrised state $\frac{\ket{\cdots \uparrow\downarrow\cdots}+\ket{\cdots \downarrow\uparrow\cdots}}{\sqrt{2}}$ because the mean-field description strongly relies on \emph{cluster decomposition properties}. This is indeed a very clear example in which ignoring the importance of cluster decomposition has dramatic consequences: the (\emph{wrong!}) initial conditions for the shift symmetrised state (\emph{i.e.} the arithmetic mean of the initial conditions for $\beta=0$ and $\beta=\pi$) are identically zero, and hence time evolution would seem to be trivial in the pre-relaxation limit. Instead, considering the equations for $\beta=0,\pi$ separately, we will show that the shift symmetrizations of local operators like $s_\ell^z$ display nontrivial dynamics (\emph{cf.} Fig. \ref{f:mz}).

\subsection{Scaling of the entanglement entropies}
We now focus on cases of local relaxation in the limit $t\sim O(\Delta^{-1})$. 
Using the results of Ref.~[\onlinecite{BF:mf}] we obtain the symbol of the mean-field Hamiltonian 
\be\label{eq:Hsym}
\mathcal H_{T}^{\rm MF}(k)=\frac{(\frac{h}{2}+m_z(T))\mathcal Q_2(k)-\gamma  m_{z,s}(T)\mathcal Q_8(k)}{\frac{1+\cos k}{4}+\gamma^2\frac{1-\cos k}{4}}\, ;
\ee
$m_z(T)=\mathrm{Tr}[\rho(T)(\sigma_1^z+\sigma_2^z)]/2$ and $m_{z,s}(T)=\mathrm{Tr}[\rho(T)(-\sigma_1^z+\sigma_2^z)]/2$ are the magnetization and the staggered magnetization respectively
\be
\ba
m_z(T)&=\int_{-\pi}^\pi\frac{\mathrm d k}{2\pi}\frac{y_2(k,T)}{\varepsilon_k^2}\, ,\\
m_{z,s}(T)&=-\gamma \int_{-\pi}^\pi\frac{\mathrm d k}{2\pi}\frac{y_8(k,T)}{\varepsilon_k^2}\, .
\ea
\ee
One can verify that $\mathcal H^{\rm MF}(k,T)$ generates the time evolution \eqref{eq:sysNeel}: $i\partial_T \Gamma^{\rm MF}(k,T)=[\mathcal H^{\rm MF}(k,T), \Gamma^{\rm MF}(k,T)]$. 
By computing the characteristic length of the Fourier transform of \eqref{eq:Hsym} we infer\cite{f:4} that the typical range of $H^{\rm MF}$ is
$r_H\sim 2/\log|\frac{|\gamma|+1}{|\gamma|-1}|$.

Being essentially a noninteracting problem, one could wonder whether analytic expressions can be obtained for the time evolution of the entanglement entropies. 
There are two main complications: first, the solution of \eqref{eq:sysNeel} is not analytically known and, second, computing the non-equilibrium time evolution of the entanglement entropies is not a simple task even in models that can be mapped to noninteracting fermions. 
The first goal is to decouple the two problems, which have different nature.   

\subsubsection{On the explicit time dependence of $H^{\rm MF}$}\label{ss:VT}%
The major obstacle to attacking the problem analytically is that the mean-field  Hamiltonian depends explicitly on the time. In addition, the time dependence is not known \emph{a priori}, being related to the solution of an infinite system of nonlinear differential equations. At first glance, this problem might appear insurmountable; in fact, it can be overcome as follows.  

We define the unitary operator
\be
V_{T}(\tilde T)=e^{i \bar H^{\rm  MF} \tilde T}U_{\rm MF}(T+\tilde T)U^\dag_{\rm MF}(T)\, ,
\ee
where  $\bar H^{\rm  MF}=H^{\rm  MF}(\infty)$ and
\be
i\partial_{T} U_{\rm MF}(T)=H^{\rm MF}(T)U_{\rm MF}(T)\, .
\ee
In other words, $V_{T}(\tilde T)$ generates mean-field time evolution from $T$ to $T+\tilde T$ followed by a reverse evolution under $\bar H^{\rm  MF}$ from $T+\tilde T$ to $T$ . 
Let us assume that the mean-field Hamiltonian relaxes sufficiently fast for the existence of the following limits
\be\label{eq:Vql}
\ba
 \exists\lim_{\tilde T\rightarrow\infty }V_{T}(\tilde T)&= V_{T}&&& \exists \lim_{T\rightarrow\infty}V_{T}(\tilde T)&=\mathrm I\, .
 \ea
\ee
We can then recast the problem into the non-equilibrium evolution under $\bar H^{\rm  MF}$ of the \emph{extrapolated density matrix}
\be\label{eq:rhotilde0}
\tilde \rho(0)=V_0\rho^{\rm MF}(0)V_0^\dag\, .
\ee
Indeed, using the existence of the limits in \eqref{eq:Vql}, we find
\be
e^{-i \bar H^{\rm  MF} T}V_0=V_T U_{\rm MF}(T)\, ,
\ee 
and hence
\be\label{eq:rhotrhoMF}
\tilde \rho(T)\equiv e^{-i \bar H^{\rm  MF} T}\tilde\rho(0)e^{i \bar H^{\rm  MF} T}=V_{T}\rho^{\rm MF}(T)V_{T}^\dag\, .
\ee
Let us analyze how $V_T$ acts on operators. It is convenient to reinterpret $V_T$ as a time evolution operator as follows:
\be
i\partial_{\tilde T}V_{T}(\tilde T)=\mathfrak H_T(\tilde T)V_{T}(\tilde T)\, .
\ee
Here $\mathfrak H_T(\tau)$ is a quasi-local effective Hamiltonian (with range  increasing with $\tau$) that approaches zero in the limit of infinite $\tau$:
\be\label{eq:Hext}
\mathfrak H_T(\tau)=e^{i \bar H^{\rm  MF}(\tau-T)}H^{\rm MF}(\tau)e^{-i \bar H^{\rm  MF}(\tau-T)}-\bar H^{\rm  MF}\, .
\ee
We denote by $v(\tau)$ the (Lieb-Robinson) maximal velocity associated with $\mathfrak H_T(\tau)$ ($v(\tau)$ is independent of $T$). At fixed time $\tau$, in an arbitrarily small time interval $\mathrm d\tau$, the spreading of local operators under a quasi-local Hamiltonian is proportional to the time interval and to the maximal velocity at which information propagates (\emph{cf.} Ref.~[\onlinecite{BHV:06}]). This means that 
$V_T(\tilde T)$ is expected to map local operators into quasi-local operators with a typical diffusion length $\xi(T,\tilde T)\sim \int_T^{\tilde T}\mathrm d \tau v(\tau)$. 
We are going to assume that the limit $\xi(T)=\lim_{\tilde T\rightarrow\infty}\xi(T,\tilde T)$ exists and is finite.
This semiclassical estimate of $\xi(T)$ is a nonincreasing function of $T$, so we find the upper bound $\xi(T)\lesssim \xi(0)$. 

Let us now consider subsystems with length $\ell\gg \xi(0)$. 
From \eqref{eq:rhotrhoMF} and our estimate of the diffusion length $\xi(0)$ it follows that the RDMs of $\tilde \rho(T)$ are unitarily equivalent to the corresponding RDMs of $\rho^{\rm MF}(T)$, up to boundary terms that extend over the typical  length $\xi(0)$. This is independent both of the subsystem length and of $T$.
On the other hand, the extensive part of the entanglement entropy is not affected by  transformations  that are unitary in the bulk, therefore, in the limit of large subsystem length, the entanglement entropies per unit length of $\rho^{\rm MF}(T)$ must be equal to the ones of $\tilde \rho(T)$. In conclusion, we mapped the problem into a sudden quench from $\tilde \rho(0)$. 

We can sketch the steps of our reasoning as follows:
\be\label{eq:mappingsq}
(H;\ket{\Psi_0})\rightarrow (H^{\rm MF}(T);\rho_{\rm GGE_0}) \rightarrow (\bar H^{\rm  MF};\tilde \rho(0))\, ,
\ee
where we used the notation $(\Lambda;\Phi)$ to indicate the dynamics generated by Hamiltonian~$\Lambda$ starting from state~$\Phi$. The first step is the mapping to a mean-field problem, which is justified by locality in the pre-relaxation limit. The second step is the reduction of the complicated mean-field dynamics to a sudden quench, which can be instead justified only for the calculation of the extensive part of the entropies. 

After simple but tedious algebra we obtain the (extrapolated) correlation matrix $\tilde\Gamma(k,0)$ of $\tilde \rho(0)$
\begin{multline}\label{eq:Gammatilde}
\tilde\Gamma(k,0)=\lim_{\tilde t\rightarrow\infty } e^{i \mathcal V^{\rm MF}(k,\infty) \tilde t}\Gamma^{\rm MF}(k,\tilde t)e^{-i \mathcal V^{\rm MF}(k,\infty) \tilde t}\\
=e^{-i\frac{\Omega_k}{2\sin k}\mathcal Q_7(k)}\Bigl[\sum_{j\in\{2,7,8\}} \frac{8 \lambda_j(k)}{\mathrm{Tr}[\mathcal Q_j^2(k)]}\mathcal Q_j(k) \Bigr]e^{i\frac{\Omega_k}{2\sin k}\mathcal Q_7(k)}
\end{multline}
with
\be\label{eq:Omega}
e^{i\Omega_k}=\frac{-2\sin\frac{k}{2}\gamma m_{z,s}+i\cos\frac{k}{2}(h+2 m_z)}{\sqrt{4\sin^2\frac{k}{2}\gamma^2 m_{z,s}^2+\cos^2\frac{k}{2}(h+2 m_z)^2}}
\ee
and
\be\label{eq:lambda}
\ba
\lambda_2(k)=&\lim_{\tilde t\rightarrow\infty}\cos(2\mathcal E_k \tilde t)\Bigl[\cos\Omega_k y_2(k,\tilde t)\\
&-\sin\Omega_k \cot\frac{k}{2}y_8(k,\tilde t)\Bigr] +\sin(2\mathcal E_k \tilde t)\frac{\varepsilon_k}{2\sin\frac{k}{2}}y_7(k,\tilde t)\\
\lambda_7(k)=&\lim_{\tilde t\rightarrow\infty}\sin(2\mathcal E_k \tilde t)\Bigl[-2\cos\Omega_k \frac{\sin\frac{k}{2}}{\varepsilon_k}y_2(k,\tilde t)\\
&+2\sin\Omega_k \frac{\cos\frac{k}{2}}{\varepsilon_k}y_8(k,\tilde t)\Bigr] +\cos(2\mathcal E_k \tilde t)y_7(k,\tilde t)\\
\lambda_8(k)=&\lim_{\tilde t\rightarrow\infty} \tan\frac{k}{2}\sin\Omega_k y_2(k,\tilde t)+\cos \Omega_k y_8(k,\tilde t)\, .
\ea
\ee
We called $m_z$ and $m_{z,s}$ the infinite time limits of $m_z(t)$ and $m_{z,s}(t)$ respectively; $\mathcal E_k$ is the dispersion relation of the two species  of quasiparticles diagonalizing \mbox{$\bar H^{\rm  MF}$}
\be\label{eq:Ek}
\mathcal E_k=\sqrt{\frac{4\sin^2\frac{k}{2}\gamma^2 m_{z,s}^2+\cos^2\frac{k}{2}(h+2 m_z)^2}{\cos^2\frac{k}{2}+\gamma^2\sin^2\frac{k}{2}}}\, .
\ee
We note that the dispersion relation \eqref{eq:Ek} becomes flat, and hence no more compatible with relaxation,  when  the magnetic field opposes  the average magnetization restricted to even or odd sites \mbox{$h=\pm 2 m_{z,s}-2 m_z$}.  

\subsubsection{Analytic expression for the entropy}  %

In noninteracting models the von Neumann entropy \mbox{$S_\ell=-{\rm Tr}\rho_\ell\log \rho_\ell$} of $\ell$ adjacent sites can be expressed in terms of the correlation matrix $\tilde \Gamma_\ell$, restricted to the corresponding subspace, as follows:
\be
S_\ell=\frac{1}{2}\mathrm{Tr}{\mathscr H}(\tilde \Gamma_\ell)\, ,
\ee
with
\be\label{eq:Hscr}
\mathscr H(x)=-\frac{1+x}{2}\log \frac{1+x}{2}-\frac{1-x}{2}\log \frac{1-x}{2}\, .
\ee
The asymptotic behaviors of determinants and traces of (block-)Toeplitz matrices like $\tilde \Gamma_\ell$ have been extensively studied. We report here a strong limit theorem for the trace of a  function $f$ of an $N\times N$ block Toeplitz matrix $\bf T$ with smooth $n\times n$ symbol $t(k)$ in the hypothesis that $f$ is analytic in an interval covering the spectrum of $\bf T$~\cite{W:trToeplitz}:
\be\label{eq:Wlimit}
\frac{1}{N}\mathrm{Tr} f({\bf T})\xrightarrow{N\rightarrow\infty} \int_{-\pi}^\pi\frac{\mathrm d k}{2\pi} \frac{1}{n} \mathrm{Tr} f(t(k))\, .
\ee
This asymptotic expression can be used to compute the extensive part of the entanglement entropies when the symbol of the correlation matrix  has no parameter comparable with $\ell$. For 4-by-4 blocks we obtain
\be
\frac{S_\ell}{\ell}\approx \int_{-\pi}^\pi\frac{\mathrm d k}{2\pi}\frac{1}{4}\mathrm{Tr}[\mathscr H(\tilde \Gamma(k))]\, .
\ee
We notice that this expression is invariant under unitary transformations on the symbol. 
Since the limit in \eqref{eq:Gammatilde} is assumed to exist, $\tilde\Gamma(k,0)$ is unitarily equivalent to $\Gamma^{\rm MF}(k,0)$, so the effective sudden quench description gives in fact the same entropy (per unit length) of the GGE of the unperturbed model
\be\label{eq:S0}
\frac{S_\ell(T=0)}{\ell}\approx \int_{-\pi}^\pi\frac{\mathrm d k}{2\pi}\frac{1}{4}\mathrm{Tr}[\mathscr H(\Gamma^{\rm MF}(k,0))]\, .
\ee

On the other hand, the scaling limit in which the rescaled time $T=\Delta t$ is comparable with $\ell$ can not be worked out using \eqref{eq:Wlimit}: the symbol depends on a parameter comparable with the matrix size.
This is the point where the mapping described in the previous section reveals its importance: having mapped the problem into a sudden quench, we are in a position to apply the method proposed in Ref.~[\onlinecite{FC:08}] and slightly generalized in Ref.~[\onlinecite{CEF1}]. 
In particular, Refs [\onlinecite{FC:08}][\onlinecite{CEF1}] considered $(2\ell)\times(2\ell)$ block Toeplitz matrices $\bf T$ with sufficiently regular 2-by-2 symbols $t^{(1)}(k,t)$ of the form
\be\label{eq:Gamma1proof}
t^{(1)}(k,t)\sim e^{i\vec \zeta_k\cdot \vec\sigma}[n^x_k\sigma^x+\vec n_k^\perp\cdot \vec\sigma e^{2i\epsilon_kt}]e^{-i\vec \zeta_k\cdot \vec\sigma}\, ,
\ee
and worked out the limit $1\ll\ell\sim t$ for the trace of powers of $\bf T$, obtaining the following result
\begin{multline}\label{eq:asympTrG}
\frac{\mathrm{Tr}[{\bf T}^{2n}(t)]}{2\ell}\rightarrow \int_{-\pi}^\pi\frac{\mathrm d k}{2\pi}\max(1-2|\epsilon'_k|\frac{t}{\ell},0)|\vec n_k |^{2n}\\
+\int_{-\pi}^\pi\frac{\mathrm d k}{2\pi}\min(1,2|\epsilon'_k|\frac{t}{\ell})(n^x_k)^{2n}\, ,
\end{multline}
where $|\vec n_k|^2=(n_k^x)^2+|\vec n^\perp_k|^2$.

In our context translation invariance is realized by a two-site shift and, in turn, the correlation matrix has (at least) a 4-by-4 symbol. 
 In fact this is not an issue because, even in this case, the symbol follows an $SU(2)$ dynamics (because of reflection symmetry), which is the main property used in the derivation of \eqref{eq:asympTrG}.
The trick is essentially to replace the Pauli matrices that appear in the 2-by-2 symbol \eqref{eq:Gamma1proof} of the original proof (we refer the interested reader to Section 3.1 of Ref.~[\onlinecite{CEF1}]) by the appropriate combinations of the $\mathcal Q_{2,7,8}$, whose square is indeed proportional to the identity, like for the Pauli matrices. The only subtlety to take care of is that now the local space has dimension $4$ and the length appearing in the expression is half of the subsystem length, being each point associated with two sites. 

For $r_H\ll \ell\sim\Delta t\sim O(\Delta^0)$ we then find that the von Neumann entropy asymptotically behaves as follows
\begin{multline}\label{eq:Sell}
\frac{S_\ell(t)}{\ell}\approx \int_{-\pi}^\pi\frac{\mathrm d k}{2\pi}\mathscr H( \Sigma_0(k))\\
+\int_{-\pi}^\pi\frac{\mathrm d k}{2\pi}\min\Bigl(4 |\mathcal E'_k| \frac{\Delta t}{\ell},1\Bigr )\Bigl[\mathscr H(\Sigma_\infty(k))-\mathscr H( \Sigma_0(k))\Bigr]
\end{multline}
The R\'enyi entropies $S_\ell^{(\alpha)}=\frac{\log{\rm Tr}\rho^\alpha}{1-\alpha}$ have the same functional form of \eqref{eq:Sell}, provided that the function $\mathscr H(x)$ is replaced with
\be
\mathscr H_\alpha(x)=\frac{1}{1-\alpha}\log\Bigl(\Bigl(\frac{1+x}{2}\Bigr)^\alpha+\Bigl(\frac{1-x}{2}\Bigr)^\alpha\Bigr)\, .
\ee
The factor of $4$ in the second line of \eqref{eq:Sell} is in place of the factor of $2$ in \eqref{eq:asympTrG} as a result of a point being associated with two sites. One immediate consequence is that the actual maximal velocity at which information propagates (in the pre-relaxation limit) is double of the velocity associated with \eqref{eq:Ek}, \emph{i.e.} \mbox{$\tilde v_M=2\max_k|\mathcal E_k'|$}.
The time at which linearity is lost is therefore given by
\be\label{eq:tM}
t_M=\frac{\ell}{4\Delta \max_k|\mathcal E'_k|}\, .
\ee
The functions $ \Sigma_{0}(k)$ and $\Sigma_{\infty}(k)$ are the eigenvalues of the symbol of the (extrapolated) correlation matrix at the initial and infinite (rescaled) times, respectively (notice that the \mbox{4-by-4} symbol of the correlation matrix has two doubly degenerate eigenvalues that differ only for the sign). In particular we have
\be
 \Sigma_0(k)=2\sqrt{\frac{\lambda_2^2(k)}{\cos^2\frac{k}{2}\varepsilon_k^2}+\frac{\lambda_7^2(k)}{\sin^2 k}+\frac{\lambda_8^2(k)}{\sin^2\frac{k}{2}\varepsilon_k^2}}=\gamma\frac{\sin\frac{k}{2}}{\varepsilon_k}\, ,
\ee
where $\lambda_j(k)$ are given in \eqref{eq:lambda} and the last equality reflects the equivalence with the unperturbed GGE \eqref{eq:S0}.

Despite Eq. \eqref{eq:Sell} depending on dynamical parameters that are not known \emph{a priori}, it is in fact an impressive result: \emph{the unknown variables do not  affect the functional form of the entropy, which follows a linear behavior until $t_M$ and then saturates to a stationary value}. This is exactly the type of universality that we were seeking in the late time evolution of the entanglement entropies. 

The physical interpretation depicted in Fig. \ref{f:1} is perfectly consistent with \eqref{eq:Sell}: $\Sigma_0$ can be interpreted as the ``cross section'' for the production of a pair of composite quasiparticles, consisting of two particles with almost the same velocity (the red and black arrows of Fig. \ref{f:1}, treated as a single entity); $\Sigma_\infty$ is instead related to the cross section associated with the constituents, which in the pre-relaxation limit are in fact distinguishable.

Importantly, the necessity of replacing $\rho^{\rm MF}(0)$ by $\tilde \rho(0)$ \eqref{eq:rhotilde0} can be interpreted as the effect of the  scattering matrix not being trivial in the presence of interactions: 
we  still describe time evolution in terms of free quasiparticles, but the interaction causes a dynamical modification of the effective cross sections entering into the problem.

Finally, we propose a minimal refinement of \eqref{eq:Sell} that combines the two scaling limits $1\ll \ell\sim t$ (computed in Ref.~[\onlinecite{FC:08}]) and $1\ll \ell\sim \Delta t$:
\begin{multline}\label{eq:Sellstar}
\frac{S_\ell(t)}{\ell}\approx \int_{-\pi}^\pi\frac{\mathrm d k}{2\pi}\min\Bigl(4 |\mathcal \varepsilon'_k| \frac{t}{\ell},1\Bigr )\mathscr H(\Sigma_0(k))\\
+\int_{-\pi}^\pi\frac{\mathrm d k}{2\pi}\min\Bigl(4 |\mathcal E'_k| \frac{\Delta t}{\ell},1\Bigr )\Bigl[\mathscr H(\Sigma_\infty(k))-\mathscr H(\Sigma_0(k))\Bigr]\, .
\end{multline}
This is the quantitative analogue of \eqref{eq:scaling}.

\subsubsection{A case study} %

\begin{figure*}[t!]
\includegraphics[width=\ratioF\textwidth]{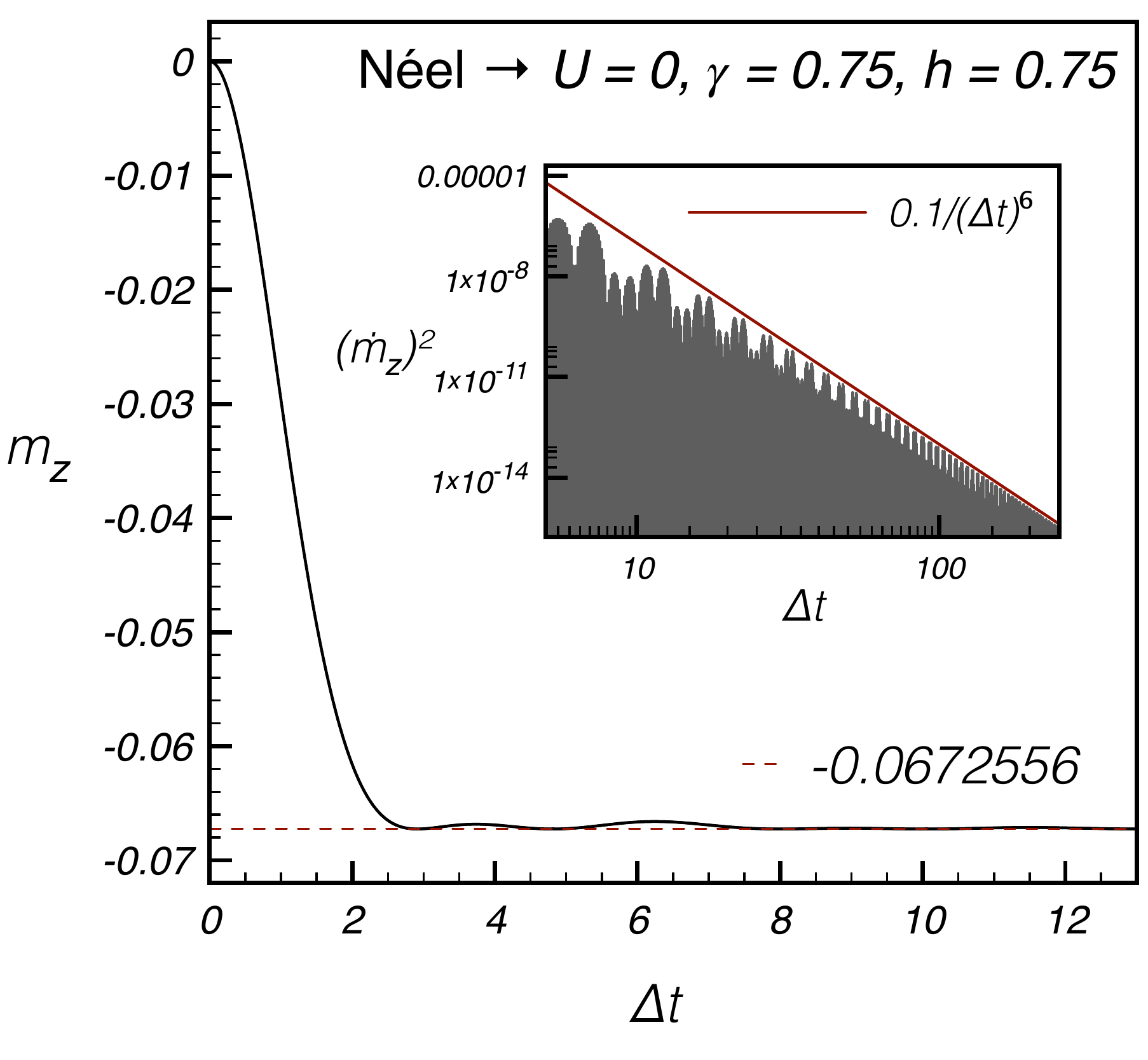}
\hspace{\Hspace}
\includegraphics[width=\ratioF\textwidth]{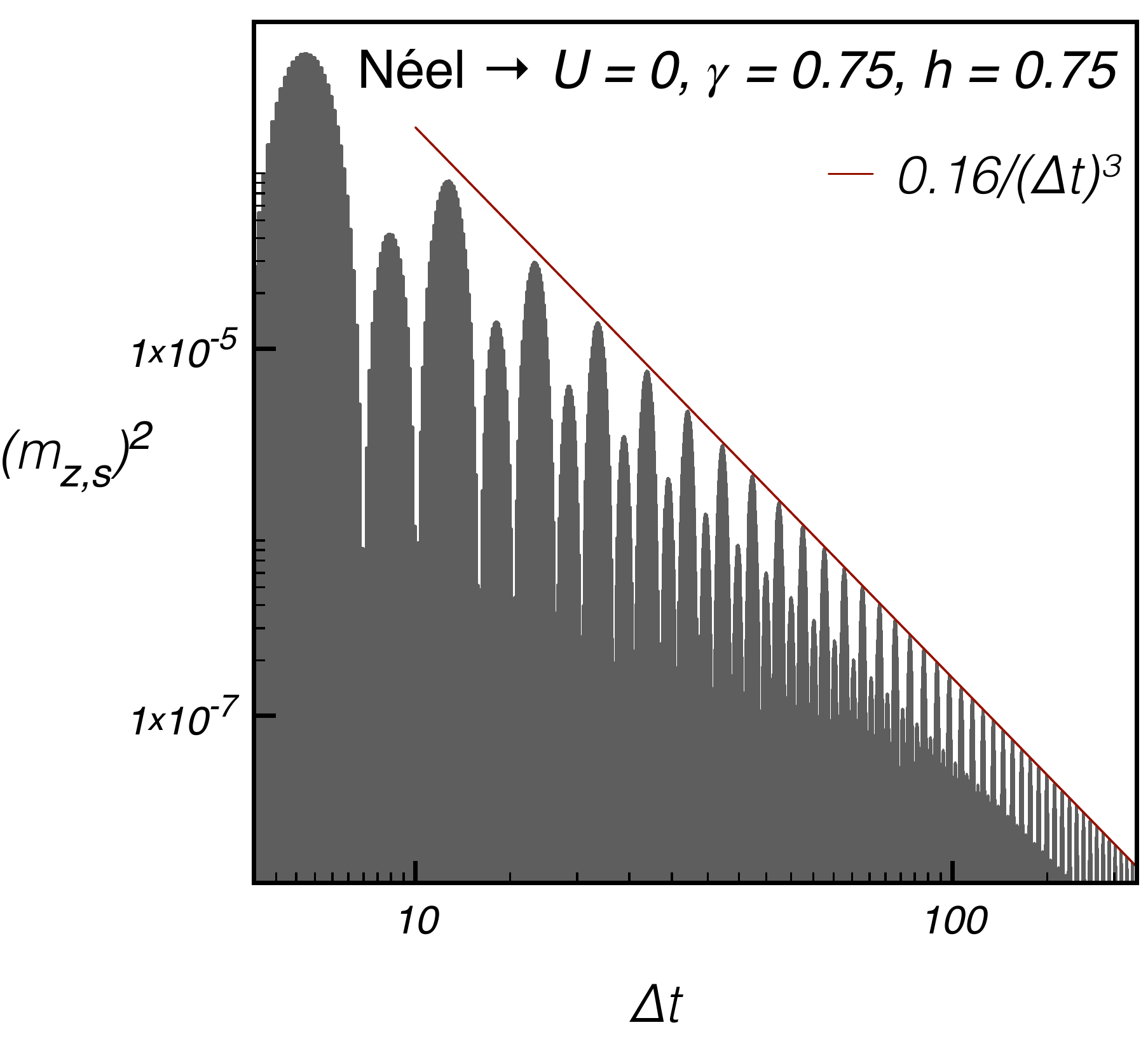}
\caption{\label{f:mz}(Left)For the quench \eqref{e:statea} displayed in the panel, the magnetization in the $z$ direction relaxes to a stationary value. The inset shows the time derivative of the magnetization approaching zero as $(\Delta t)^{-3}$. (Right) The staggered magnetization decays as $(\Delta t)^{-\frac{3}{2}}$. For the sake of clarity, in the plots in log-log scale the regions under the curves are shadowed.
}
\end{figure*}
\begin{figure}[tbp]
\includegraphics[width=\ratioF\textwidth]{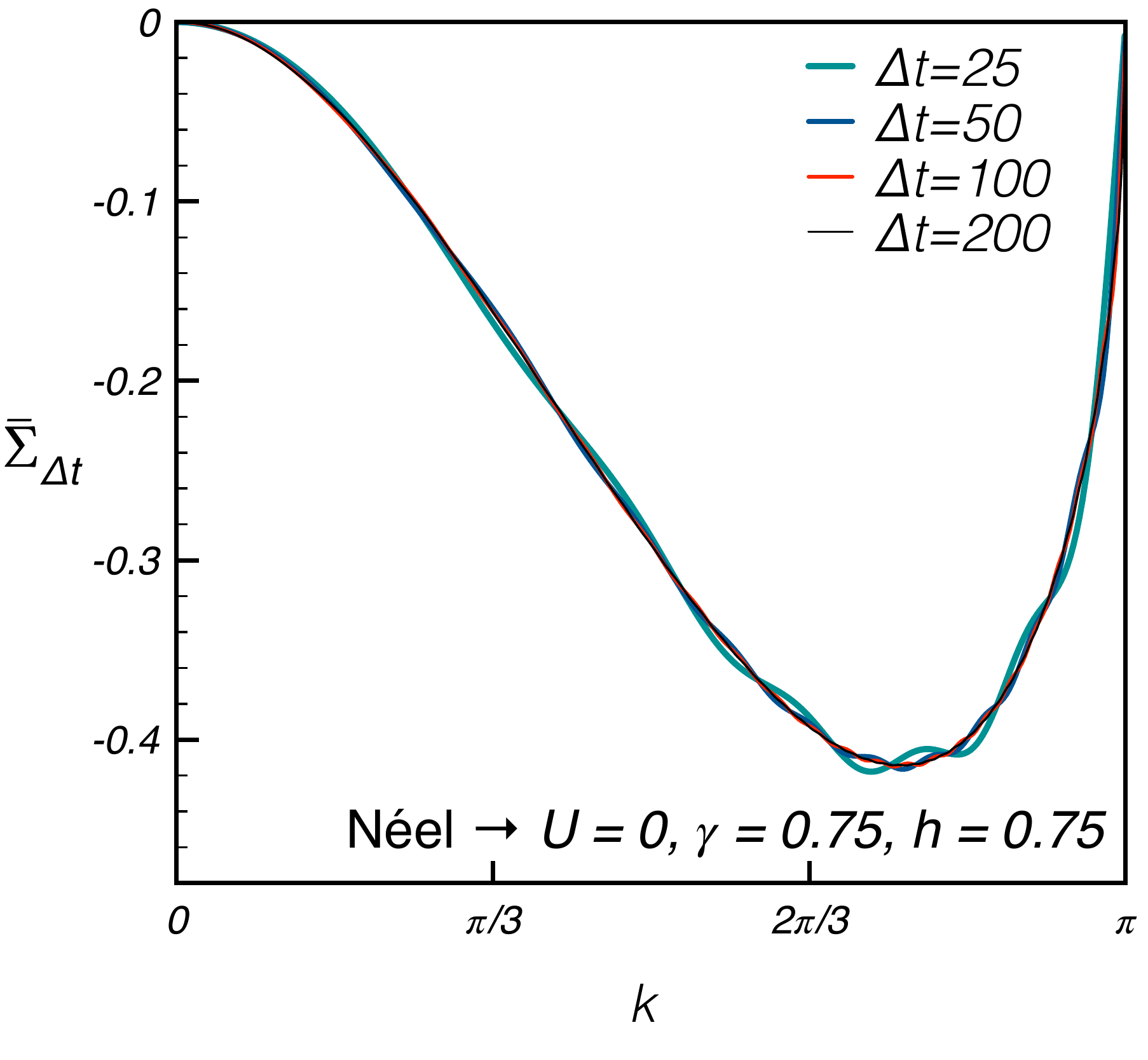}
\caption{\label{f:sigmainf}The quantity $\bar \Sigma_{\Delta t}(k)\equiv \frac{2 y_2 (k,\Delta t)}{\cos\frac{k}{2}\varepsilon_k}$ as a function of the momentum $k$ for $4$ times after the quench \eqref{e:statea} displayed in the panel. The curves become thinner as the time gets larger. Consistently with \eqref{eq:lambda} when one-site shift invariance is restored, the limit of infinite time for $y_2(k,\Delta t)$ seems to exist.  
}
\end{figure}
\begin{figure*}[t!]
\begin{center}
\includegraphics[width=\ratioF\textwidth]{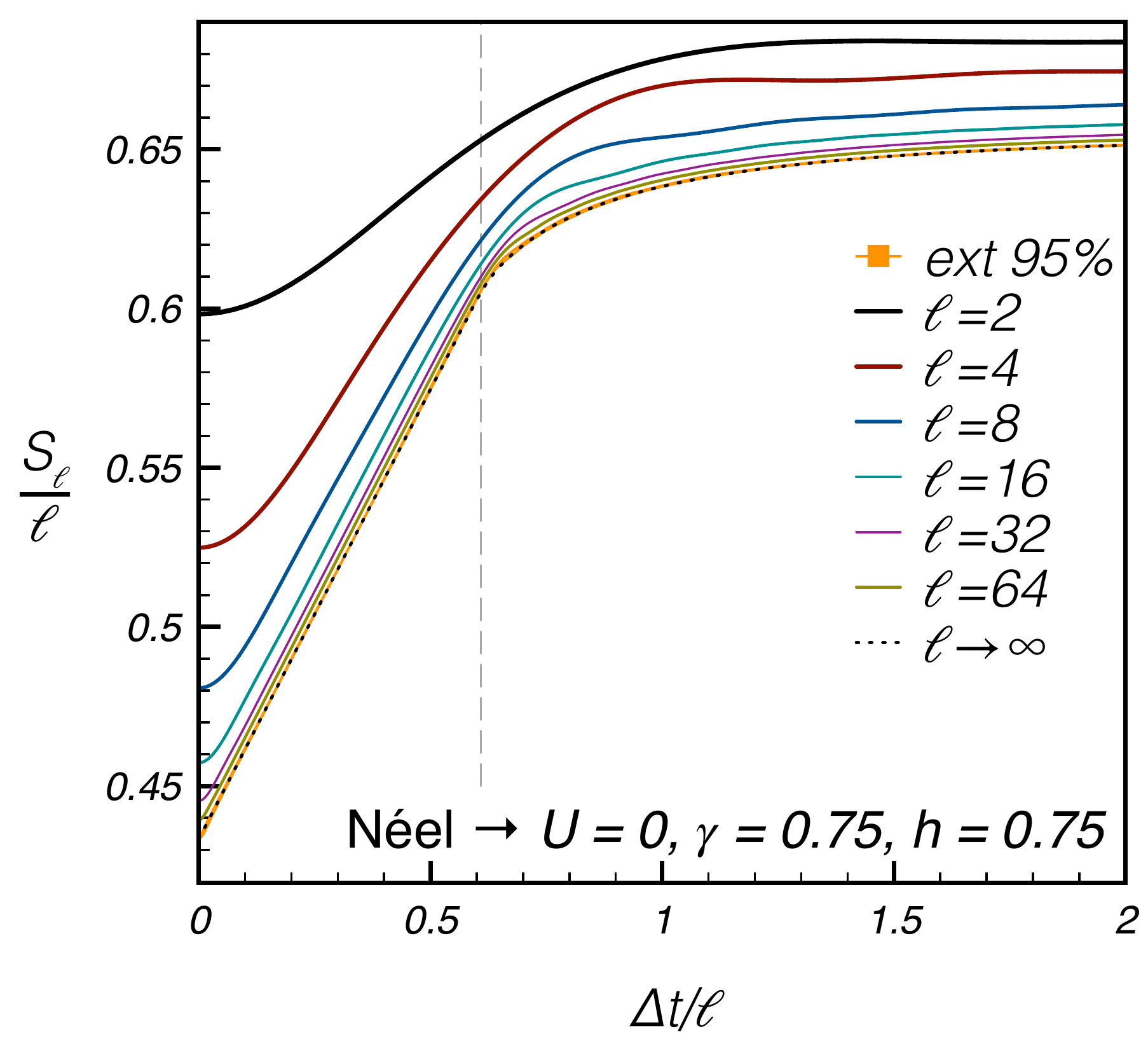}
\hspace{\Hspace}
\includegraphics[width=\ratioF\textwidth]{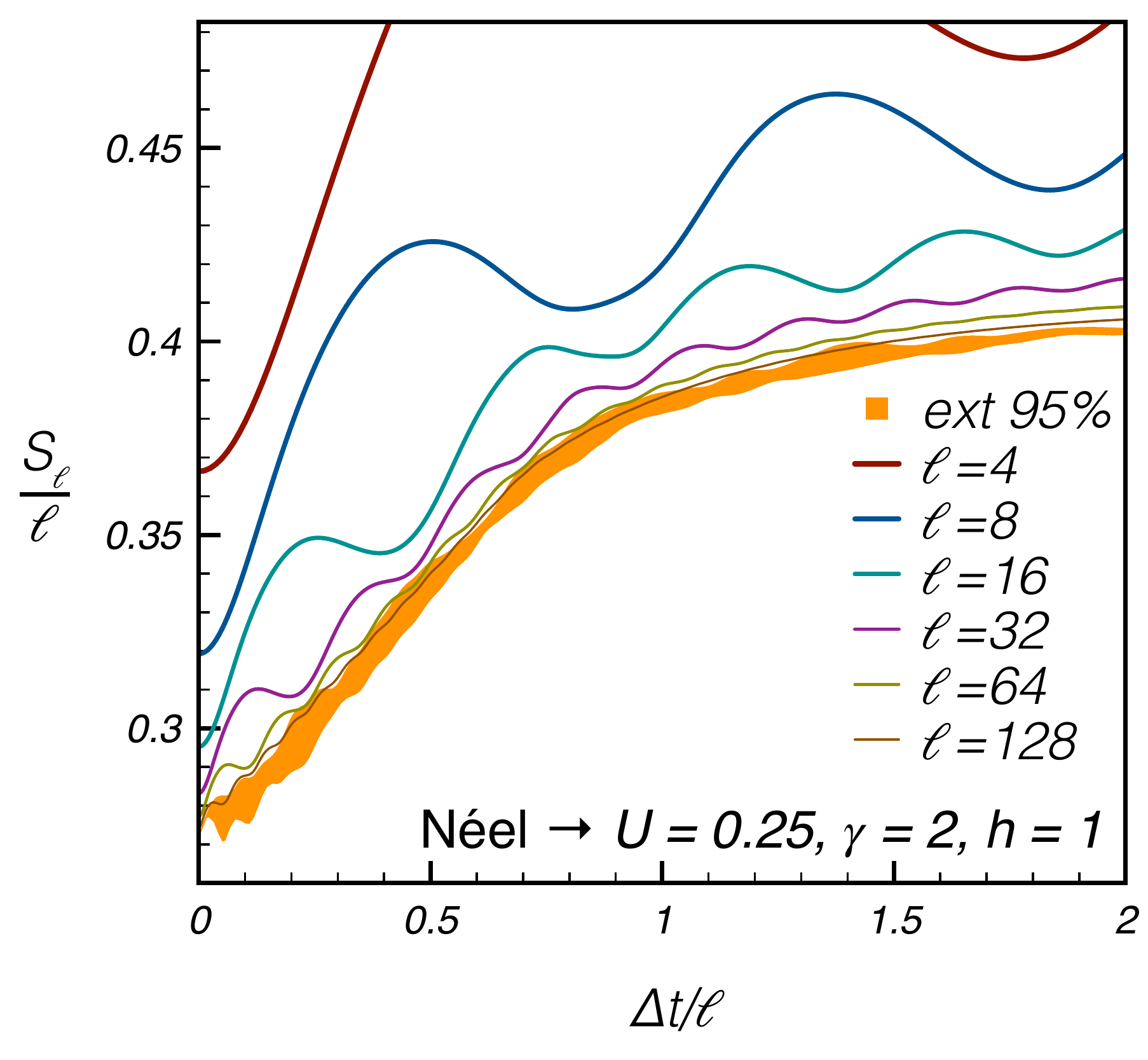}
\end{center}\caption{\label{f:Sell}{
The von Neumann entropy per unit length as a function of the rescaled time per unit length for various intervals within mean-field. At any given rescaled time, the extrapolation is done by finding the best parabolic fit of the data, weighted by $\ell^{6}$ (being too close to the asymptotic values, the data for $\ell=128$ are not displayed in the left panel). The corresponding confidence interval at $95\%$ is shown as an orange region (in the left panel the width of the region has been increased to make it visible).  
 (Left) For the quench \eqref{e:statea} the agreement with formula \eqref{eq:Sell} is excellent. The gray dashed vertical line is \eqref{eq:tM}. (Right) Another quench \eqref{e:stateb} that leads to local relaxation (in the pre-relaxation limit) though one-site shift invariance is not restored and there are more pronounced oscillations; data are still compatible with the scaling law \eqref{eq:scaling}.
} 
}
\end{figure*}

In order to use \eqref{eq:Sell}, we need another ingredient, namely the symbol of the (extrapolated) correlation matrix that emerges at infinite (rescaled) time. 
We will consider an example in which one-site shift invariance is restored, \emph{i.e.} the staggered magnetization approaches zero $m_{z,s}= 0$ and the magnetization relaxes to a stationary value. By inspecting \eqref{eq:Gammatilde}\eqref{eq:Omega}\eqref{eq:lambda} we find $\cos\Omega_k=0$,  $\sin\Omega_k=\mathrm{sgn}(h+2 m_z)$, $y_2(k,t)$ becoming stationary and $y_{7,8}(k,t)$ keeping oscillating with frequency $2\mathcal E_k $.

First of all we evaluate the large time asymptotics of the maximal velocity $v(\tau)$ of $\mathfrak H_0(\tau)$ \eqref{eq:Hext}, so as to check whether  the sudden quench description from the extrapolated density matrix \eqref{eq:rhotilde0} can be used. 
The dispersion relation $\mathfrak E(k,\tau)$ of $\mathfrak H_0(\tau)$ is readily computed and reads as
\be
\mathfrak E(k,\tau)=2\sqrt{\frac{\sin^2\frac{k}{2}\gamma^2 m_{z,s}^2(\tau)+\cos^2\frac{k}{2}(m_z(\tau)-m_z)^2}{\cos^2\frac{k}{2}+\gamma^2\sin^2\frac{k}{2}}}\, .
\ee
If $m_{z,s}(\tau)$ and  $m_z(\tau)-m_z$ approach zero with different power laws, at long times we obtain
\be
v(\tau)\sim 2\max(|m_{z,s}(\tau)|,|m_{z}(\tau)-m_z|)\, .
\ee
Thus, the typical diffusion length of $V_T$, namely \mbox{$\xi(0)\sim\int_0^\infty \mathrm d\tau v(\tau)$}, is finite if the mean-field parameters relax faster than $1/\tau$. This will be the self-consistency condition to be checked \emph{a posteriori}.

Using that one-site shift invariance is restored, the symbol of the long time limit of the (extrapolated) correlation matrix is given by the one-site shift invariant part of $\tilde \Gamma(k,0)$ \eqref{eq:Gammatilde}:
\be
\tilde \Gamma(k,\infty)=\mathrm{sgn}(h+2 m_z) \frac{2\lambda_8(k)}{\sin k\, \varepsilon_k^2}\mathcal Q_2(k)=\frac{2y_2(k,\infty)}{\cos^2\frac{k}{2}\, \varepsilon_k^2}\mathcal Q_2(k)\, .
\ee
From this we identify 
\be\label{eq:sigmainf}
\Sigma_\infty(k)=\frac{2 y_2(k,\infty)}{\cos\frac{k}{2}\varepsilon_k}\, .
\ee
\begin{remark}
Since the mean-field Hamiltonian depends explicitly on the time, there are no simple shortcuts to compute the long time limit of dynamical quantities like \eqref{eq:sigmainf}. In particular
$\Sigma_\infty(k)$ can not be obtained from the correlation matrix of the GGE of the unperturbed model only knowing the mean-field Hamiltonian at late times.  If we had approximated the dynamics by a sudden quench starting from the GGE of the unperturbed Hamiltonian (\emph{i.e.} $\Gamma^{\rm MF}(k,0)$ instead of $\tilde \Gamma(k,0)$), we would have found the late time correlation matrix vanishing, namely $\Sigma_{\infty}(k)=0$ and hence saturation of the entropies to the maximal  values allowed\cite{f:6}. 
Such totally incoherent state $\rho_\ell \propto \1$ is equivalent to the GGE constructed with the one-site shift invariant conservation laws of the unperturbed model which in this case is also clearly equivalent to the GE.  
This is another example of the puzzling situation, pointed out in Ref.~[\onlinecite{BF:mf}], where one-site shift invariance is restored in the pre-relaxation limit but the emerging stationary state is not constrained only by a deformation of the one-site shift invariant conservation laws of the unperturbed model. This could suggest that some extra degeneracy in the spectrum survives the leading order of perturbation theory. However, apparently this degeneracy can not be associated with the non-abelian set of local conservation laws of the XY model, forming its one-site shift invariant charges an abelian subset. Ref.~[\onlinecite{BF:mf}] proposed that this could signal issues in the limit $\Delta\rightarrow 0$ of quasi-local (quasi-)conserved operators, but, in fact, this is still an open question. 
\end{remark}

\subsection{Numerical analysis within mean-field}\label{ss:withinMF}%

\begin{figure*}[t!]
\begin{center}
\includegraphics[width=\ratioF\textwidth]{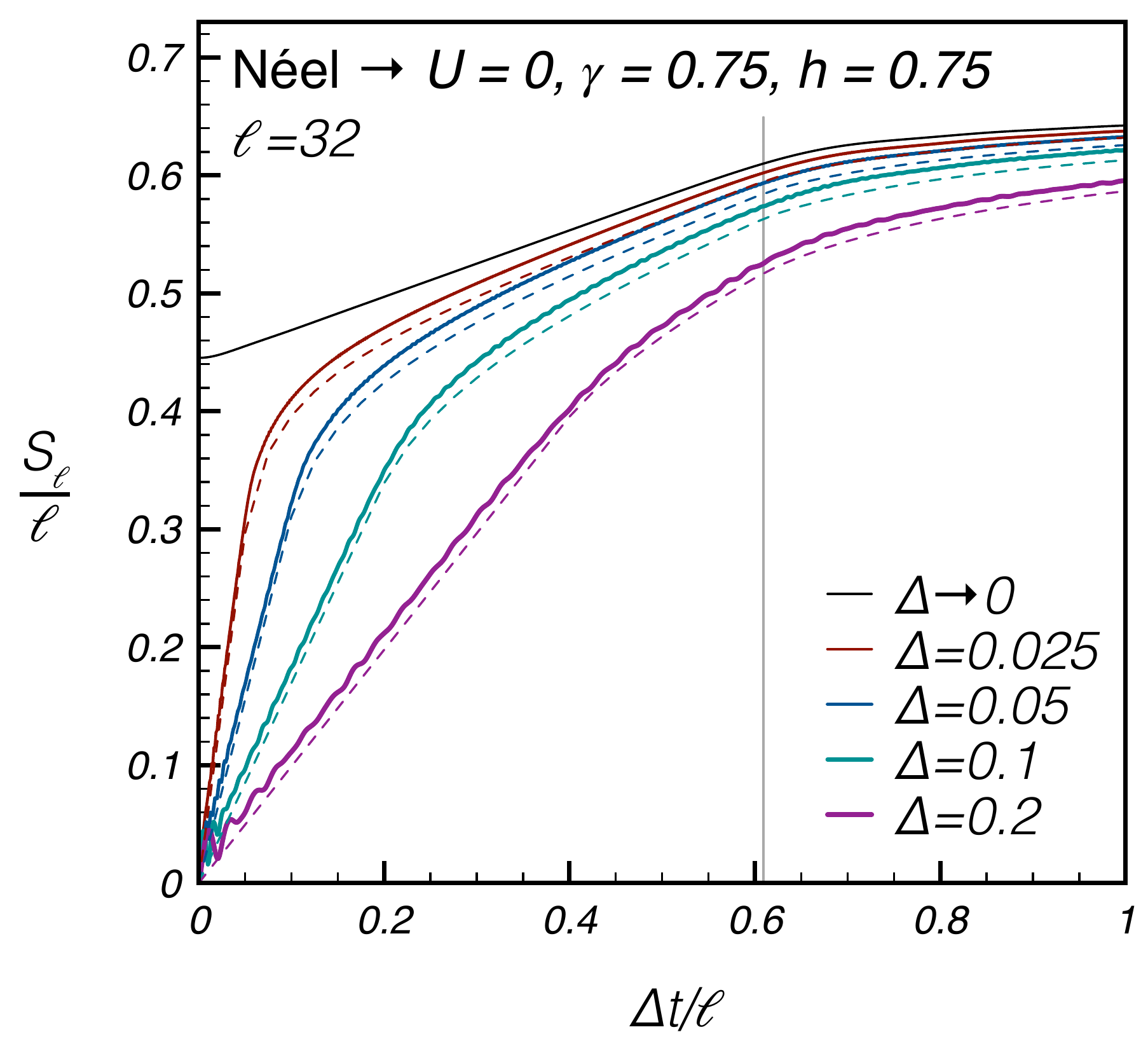}
\hspace{\Hspace}
\includegraphics[width=\ratioF\textwidth]{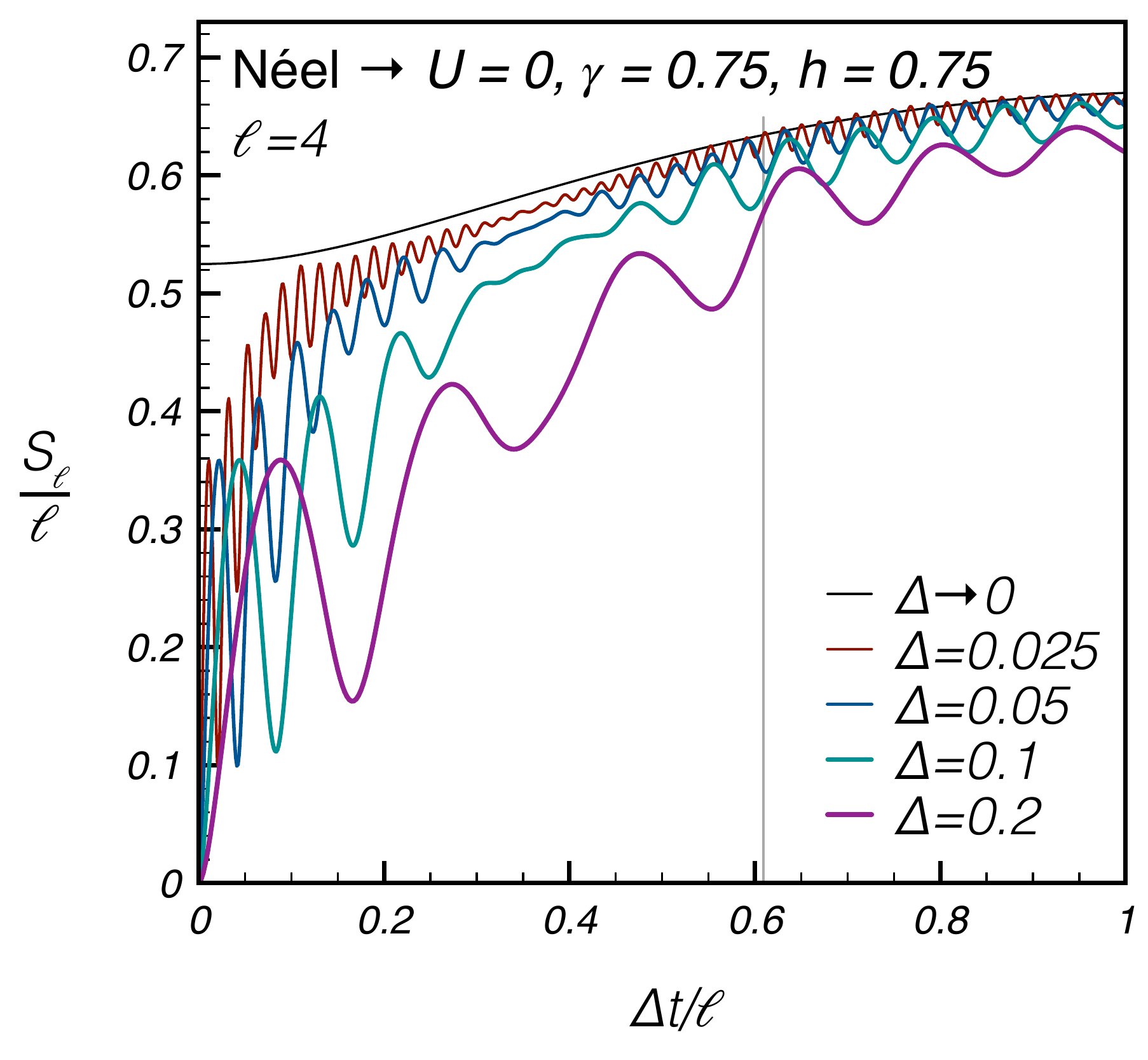}
\end{center}\caption{\label{f:Sintell}{
The time evolution of the von Neumann entropy per unit length within the intermediate mean-field \eqref{eq:mf0} as a function of the rescaled time per unit length for the quench \eqref{e:statea} displayed in the panels and various values of $\Delta$.  The black curves corresponding to $\Delta\rightarrow 0$ are displayed in
the left panel of Fig.~\ref{f:Sell} in purple (left) and red (right). The gray vertical line is \eqref{eq:tM}. In the left panel the dashed lines are the refined predictions \eqref{eq:Sellstar}. 
} 
}
\end{figure*}

\begin{figure*}[t!]
\begin{center}
\includegraphics[width=\ratioF\textwidth]{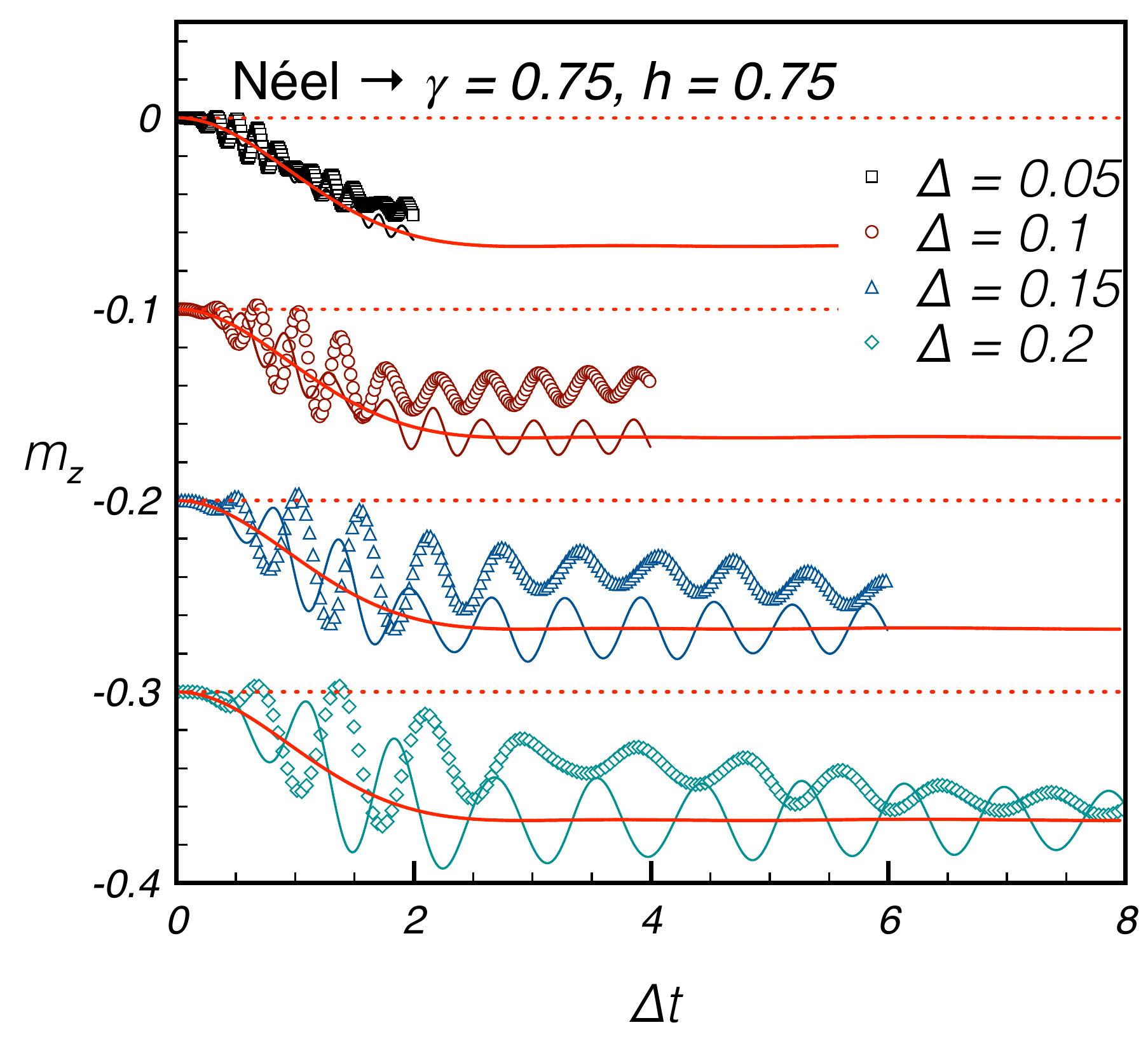}
\hspace{\Hspace}
\includegraphics[width=\ratioF\textwidth]{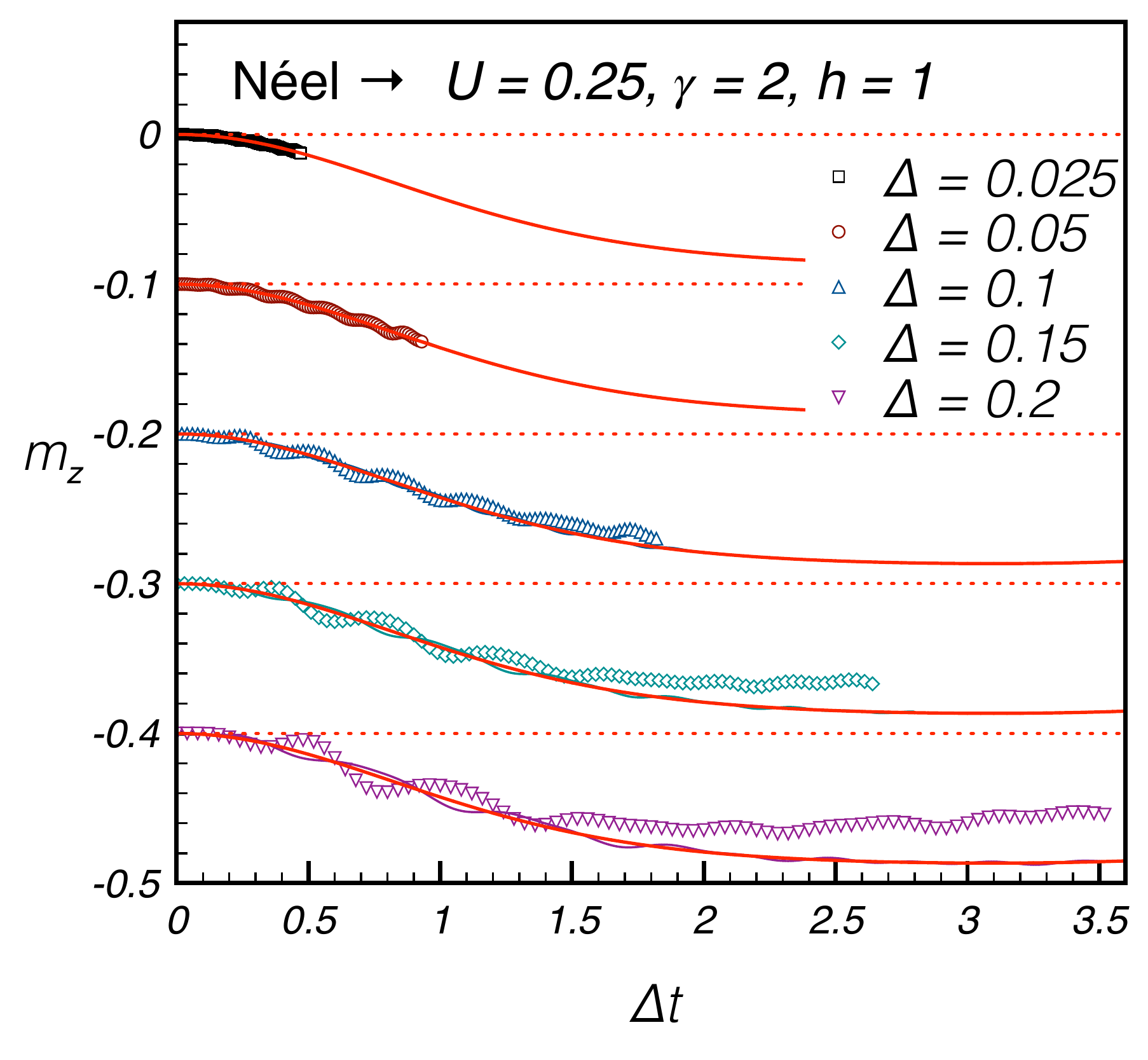}
\end{center}\caption{Evolution of the average magnetization
 $m_z$ after the quenches \eqref{e:statea} and \eqref{e:stateb} displayed in the panels for various ``small'' values of $\Delta$.
The symbols are the iTEBD numerical results and the thin colored lines correspond to mean-field (MF) time evolution \eqref{eq:mf0} starting from the initial state. The thick red lines are instead the MF predictions \eqref{eq:rho0} starting form the unperturbed GGE (dotted lines).
For the sake of clarity, in the left (right) panel  both  datasets and MF curves corresponding to $\Delta > 0.05$ ($\Delta > 0.025$) are vertically shifted (the red curves are in fact coincident!).
}\label{f:mz_Neel}
\end{figure*}

\subsubsection{Prediction \emph{vs} numerical solution.}
Formula \eqref{eq:Sell} is written in terms of quantities computed at late times after the quench, which must be extracted from the numerical solution of the equations.

In Fig \ref{f:mz} we show the time evolution of the  magnetization, its time derivative and the staggered magnetization in the pre-relaxation limit for a particular quench with $\gamma=0.75$, $h=0.75$ and $U=0$. The numerical data leave no doubt the parameters of the mean-field Hamiltonian time relax and one-site shift invariance is finally restored. In particular, our best estimate of the late time magnetization is $m_z\approx -0.0672556$. We notice that both mean-field parameters relax faster than $1/\Delta t$, corroborating the validity of the mapping \eqref{eq:mappingsq} to a sudden quench from the extrapolated density matrix \eqref{eq:rhotilde0}.
The asymptotic value of $m_z$ determines the time $t_M$ \eqref{eq:tM} at which the entire subsystem becomes (substantially) entangled with the rest, but the full time evolution of the entanglement entropies requires also the knowledge of $\Sigma_\infty(k)$.

Fig.~\eqref{f:sigmainf} shows the quantity $\bar \Sigma_T(k)\equiv \frac{2 y_2 (k,T)}{\cos\frac{k}{2}\varepsilon_k}$ as a function of the momentum $k$ for various times\cite{f:5}. 
The limit $\lim_{T\rightarrow\infty}\bar \Sigma_T(k)=\Sigma_\infty(k)$ is reached rather quickly, at least for momenta not too close to $\pm \pi$. 

Having checked that the parameters $\gamma=0.75$ and \mbox{$h=0.75$} are compatible with relaxation and restoration of one-site shift invariance, we can now compare the mean field solution with the asymptotic formula \eqref{eq:Sell}. In Fig.~\ref{f:Sell} the time evolution of the entanglement entropy per unit length as a function of the rescaled time $\Delta t$ is plotted for various subsystem lengths. 
At any given rescaled time, the extrapolation is done by finding the best parabolic fit in $1/\ell$, weighted by $\ell^{6}$ (the inverse of the square of the leading correction neglected). 

The excellent agreement between the left panel of Fig.~\ref{f:Sell} and \eqref{eq:Sell} confirms the conjecture that $V_{T}$, \eqref{eq:rhotrhoMF}, is only responsible for a quasi-local mixing that does not affect the extensive part of the entanglement entropy. 
We also notice that the value of the entropy per unit length associated with the second plateau is visibly smaller than $\log 2$, which would have been obtained neglecting the mean-field dynamics at intermediate rescaled times (\emph{i.e.} approximating $V_{T}$, \eqref{eq:Vql}, by the identity).

Figure \ref{f:Sell} exhibits the scaling behavior claimed in \eqref{eq:scaling}.  

\subsubsection{Intermediate mean-field}\label{ss:intMF} %

\begin{figure*}[t!]
\begin{center}
\includegraphics[width=\ratioF\textwidth]{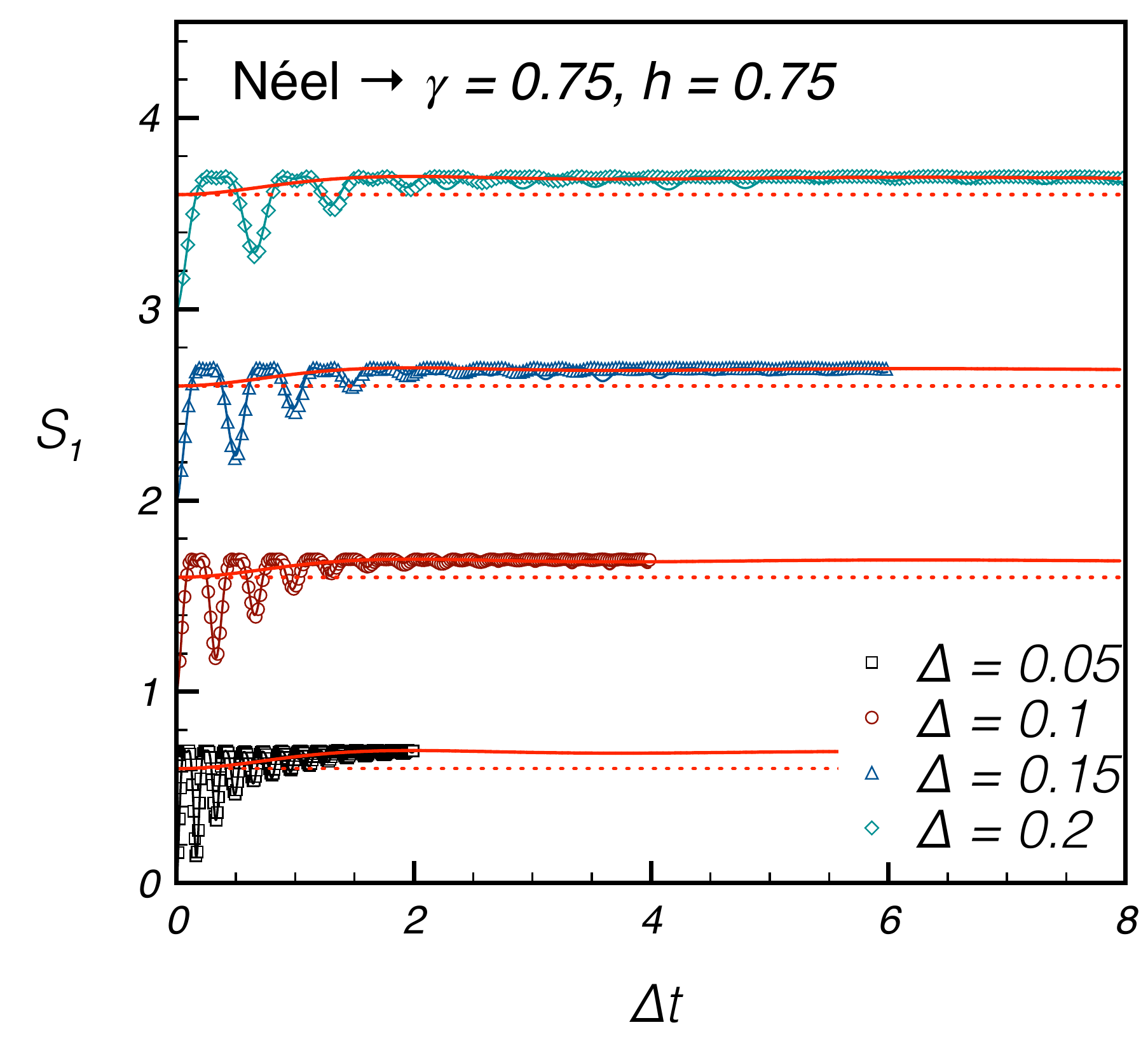}
\hspace{\Hspace}
\includegraphics[width=\ratioF\textwidth]{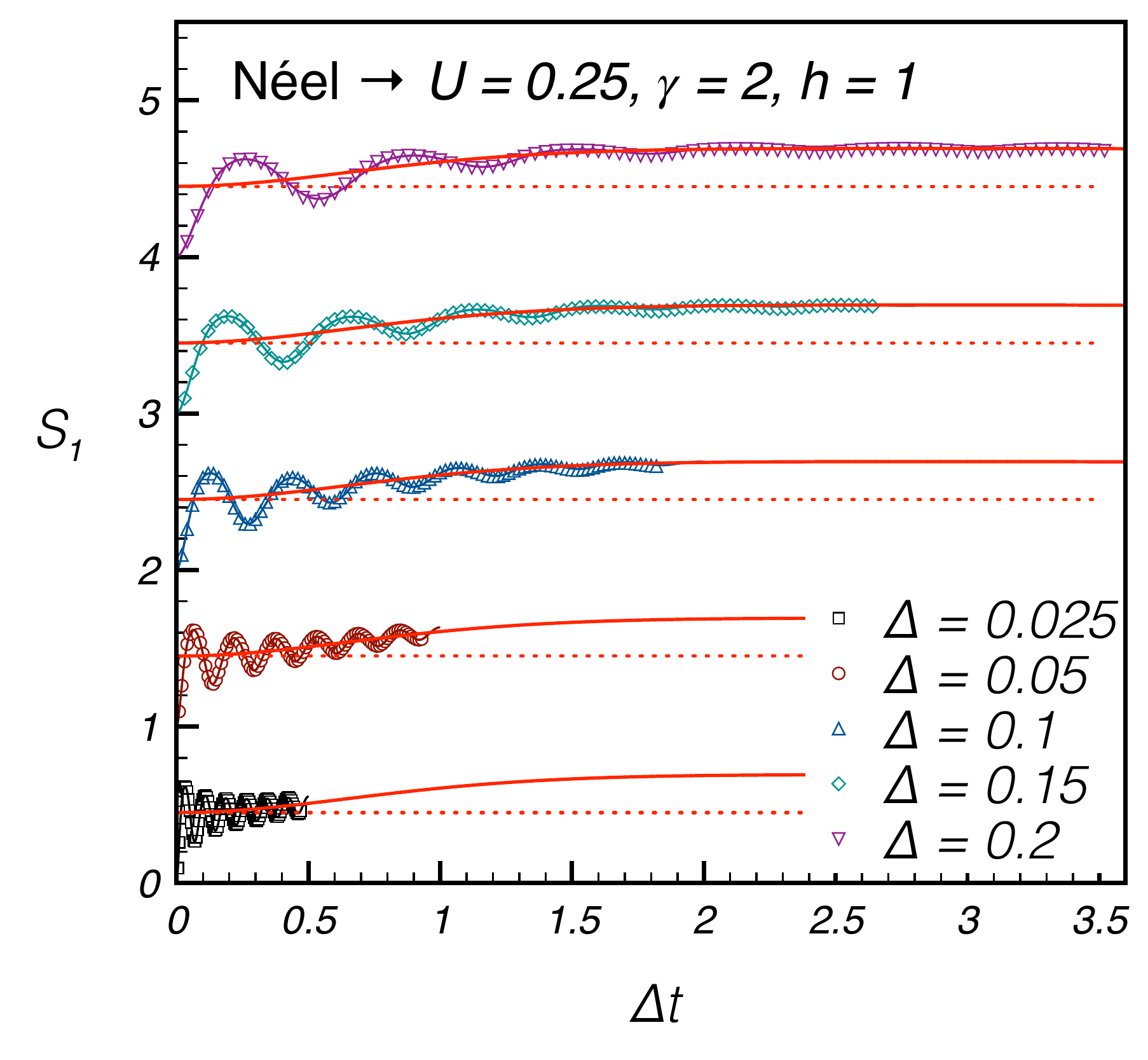}\\
\includegraphics[width=\ratioF\textwidth]{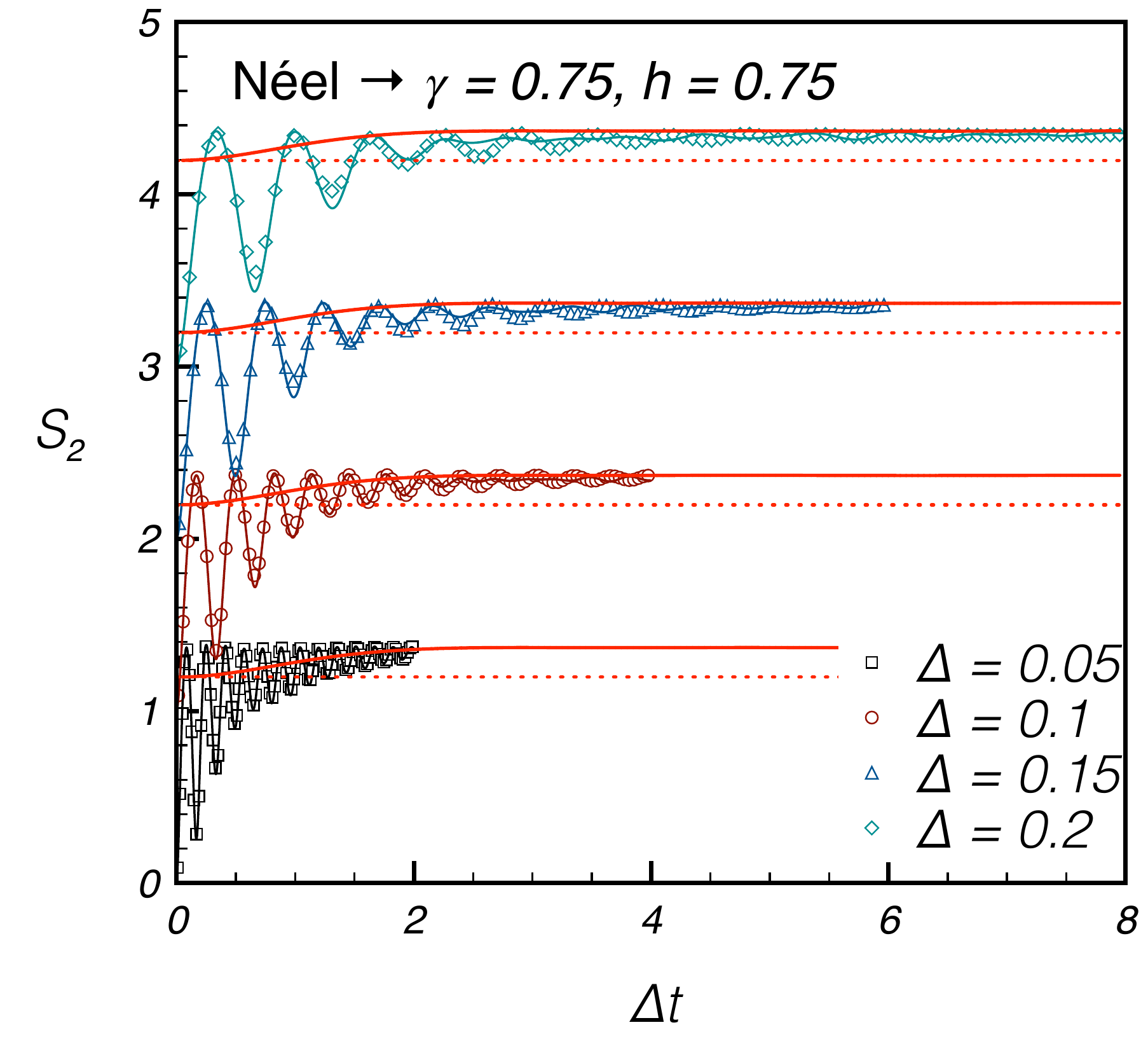}
\hspace{\Hspace}
\includegraphics[width=\ratioF\textwidth]{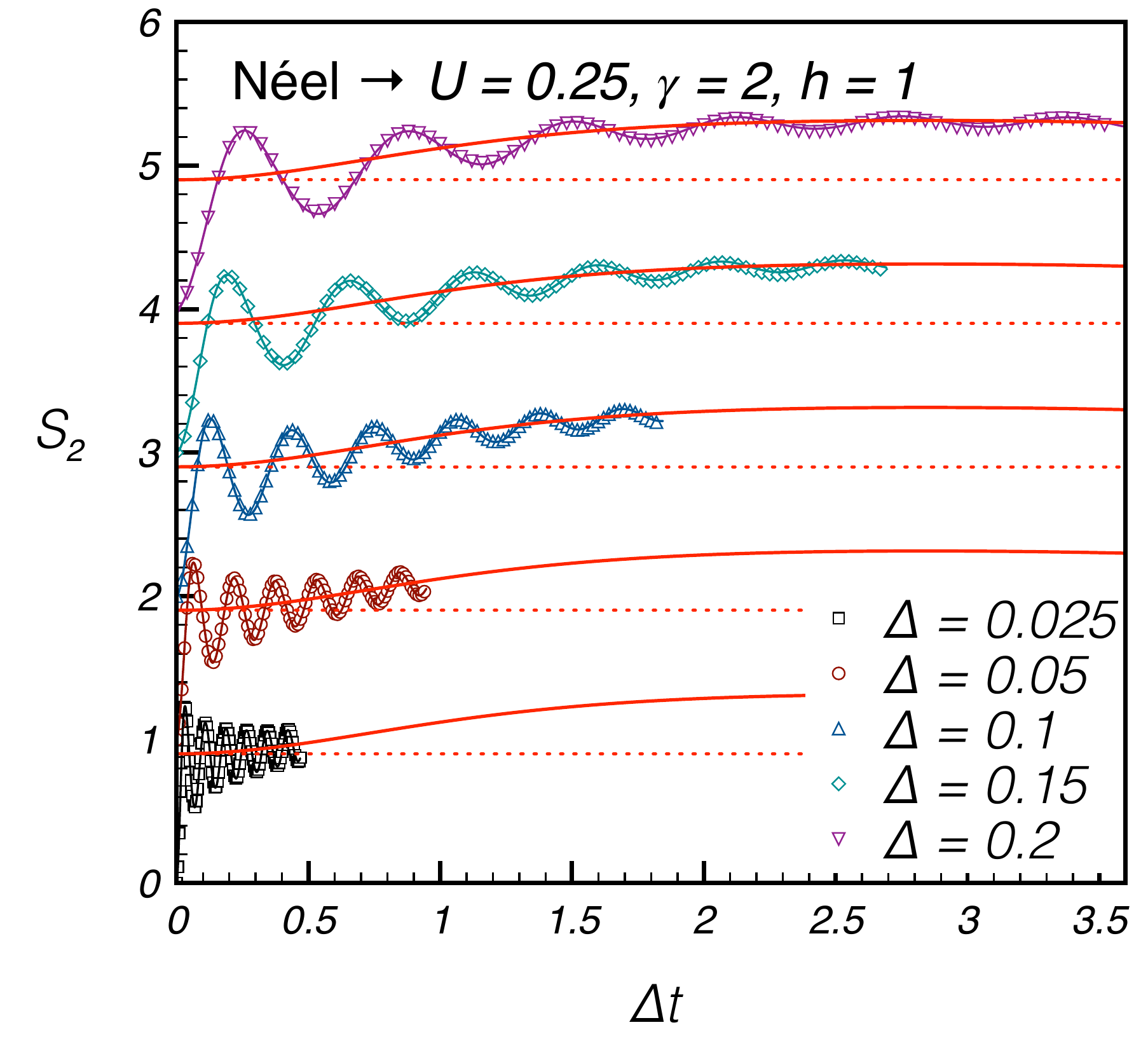}
\end{center}\caption{Evolution of the von Neumann entropy for a subsystem with  
$\ell = 1$ (top) and $\ell=2$ (bottom) sites after the quenches \eqref{e:statea} and \eqref{e:stateb} displayed in the panels for various ``small'' values of $\Delta$. The notations are the same as in Fig.~\ref{f:mz_Neel}.
}\label{f:S1_Neel}\label{f:S2_Neel}
\end{figure*}

\begin{figure*}[t!]
\begin{center}
\includegraphics[width=\ratioF\textwidth]{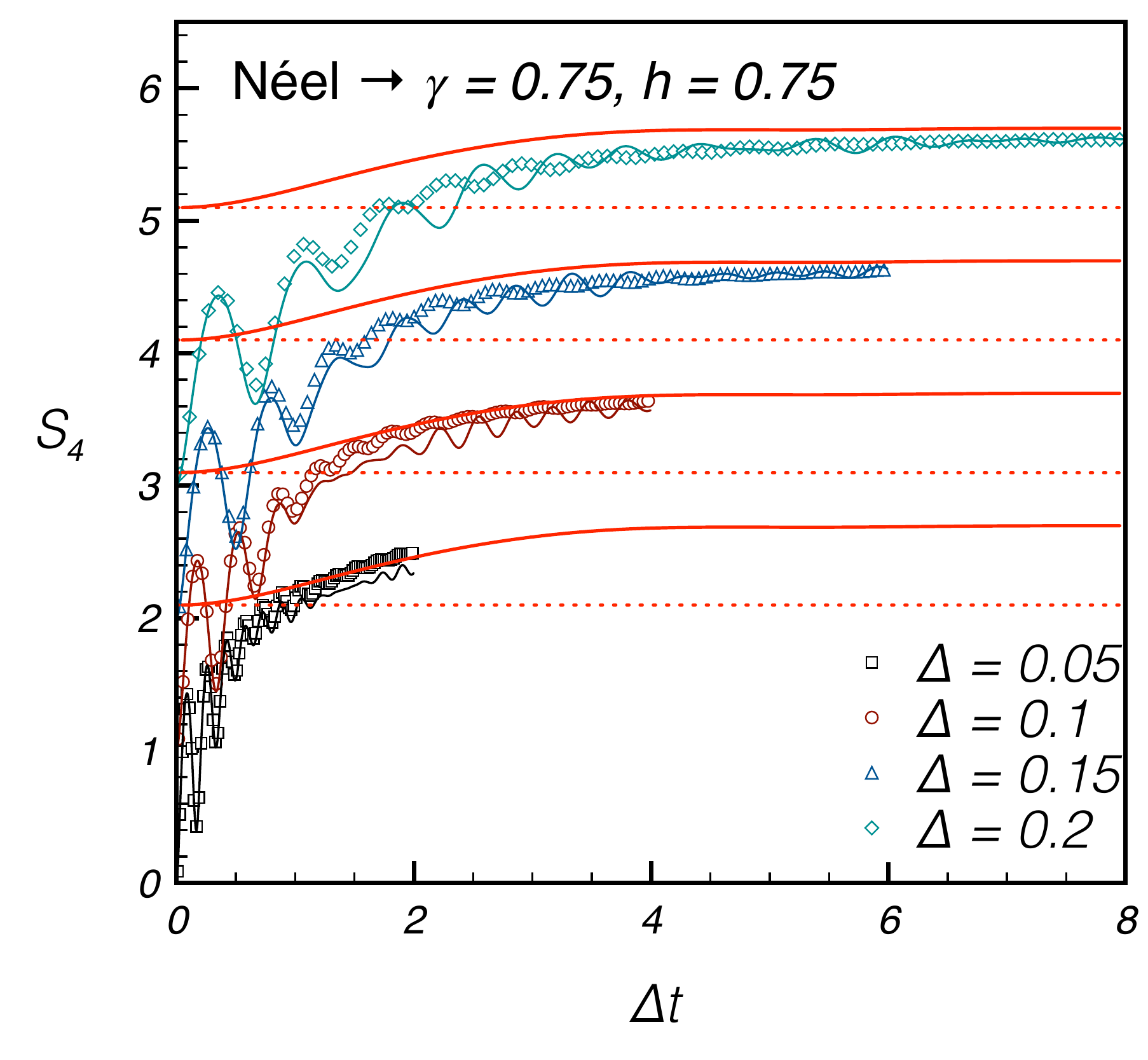}
\hspace{\Hspace}
\includegraphics[width=\ratioF\textwidth]{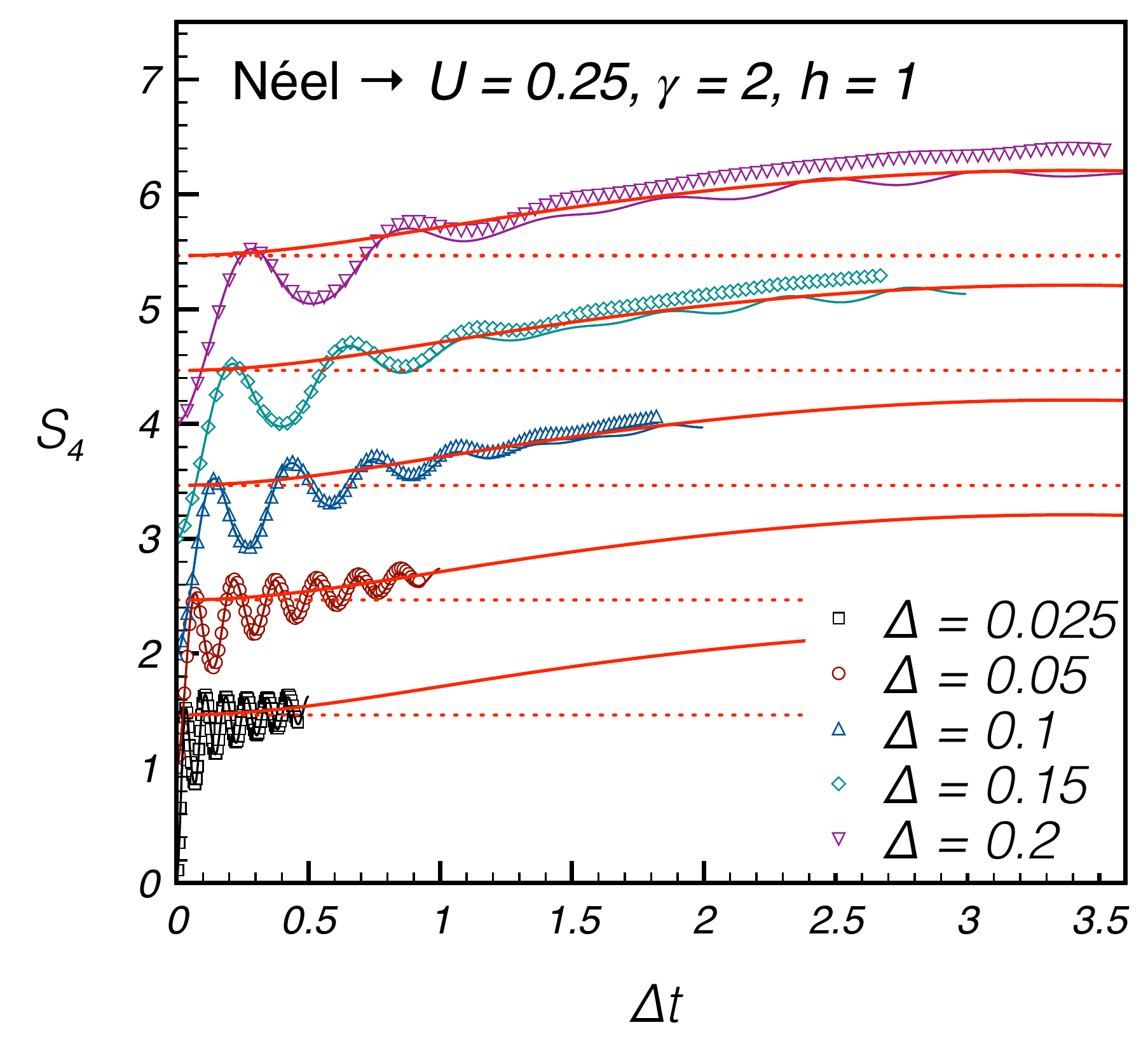}\\
\includegraphics[width=\ratioF\textwidth]{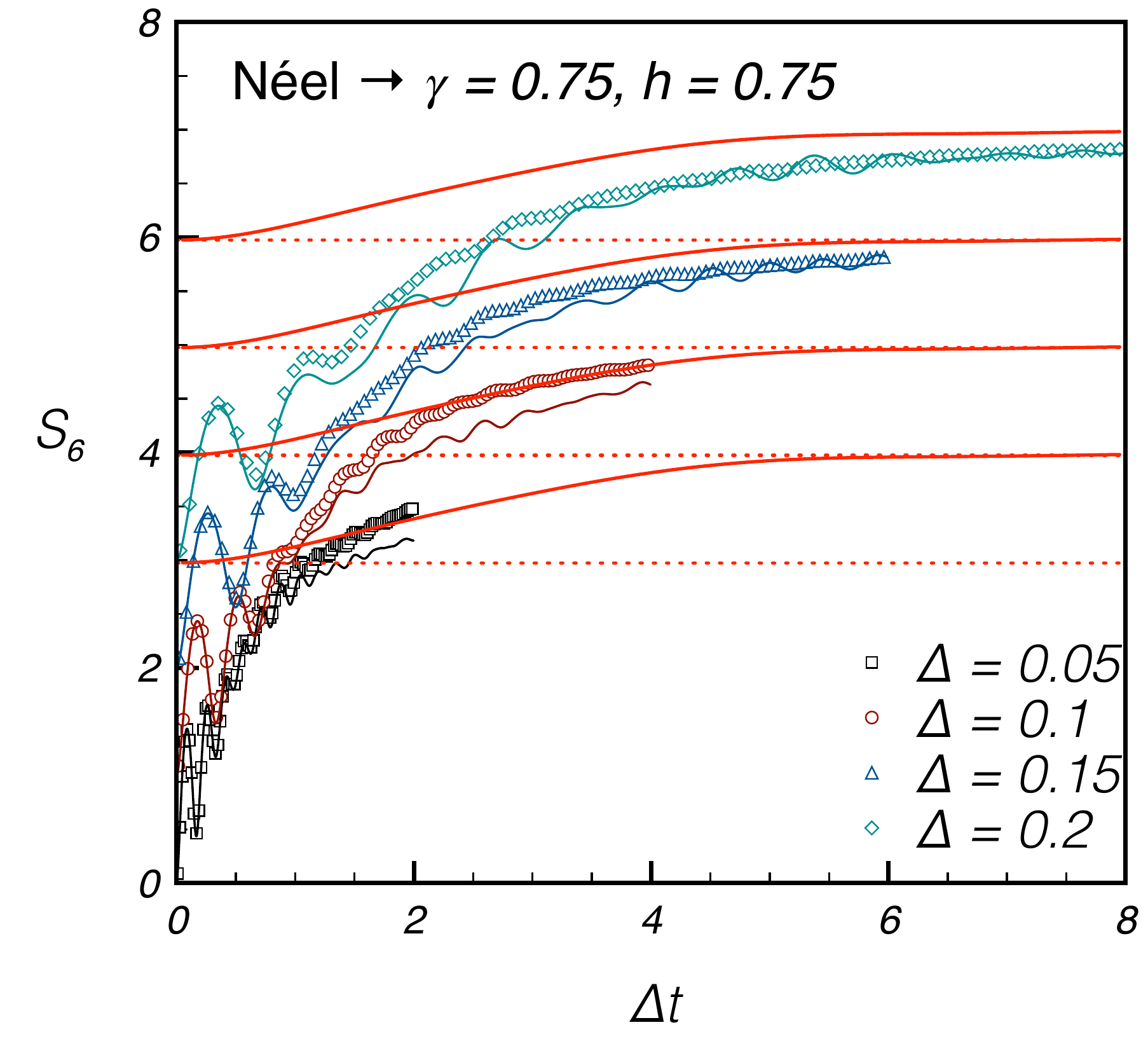}
\hspace{\Hspace}
\includegraphics[width=\ratioF\textwidth]{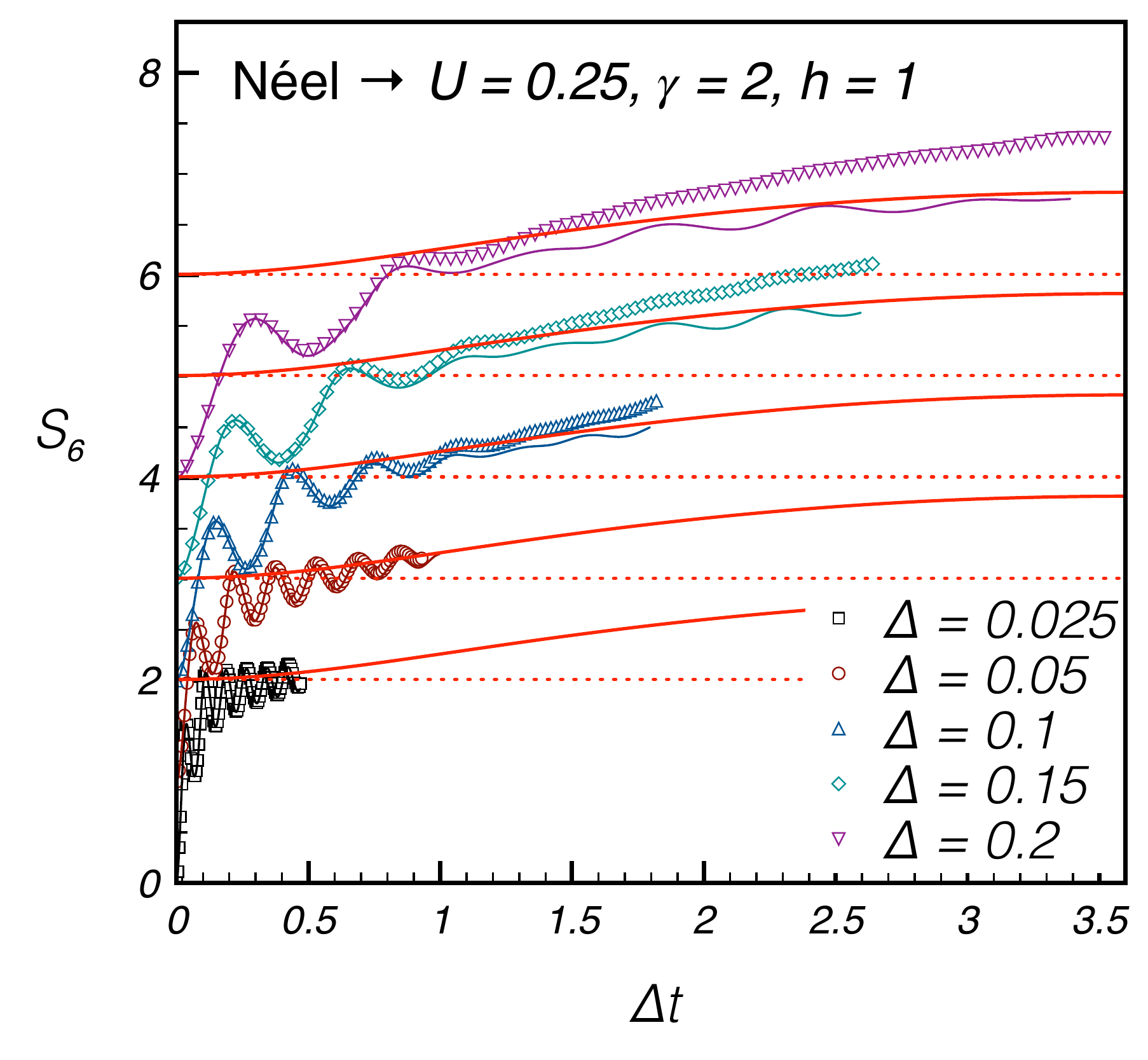}
\end{center}\caption{The same as in Figure \ref{f:S1_Neel} for a subsystem with $\ell = 4$ (top)  and $\ell=6$ (bottom) sites.
}\label{f:S4_Neel}\label{f:S6_Neel}
\end{figure*}

\begin{figure*}[t!]
\begin{center}
\includegraphics[width=\ratioFthree\textwidth]{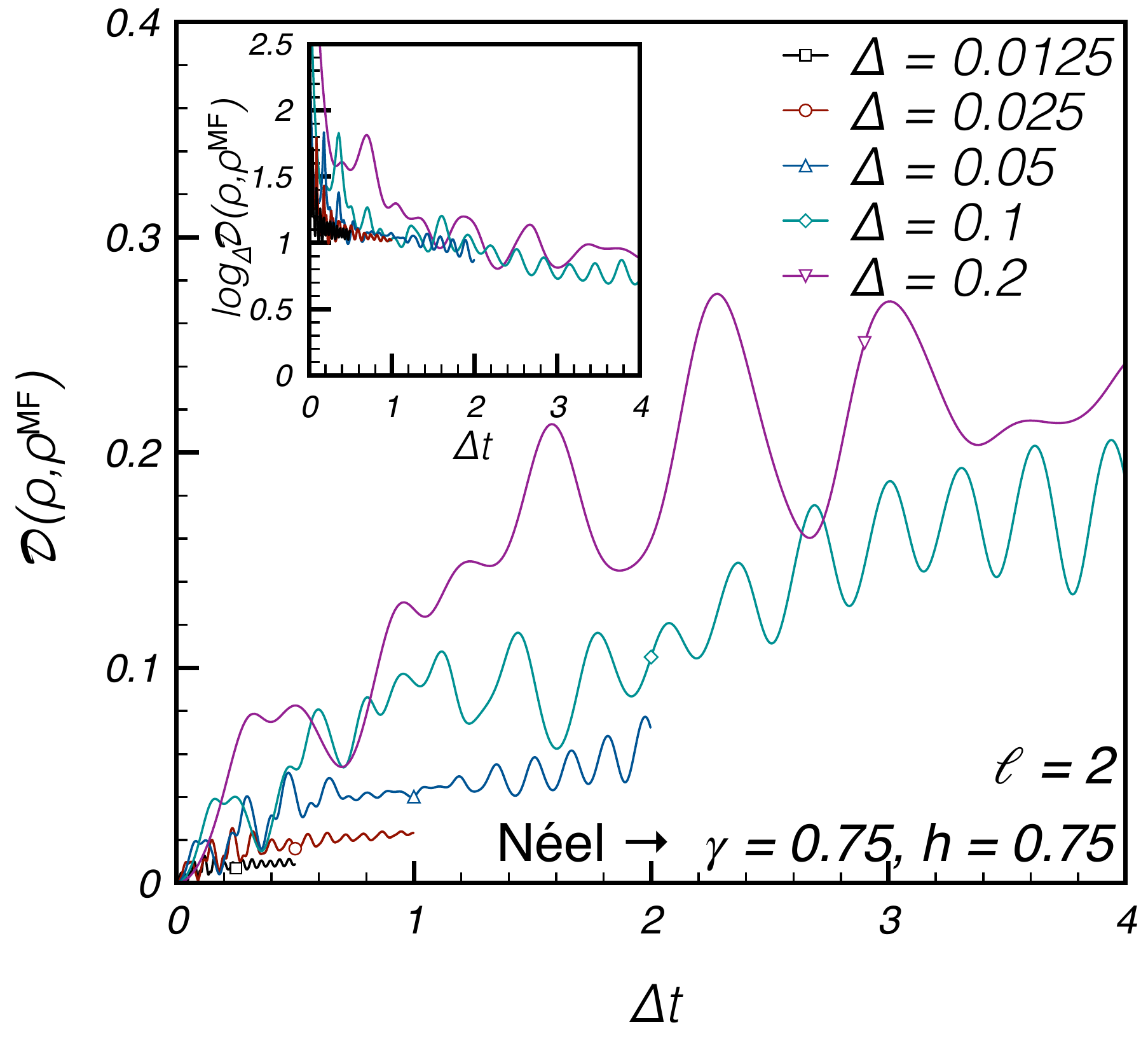}
\includegraphics[width=\ratioFthree\textwidth]{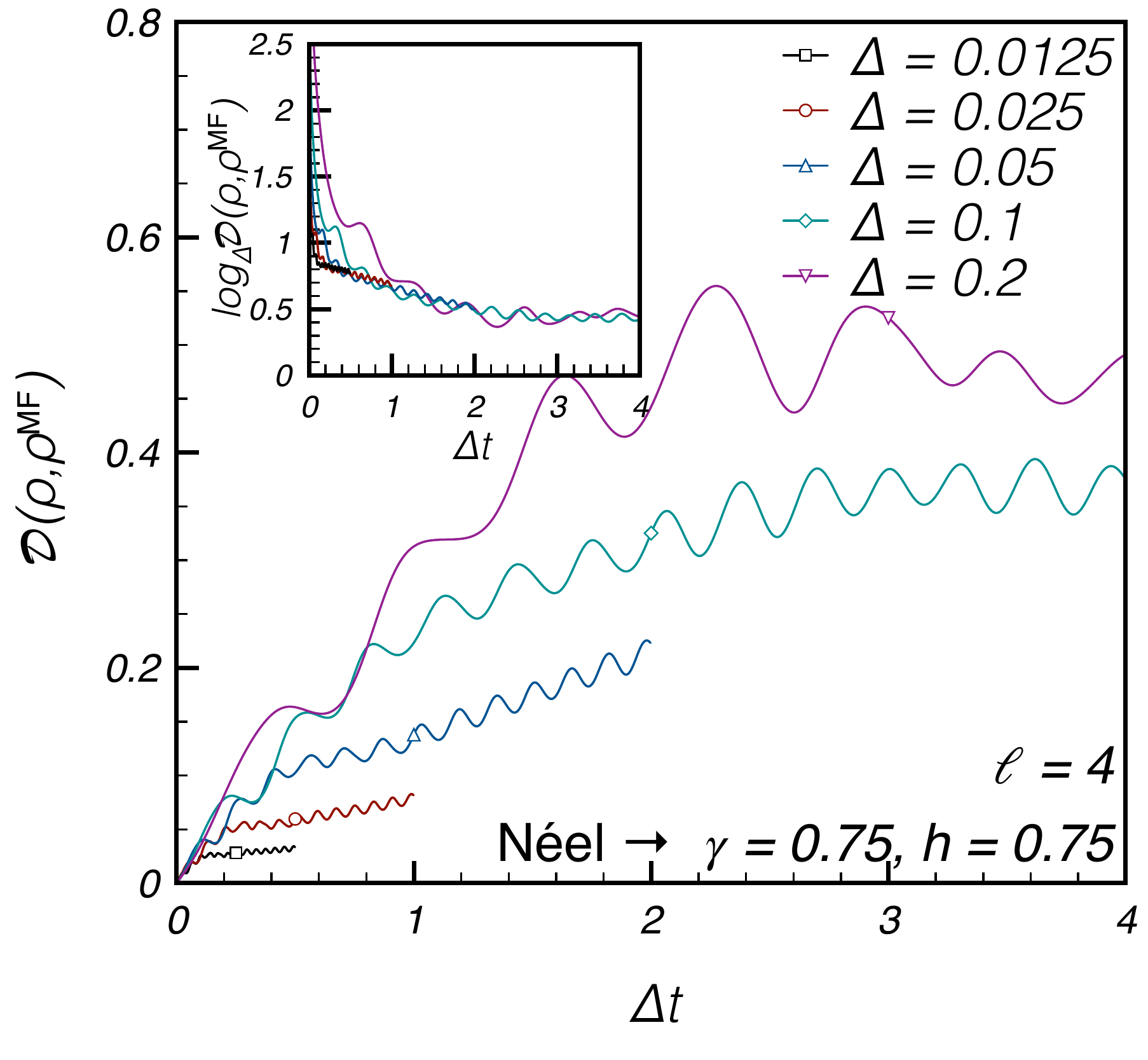}
\includegraphics[width=\ratioFthree\textwidth]{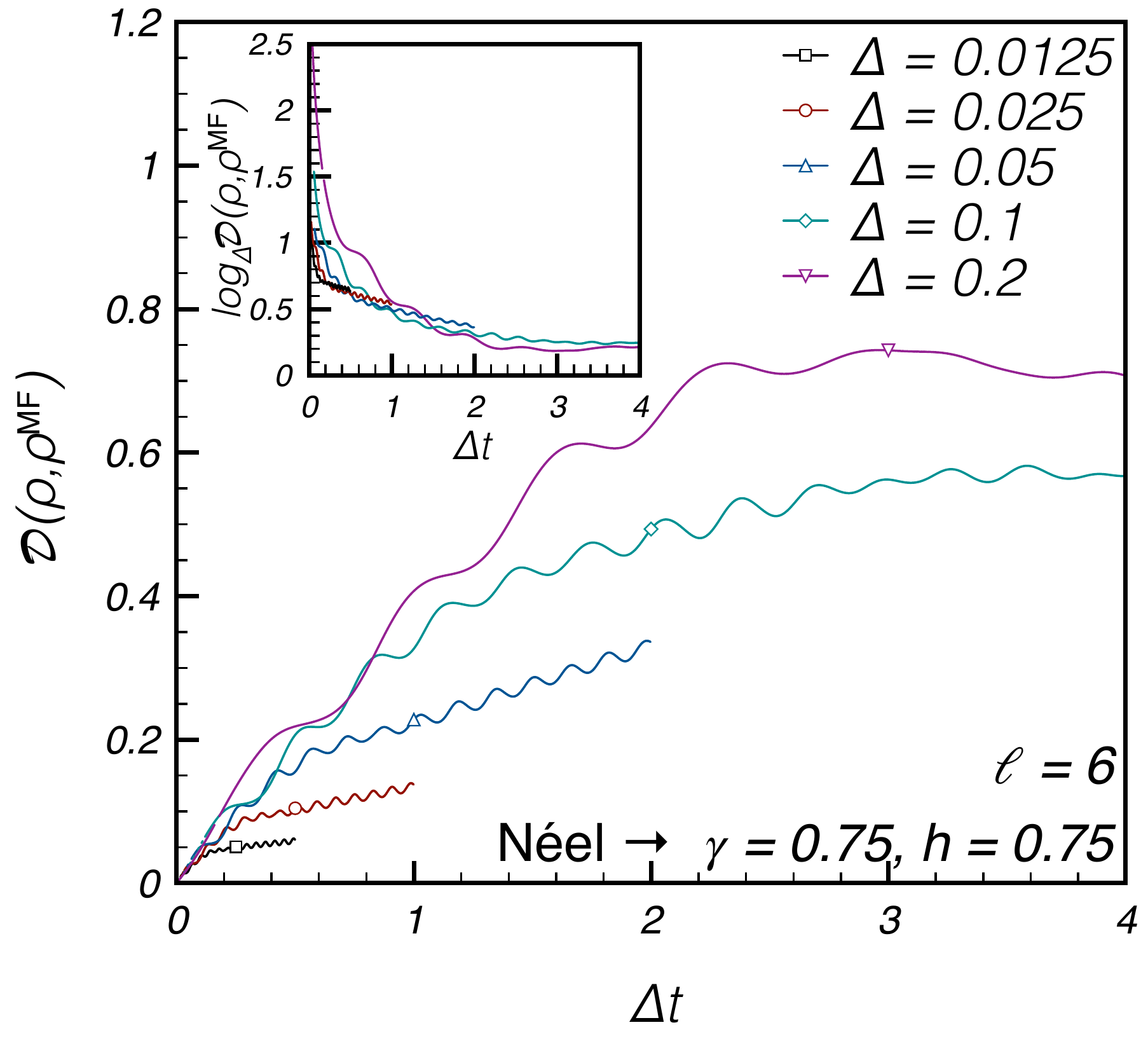}\\
\includegraphics[width=\ratioFthree\textwidth]{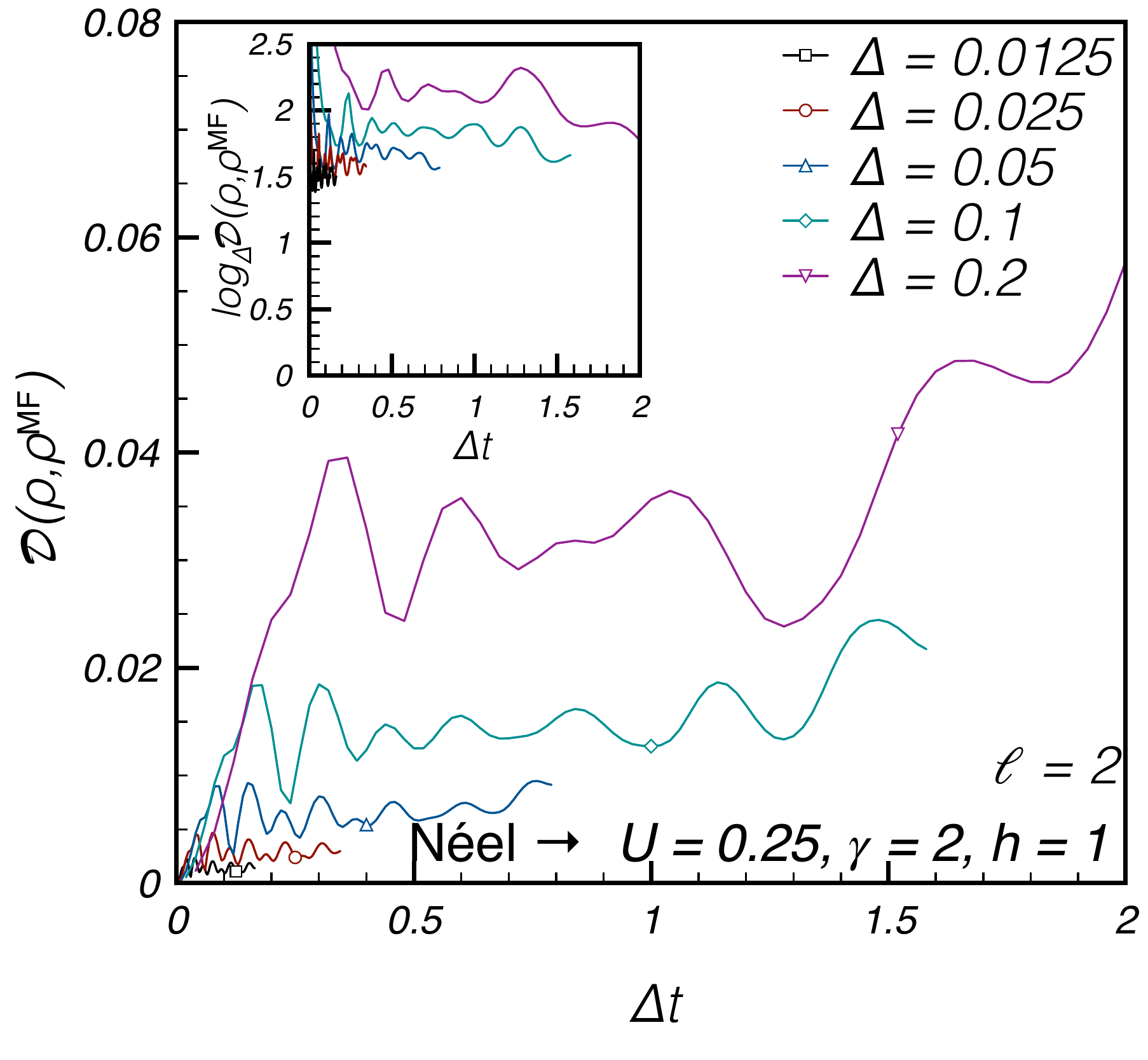}
\includegraphics[width=\ratioFthree\textwidth]{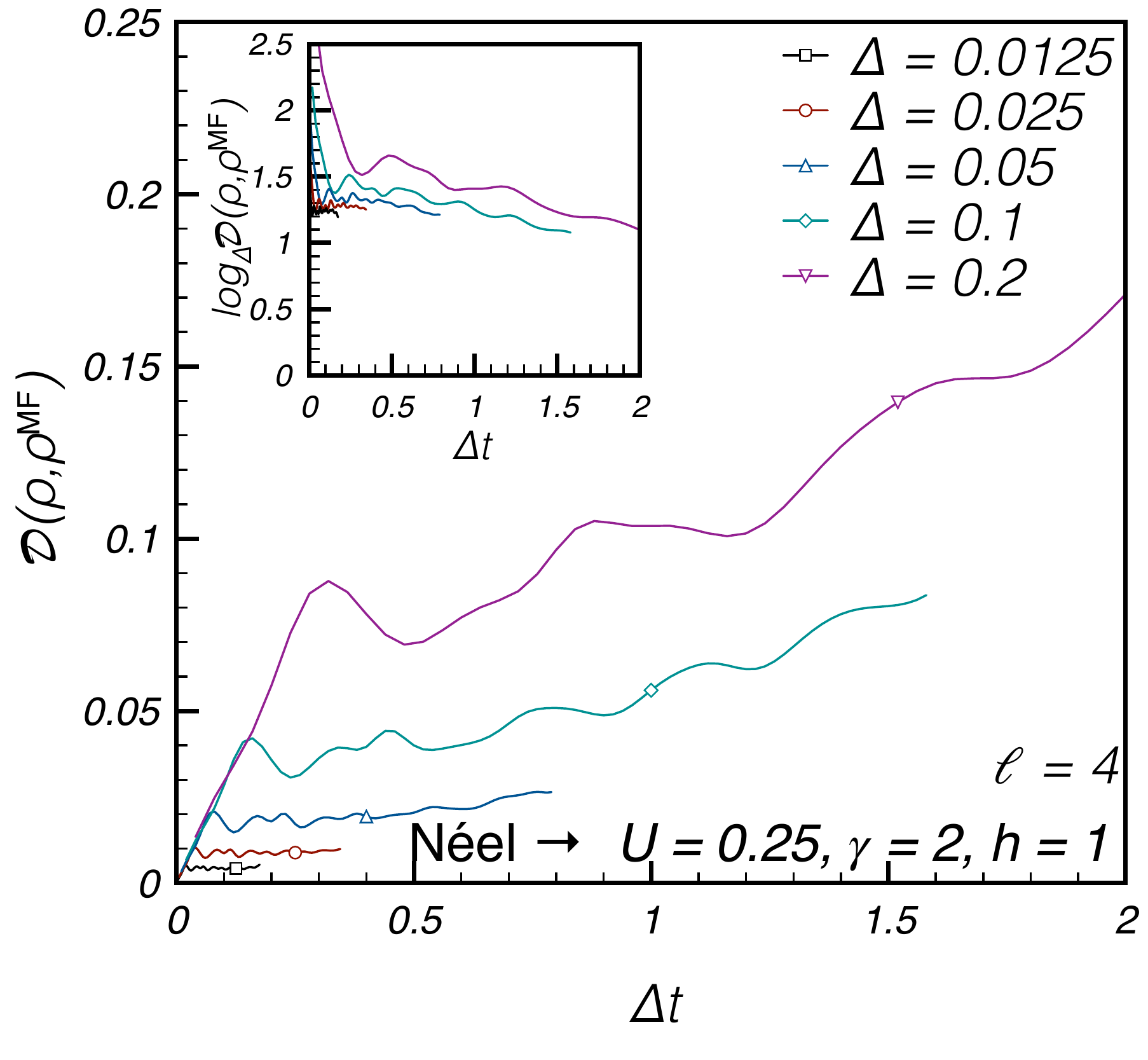}
\includegraphics[width=\ratioFthree\textwidth]{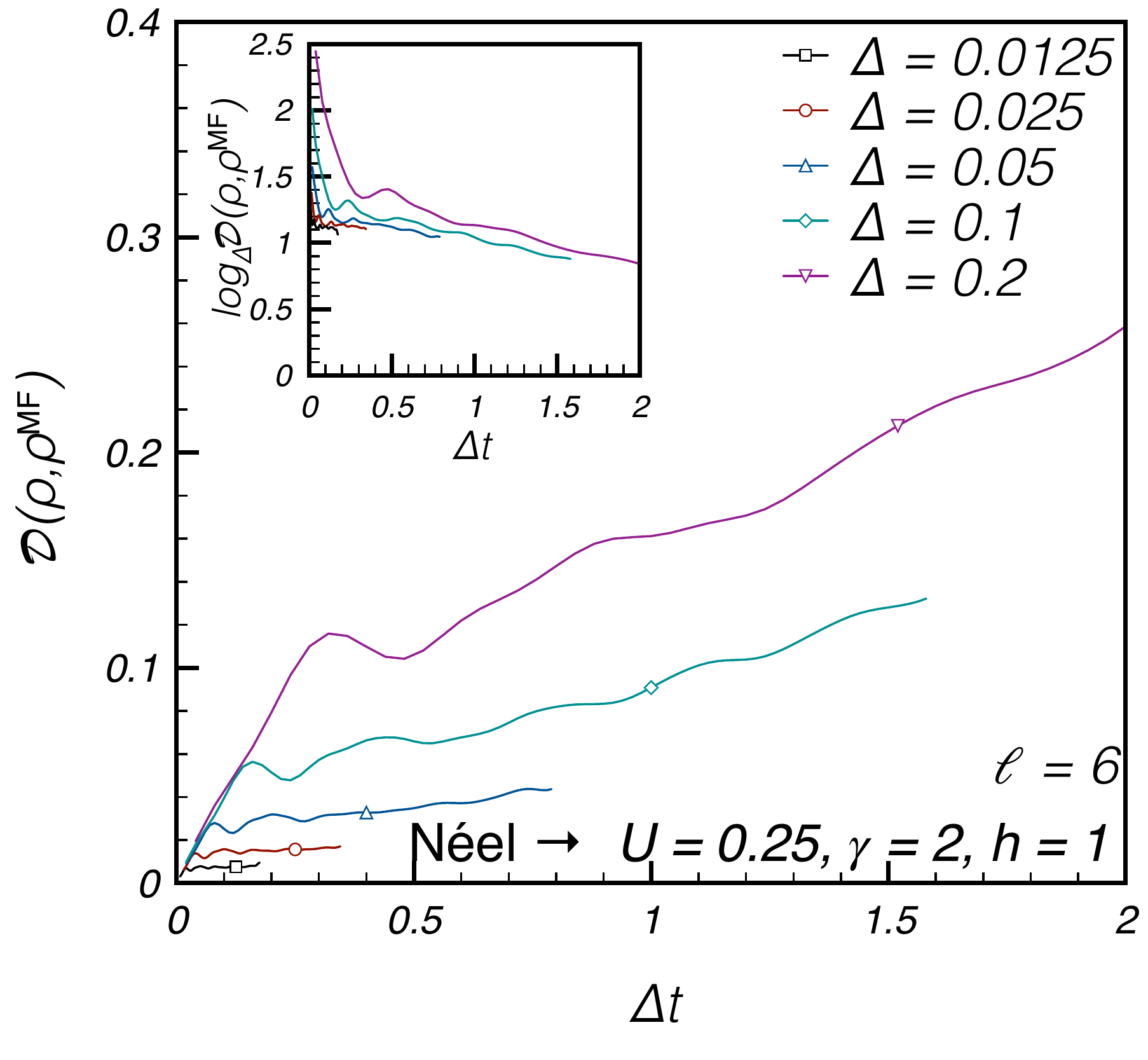}
\end{center}\caption{(Main) The normalized Frobenius distance (see the supplemental material~[\onlinecite{SM}]) 
between the exact RDM and the one obtained via the intermediate MF description~\eqref{eq:mf0}
for subsystems with a number of sites $\ell = 2,\, 4,\,6$ for the quenches \eqref{e:statea} and \eqref{e:stateb}
displayed in the panels and  various ``small'' values of $\Delta$.
As $\Delta$ gets smaller, at fixed $\Delta t$  the distance is reduced. 
(Inset) The logarithm of the distance in base $\Delta$ is shown. 
The data suggest the logarithm is approaching a finite value, that is to say the distance decays to zero as a power-law in $\Delta$.  
}\label{f:Dist_Neel}
\end{figure*}

The results of the previous sections, culminating in \eqref{eq:Sell} and Fig.~\ref{f:Sell}, rely on the mean-field description, which is supposed to emerge in the limit $\Delta\ll 1$ at times $t\sim \Delta^{-1}$. In the pre-relaxation limit it was legitimate to replace the initial state by the GGE of the unperturbed model, however in actual numerical simulations this introduces relevant corrections that can spoil the agreement with the asymptotic values. 
Here we discuss the extent of such corrections by considering an intermediate mean-field approximation, where the initial state is \emph{not} replaced by $\rho^{\rm MF}(0)$, as instead done in \eqref{eq:rho0}. We consider the following system:
\be\label{eq:mf0}
\ba
& \braket{\mathcal O}_{\rm MF}(t)=\braket{\Psi_{\rm MF}(t)| \mathcal O|\Psi_{\rm MF}(t)}\\
&i \partial_{t}\ket{\Psi_{\rm MF}(t)}= [H_0+\Delta H^{\rm MF}(\Delta t)]\ket{\Psi_{\rm MF}(t)}\\
&\ket{\Psi_{\rm MF}(0)}=\ket{\Psi_0}\, .
\ea
\ee
Differently from \eqref{eq:rho0}, we have not written the equations in terms of the rescaled time $T=\Delta t$. This choice reflects the fact that the mean-field time evolution of $\ket{\Psi_0}$ includes contributions that vary in a timescale $\sim\Delta^0$. The immediate advantage is that such description is practically undistinguishable from exact time evolution at times $t\ll \Delta^{-1}$, when the Hamiltonian can be approximated by $H_0$. Then, in the pre-relaxation limit, it becomes equivalent to \eqref{eq:rho0}\cite{BF:mf}. We do not expect any improvement in the corrections $o(\Delta^0)$ at times $t\sim\Delta^{-1}$, which are in fact  beyond mean-field, but the intermediate mean-field is useful to recognize the effects of the preliminary relaxation to the first plateau. 
Incidentally, we point out that we worked out system \eqref{eq:rho0} instead of \eqref{eq:mf0} because in the latter the number of differential equations is doubled and half of the information is useless in the pre-relaxation limit, making the solution of  \eqref{eq:mf0} less indicated for analytic investigations. In addition, we are able to compute the time evolution \eqref{eq:mf0} only if the initial state is a Slater determinant, whereas in principle \eqref{eq:rho0} can be used also quenching from an interacting model.  

The left panel of Fig.~\ref{f:Sintell} shows the time evolution of the entanglement entropy per unit length under \eqref{eq:mf0}  for various values of $\Delta$ and fixed subsystem length $\ell=32$. The scaling behavior of \eqref{eq:scaling}, including the preliminary linear growth in the space-time scaling limit \mbox{$1\ll\ell\leq t/(2v_M)$}, is completely revealed.
In addition, the data are in good agreement with the prediction \eqref{eq:Sellstar}. It is however important to note that the actual corrections to the mean-field solution could spoil the physical meaning of the plot. The larger $\Delta$ is and the shorter the time window that can be described by mean-field: it is unreasonable to expect agreement with exact dynamics when $\Delta t\sim \Delta^{-1}$!
For the subsystem length considered in the right panel of Fig.~\ref{f:Sintell} ($\ell=4$) these potential problems should be less severe. Nevertheless, oscillations apart, the behavior is essentially the same as in the left panel.

\subsection{Mean field \emph{vs} iTEBD simulations}\label{ss:comparison}%

This section is devoted to comparing the mean-field predictions  with the exact quench dynamics induced by Hamiltonian~\eqref{eq:H} and obtained via iTEBD numerical simulations for quenches starting from the N\'eel state (some details of the numerical simulations are reported in the supplemental material~[\onlinecite{SM}]).
We consider two different quenches, with parameters: 
\begin{enumerate}[(a)]
\item \label{e:statea}$\gamma =0.75,\, h=0.75,\, U=0$; 
\item \label{e:stateb}$\gamma =2,\, h=1,\, U=0.25$.
\end{enumerate}
In both cases we expect relaxation in the pre-relaxation limit but in \eqref{e:stateb} one-site shift invariance is not restored. 
In the iTEBD simulations the (small) parameter $\Delta$ has been chosen in the interval $[0.0125,0.2]$.
The computational complexity does not change much with the interaction strength, making it extremely difficult to reach large
rescaled times $\Delta t$ for the smallest values of $\Delta$. This is the reason why we have also examined quenches with values of the interaction that can not be considered strictly small,
e.g. $\Delta = 0.1$ or $0.2$. Nonetheless, we would emphasize from the very beginning that, 
even for these values of $\Delta$, we expect  an overall trend that should at least qualitatively agree with mean field.  In other words, even if the crossover is mathematically defined only in the pre-relaxation limit, the effects should remain visible also for finite values of $\Delta$.

Fig \ref{f:mz_Neel} shows the time evolution of the magnetization $m_z$. The iTEBD numerical data (symbols in the plots) exhibit deviations from mean-field that are reduced as $\Delta$ gets 
closer and closer to zero. This is exactly what we expect: a finite value 
of the interaction strength necessarily introduces some corrections to mean-field
dynamics.
The magnetization moves away from the ${\rm GGE}_0$ value  $m_z=0$ and, up to subleading corrections, data for different $\Delta $ and fixed $\Delta t$ collapse to a curve that is compatible with the mean-field prediction. This is a conclusive confirmation that the pre-relaxation limit is nontrivial and apparently described by mean-field. In addition, the
magnetization seems to approach a nonzero value for arbitrarily small $\Delta$. Since the energy density is $-\frac{\Delta}{8}$ and, in turn, in the limit $\Delta\rightarrow 0$ the thermal expectation value approaches zero, one can infer just from numerics that the second plateau emerging in the pre-relaxation limit is different from the infinite time limit. It corresponds to the highly nontrivial pre-thermalization plateau that, as mentioned in the introduction, is generally developed after a pre-relaxation plateau when the unperturbed model is non-abelian integrable.

In Figs \ref{f:S1_Neel} 
and \ref{f:S6_Neel} we
show the time
evolution of the entanglement entropy for subsystems with $\ell = 1,\,
2,\, 4,$ and $6$ lattice sites, respectively.
In particular we compare the iTEBD data with both the mean-field prediction starting
from the ${\rm GGE}_0$ (thick red lines in
the figures) and the intermediate mean-field 
\eqref{eq:mf0} (thin colored lines). 
 Let us analyze more attentively the behavior of the entropies. Generally,
$S_{\ell}$ experiences an almost linear growth (apart from oscillations) up to a
time \mbox{$t^{*} \simeq \ell/(2 v_M)$}, where the maximal velocity is given by $v_M=|\gamma-1|+o(\Delta^0)$; after that, the entropy approximately
 reaches the corresponding ${\rm GGE}_0$ value. This transient
represents the well-known 
Calabrese-Cardy regime \cite{CC:05} of linearity after
a global quench. 
Obviously,  in the limit $\Delta\to 0$ at fixed rescaled time $\Delta t$, 
this is squeezed into an infinitesimal slice at zero rescaled time (\emph{cf}. Fig. \ref{f:Sintell}), so the entropy in the ${\rm GGE}_0$ becomes the starting value of the entropy in the mean-field description. 
However, the iTEBD simulations are done at fixed $\Delta$ and the larger the interaction and the
subsystem size are and  the more evident the transient is.

We observe a good agreement between
numerical simulations and mean-field,
within the limits of corrections that tend to zero as $\Delta\rightarrow 0$.
Nevertheless, some comments are in order: the interplay between $\ell$ and $\Delta$ determines the size of the corrections to mean-field. 
We expect mean-field to be valid in a
regime $\Delta t  \ll \Delta^{-\alpha}$ and $\ell \ll \Delta^{-\beta}$, with $\alpha$ and $\beta$ some unknown exponents.
The results of Ref.~[\onlinecite{BEGR:preT}] suggest that thermalization in generic models is expected in a timescale $\sim\Delta^{-2}$, so we can set $\alpha=1$; concerning $\beta$, the emergent mean-field scaling regime $\Delta t\sim \ell$ points to a correspondence between $\ell$ and $\Delta t$, so we assume $\beta\sim\alpha=1$.  
These estimations allow us to understand why for $S_{6}$ the
data show larger deviations from mean-field: for $\Delta = 0.2$ (or $0.1$), one has
$\Delta \ell \sim 1$, which could be outside  
the region of validity of mean-field. On the other hand, $S_{1}$ and $S_2$ show excellent agreement
with the theoretical predictions.

In order to establish the validity of the mean-field dynamics we also carried out a meticulous analysis
of the scaling properties of the normalized Frobenius
distance $\mathcal{D}(\rho_{\ell},\rho_{\ell}^{\rm MF})=||\rho_{\ell}-\rho_{\ell}^{\rm MF}||/(||\rho_{\ell}||^2+||\rho_{\ell}^{\rm MF}||^2)^{1/2} $  considered in Ref.~[\onlinecite{FE:13}] (see also the supplemental material~[\onlinecite{SM}]) between the iTEBD reduced density matrices
and those obtained via \eqref{eq:mf0}.

In Fig \ref{f:Dist_Neel} we show the time evolution of  $\mathcal{D}(\rho_{\ell},\rho_{\ell}^{\rm MF})$ for different  values of $\ell$ and $\Delta$. 
 In the regime of validity of mean-field
$ \Delta\ell/(2|\gamma-1|) \lesssim\Delta t \lesssim \Delta^{-1}$ (the lower bound is for the mean-field \eqref{eq:rho0}), the distance
$\mathcal{D}(\rho_{\ell},\rho_{\ell}^{\rm MF})$
is reduced in magnitude as $\Delta$ becomes smaller and smaller. Moreover, for fixed $\Delta$, the distance increases as the 
subsystem length becomes larger.

If the assumptions behind \eqref{eq:rho0} are satisfied, for $\Delta\to 0$ the distance between the two reduced density matrices should vanish. Showing the base-$\Delta$ logarithm of the distance for a sequence of $\Delta$ at fixed ratio between the one and the next ($\Delta=0.2\cdot  2^{-n}$) is the way that we chose to bring it out: a nonzero limiting distance would result in the logarithm to approach zero as \mbox{$\sim 1/(n+2.3)$}, which can be fairly recognized by sight (instead, the logarithm remains nonzero if the distance scales as a power law in $\Delta$).
Our impression based on the analysis of the insets of Fig. \ref{f:Dist_Neel} is that the data are more compatible with the distance approaching zero as a power law in $\Delta$ instead of reaching a nonzero value that might have signaled some issues in the hypotheses behind \eqref{eq:rho0}.

\begin{figure*}[t!]
\begin{center}
\includegraphics[width=\ratioF\textwidth]{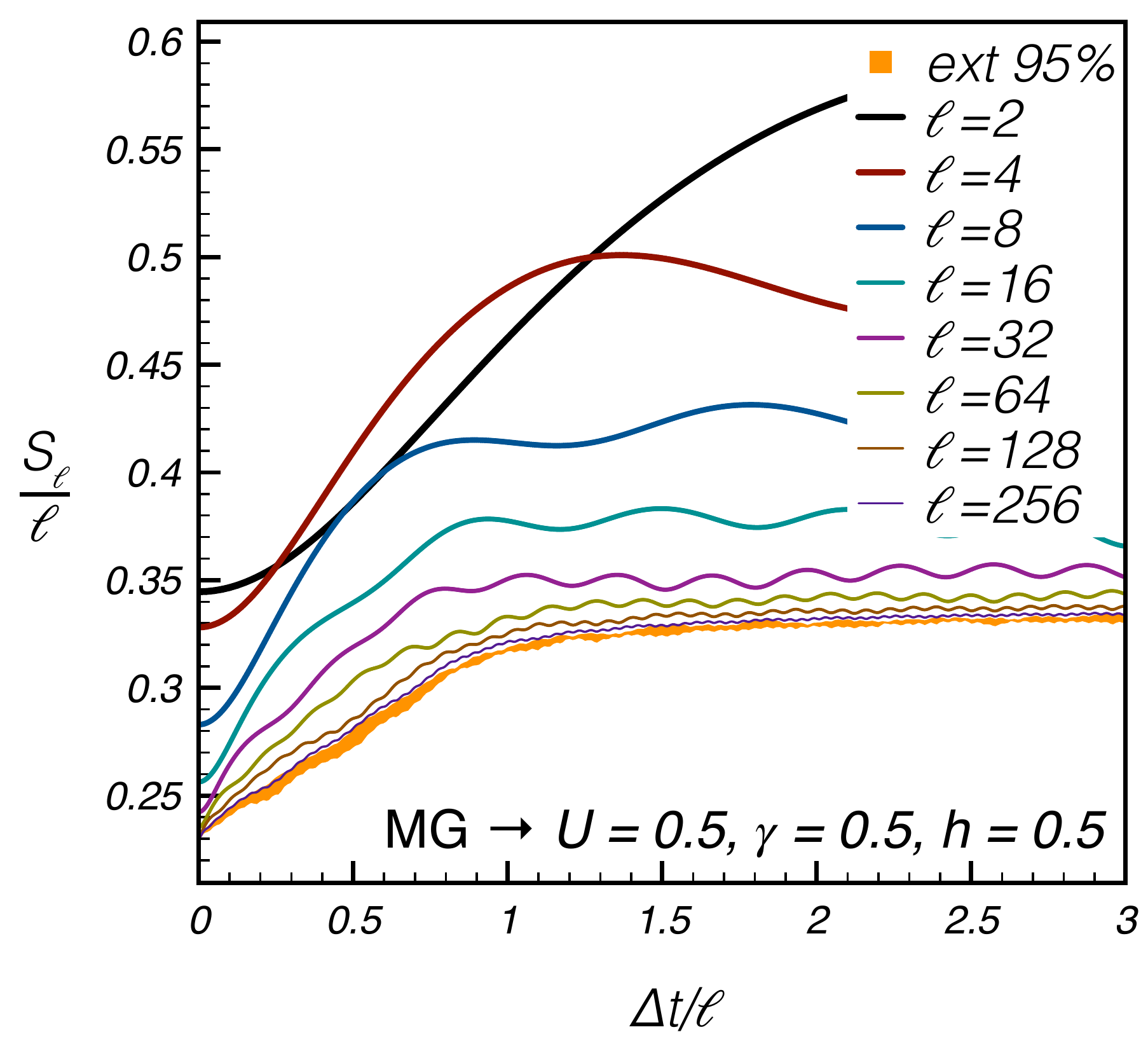}\hspace{\Hspace}
\includegraphics[width=\ratioF\textwidth]{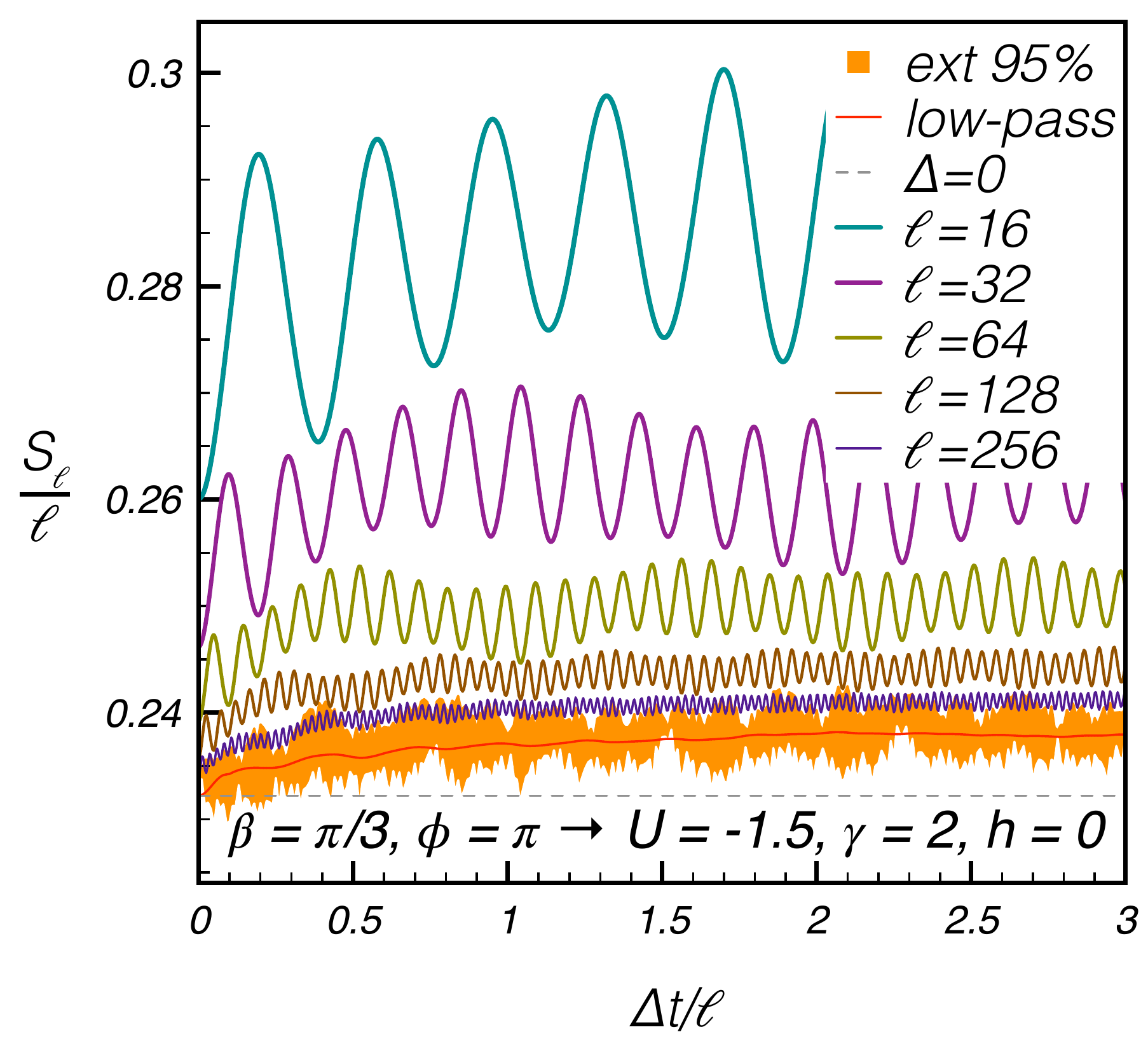}
\end{center}\caption{Evolution of the entanglement entropy per unit length within mean-field \eqref{eq:rho0} for various lengths for the quenches \eqref{e:statec} and \eqref{e:stated} displayed in the panels. In both cases, expectation values of local observables show persistent oscillatory behavior. (Left) In the pre-relaxation limit the entropy per unit length exhibits the same scaling behavior \eqref{eq:scaling} as in the cases of local relaxation. (Right) Persistent oscillations are an indication of less effective dephasing mechanisms: for large subsystems the entropy per unit length does not grow significantly.  The red curve (low-pass) is an extrapolation based on the average of the upper and lower envelopes. The accuracy of the extrapolations is clearly spoiled by the sizable oscillations.
}\label{f:SMFosc}
\end{figure*}

\section{Persistent oscillations}\label{s:oscillations}%

\begin{figure*}[t!]
\begin{center}
\includegraphics[width=\ratioF\textwidth]{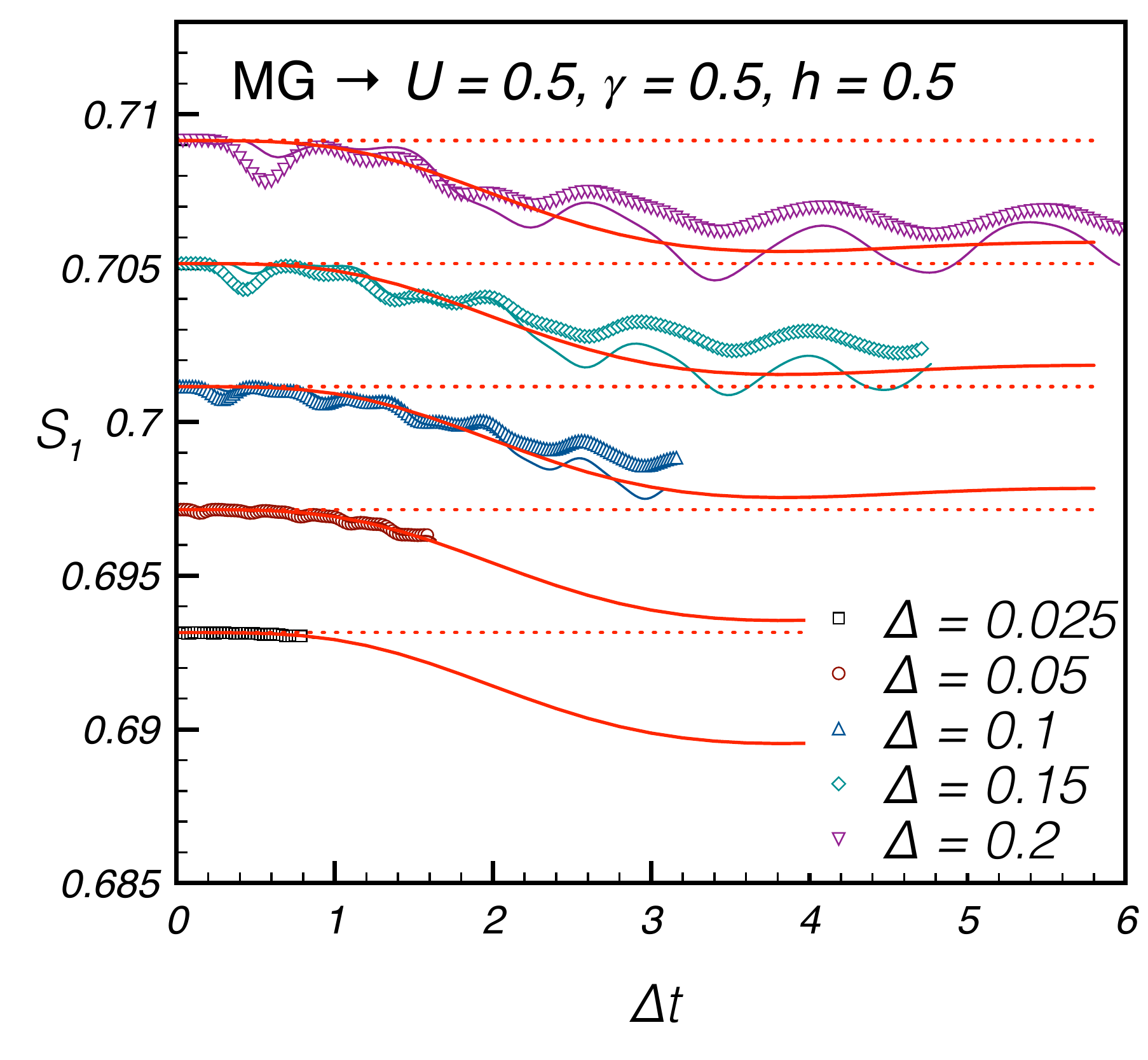}
\hspace{\Hspace}\includegraphics[width=\ratioF\textwidth]{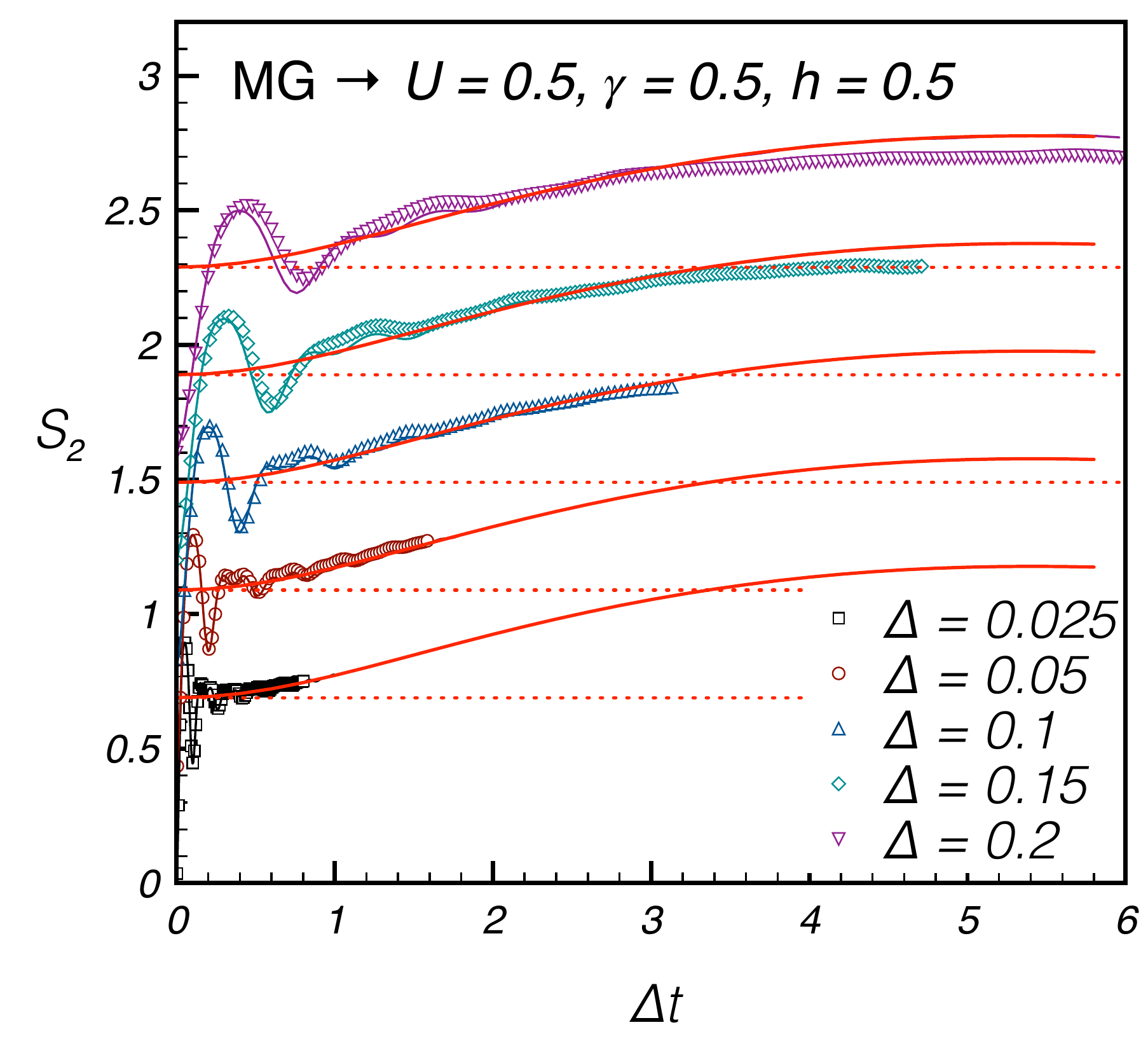}\\
\includegraphics[width=\ratioF\textwidth]{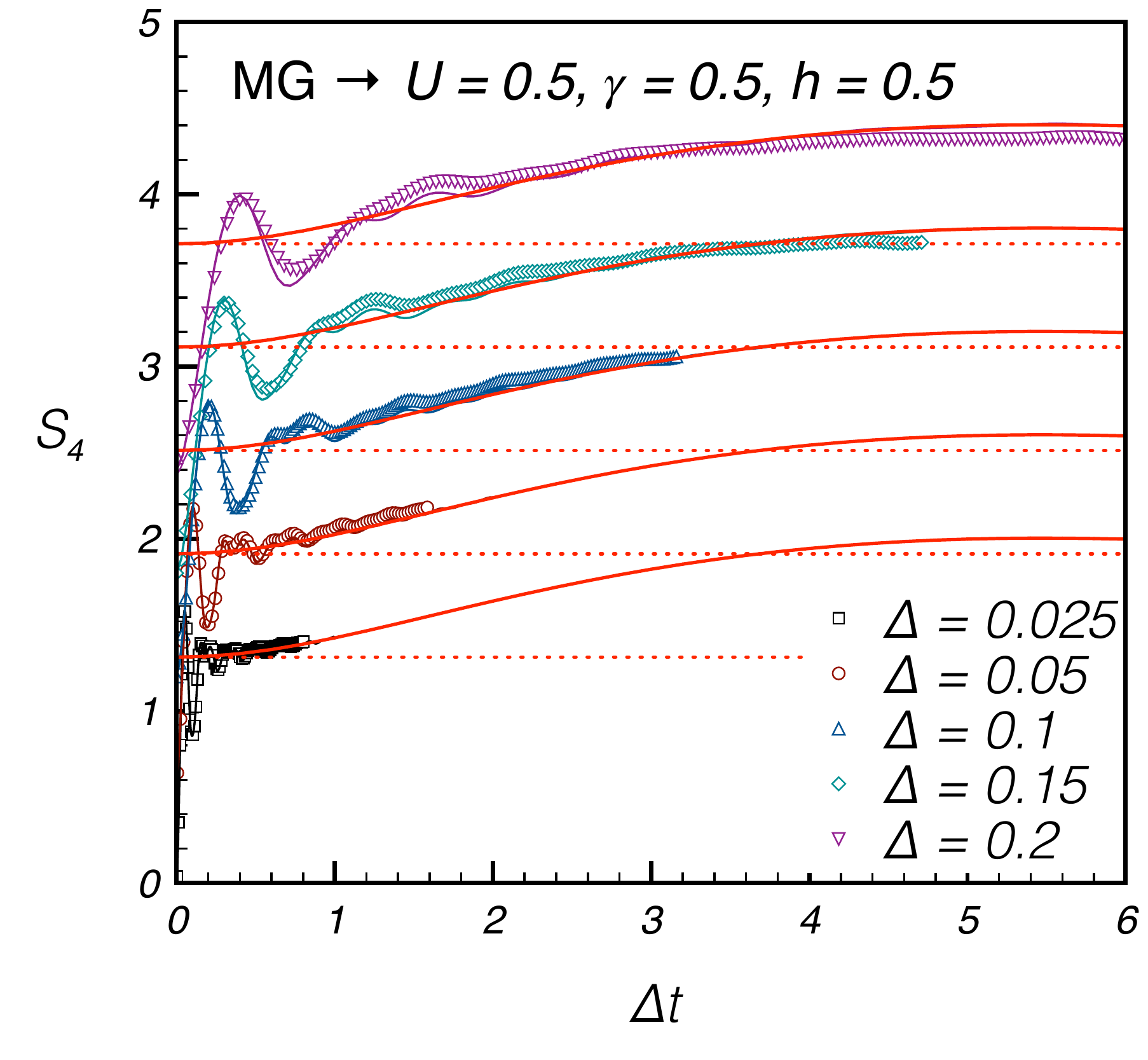}\hspace{\Hspace}\includegraphics[width=\ratioF\textwidth]{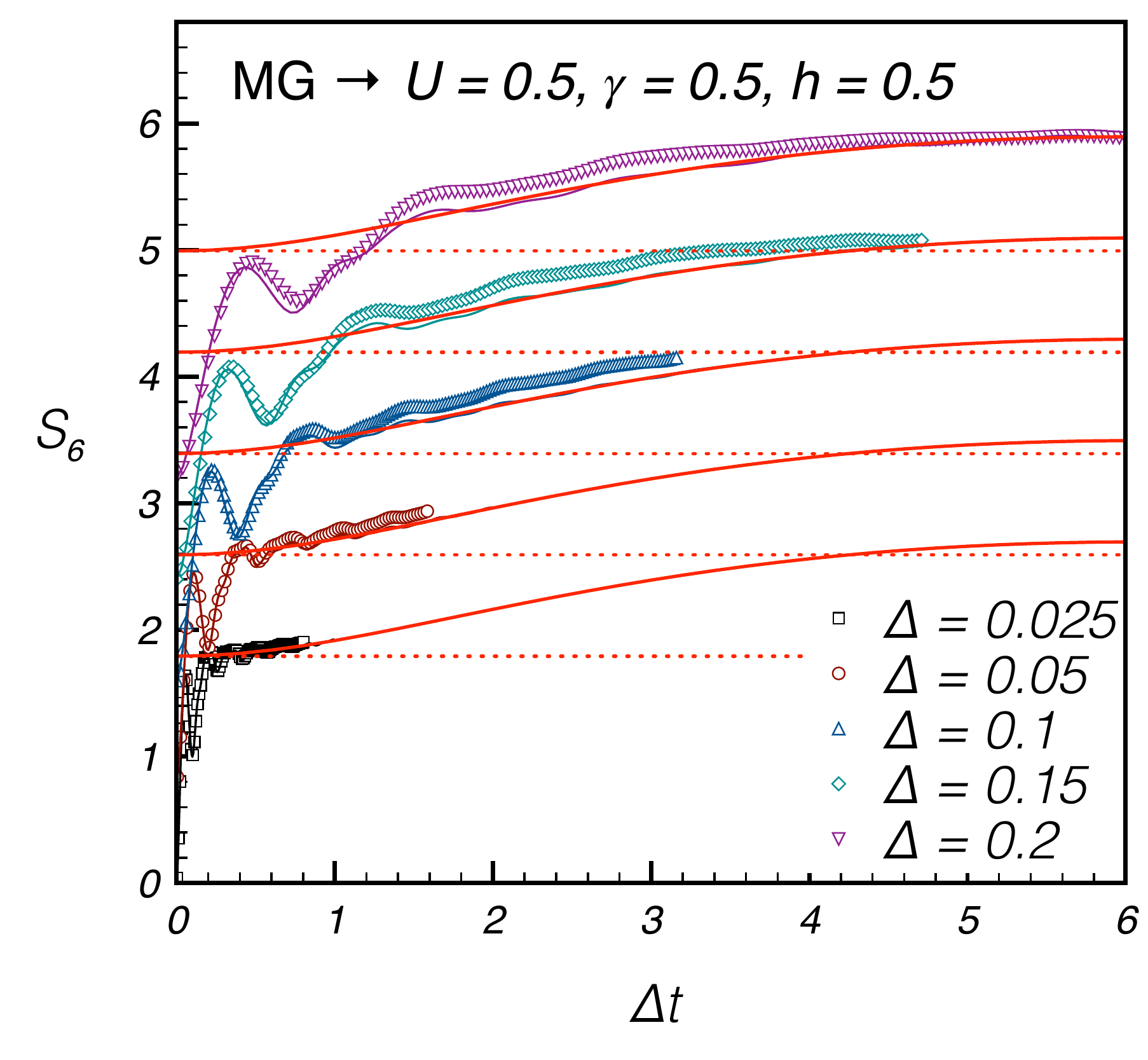}
\end{center}\caption{Evolution of the von Neumann entropies for the quench \eqref{e:statec} displayed in the panels. 
For the sake of clarity all datasets with $\Delta > 0.025$ have been vertically shifted. 
The notations are the same as in Fig. \ref{f:S1_Neel}.
}\label{f:S_MG}
\end{figure*}

As briefly discussed in Section \ref{p:osc}, there are situations in which the solution of the mean-field equations displays persistent oscillatory behavior at arbitrarily long times.  
In Ref.~[\onlinecite{BF:mf}] persistent oscillations have been interpreted as manifestations of localized (quasi-)excitations. Following that point of view, the oscillatory behavior is not expected to modify the extensive part of the entanglement entropy. 
We test this conjecture on two quenches: 
\begin{enumerate}[(a)]
\setcounter{enumi}{2}
\item \label{e:statec}Time evolution of the ground state of the Majumdar-Ghosh Hamiltonian ($\beta=\frac{\pi}{2}$ and $\phi=\pi$ in \eqref{eq:state}) under the Hamiltonian \eqref{eq:H} with $\gamma=0.5$, $h=0.5$ and \mbox{$U=0.5$}. 
\item \label{e:stated}Initial state that breaks reflection symmetry, namely  $\beta=\frac{\pi}{3}$ and $\phi=\pi$ in  \eqref{eq:state}, with Hamiltonian parameter $\gamma=2$, $h=0$ and $U=-1.5$. 
\end{enumerate}
We briefly comment on the mean-field solution of the dynamics. 
In case \eqref{e:statec} the nontrivial part of the dynamics involves the families of charges with symbols \eqref{eq:Q} proportional to $\mathcal Q_2(k)$, $\mathcal Q_5(k)$, and $\mathcal Q_6(k)$. Instead, in case \eqref{e:stated} the relevant symbols are $\mathcal Q_4(k)$, $\mathcal Q_5(k)$, and $\mathcal Q_8(k)$. Together with the quenches from the N\'eel state (where the symbols involved were $\mathcal Q_2(k)$, $\mathcal Q_7(k)$ and $\mathcal Q_8(k)$), these are representative of all the quenches starting from  states of the form \eqref{eq:state} in which the mean-field dynamics is reduced to three families of local conservation laws. 
We do not report the corresponding systems of differential equations, which however can be found in Ref.~[\onlinecite{BF:mf}].

\subsection{Numerical analysis within mean-field}\label{ss:osc_withinMF}  %

Figure \ref{f:SMFosc} shows the time evolution of the von Neumann entropy for the  quenches \eqref{e:statec} and \eqref{e:stated}.  
In both cases the oscillations enter as subleading corrections to the entropies per unit length, as it was conjectured in Section \ref{p:osc}.  
In the left panel the nontriviality of the scaling limit is manifest. 
In addition, the data are compatible with the asymptotic formula \eqref{eq:scaling}.   
We notice however that the increase in the entropy is modest. This is not unexpected since the presence of persistent oscillations signals that the dephasing mechanisms behind the relaxation process are less effective. As a consequence, the entropy growth is restrained. 
This effect can be so important that the extensive part of the entropy could appear almost independent of time, as in the right panel of Figure \ref{f:SMFosc}. In the limit of large subsystem length the entropy practically does not move from the value associated with the unperturbed GGE. This time, even though very large subsystems are considered, it is difficult to identify the linear part of the time evolution. However, an extrapolation with confidence level $95\%$ suggests that the scaling limit is likely to be different from the value corresponding to the unperturbed GGE.

Since we have not analytically computed the asymptotics of the entropy for quenches with persistent oscillations and our intuition is only based on the qualitative picture of Section \ref{s:picture}, we can not rule out the emergence of other scaling behaviors when there are persistent oscillations and the initial state is sufficiently general. 
Specifically, the mapping of the mean-field time evolution into a sudden quench described in Section \ref{ss:VT} can not be used without modifications and, for example, we can not exclude that the limit $\Delta t\rightarrow 0$ in the asymptotic expression could be different from the value corresponding to the GGE of the unperturbed model. The right panel of Fig. \ref{f:SMFosc} suggests indeed this might be happening. 
Nevertheless, all the cases that we considered are compatible with the scaling law~\eqref{eq:scaling}.

\subsection{Mean field \emph{vs} iTEBD simulations}\label{ss:osci_comparison}

\begin{figure*}[t!]
\begin{center}
\includegraphics[width=\ratioF\textwidth]{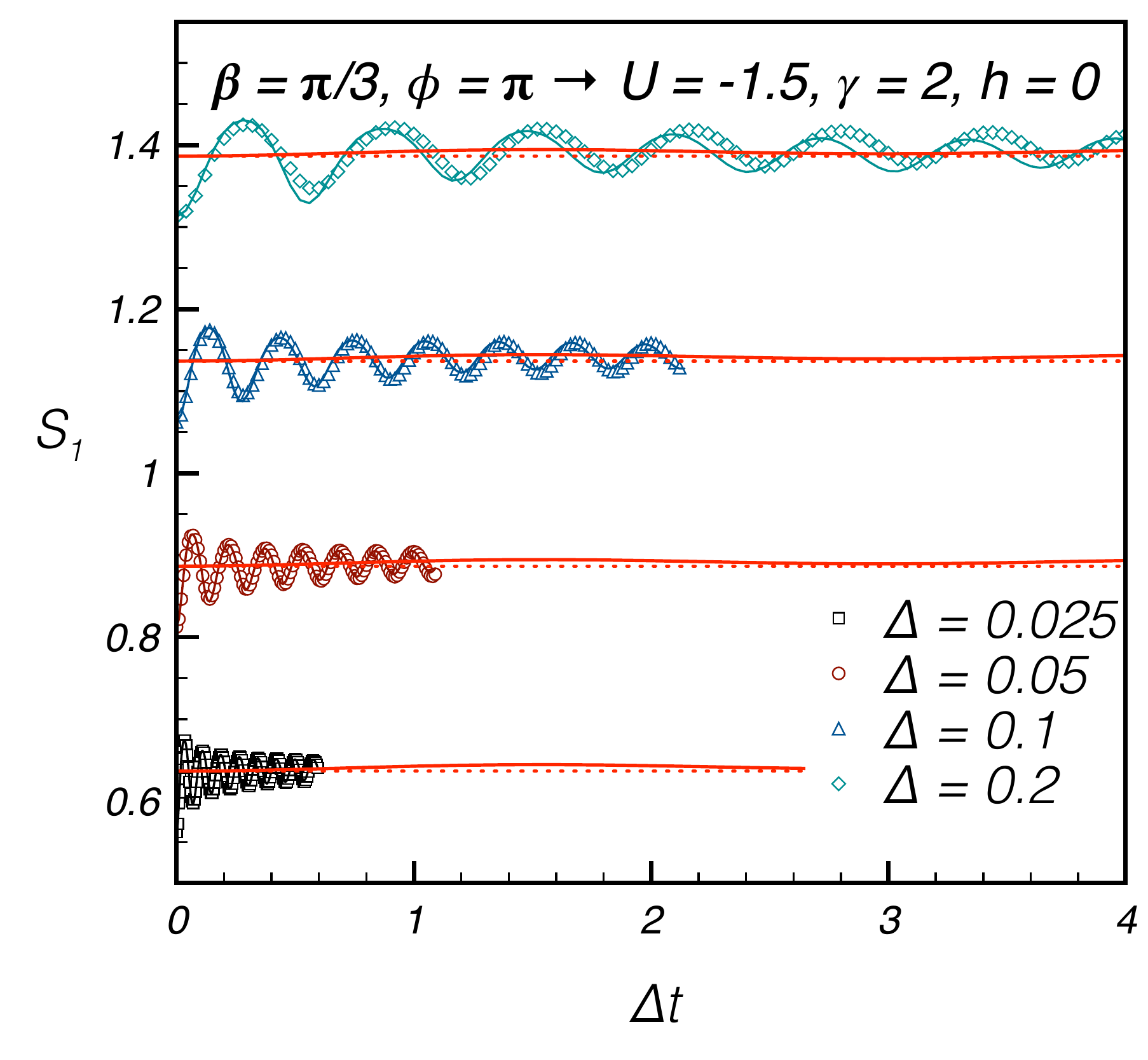}\hspace{\Hspace}\includegraphics[width=\ratioF\textwidth]{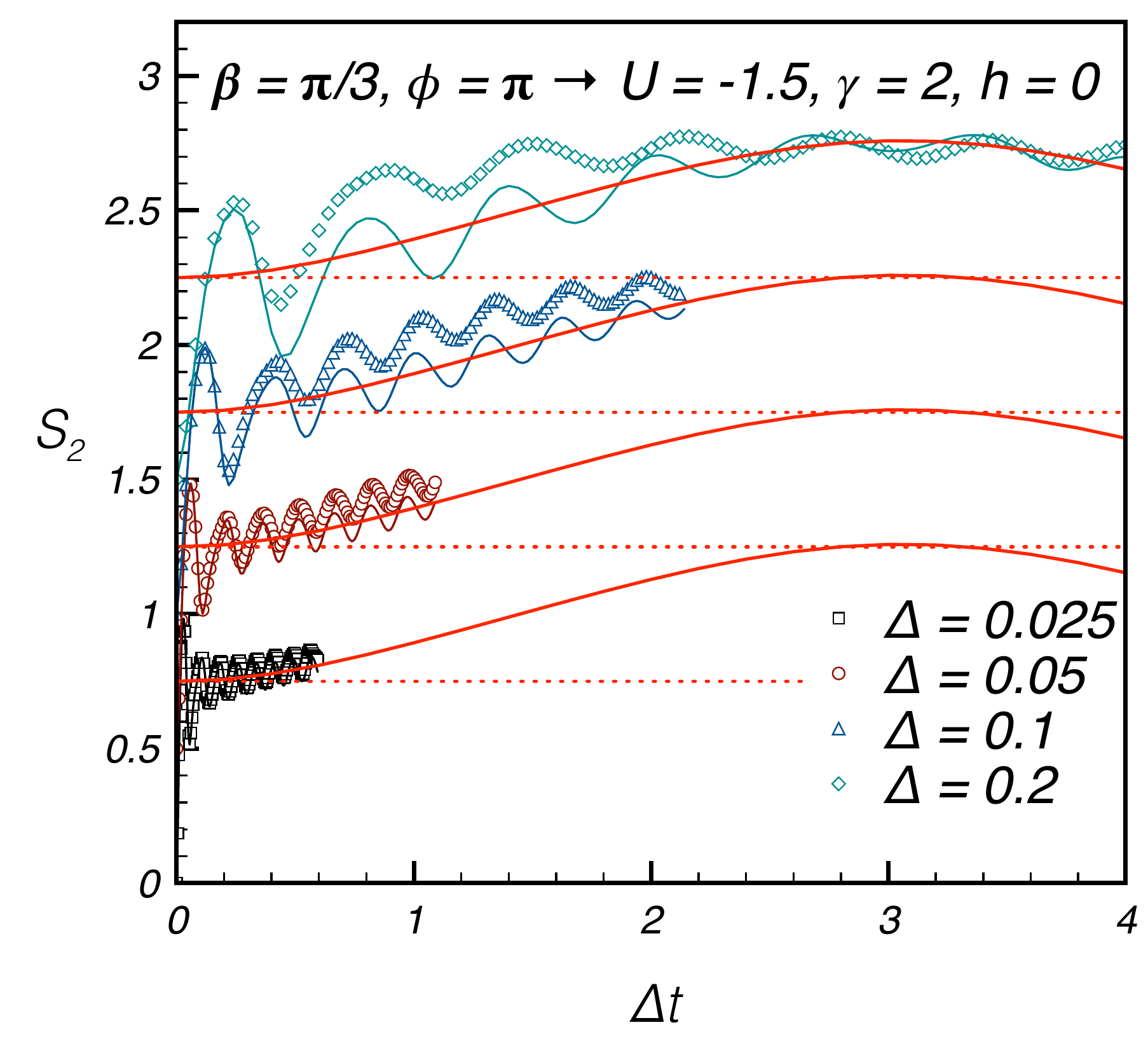}\\
\includegraphics[width=\ratioF\textwidth]{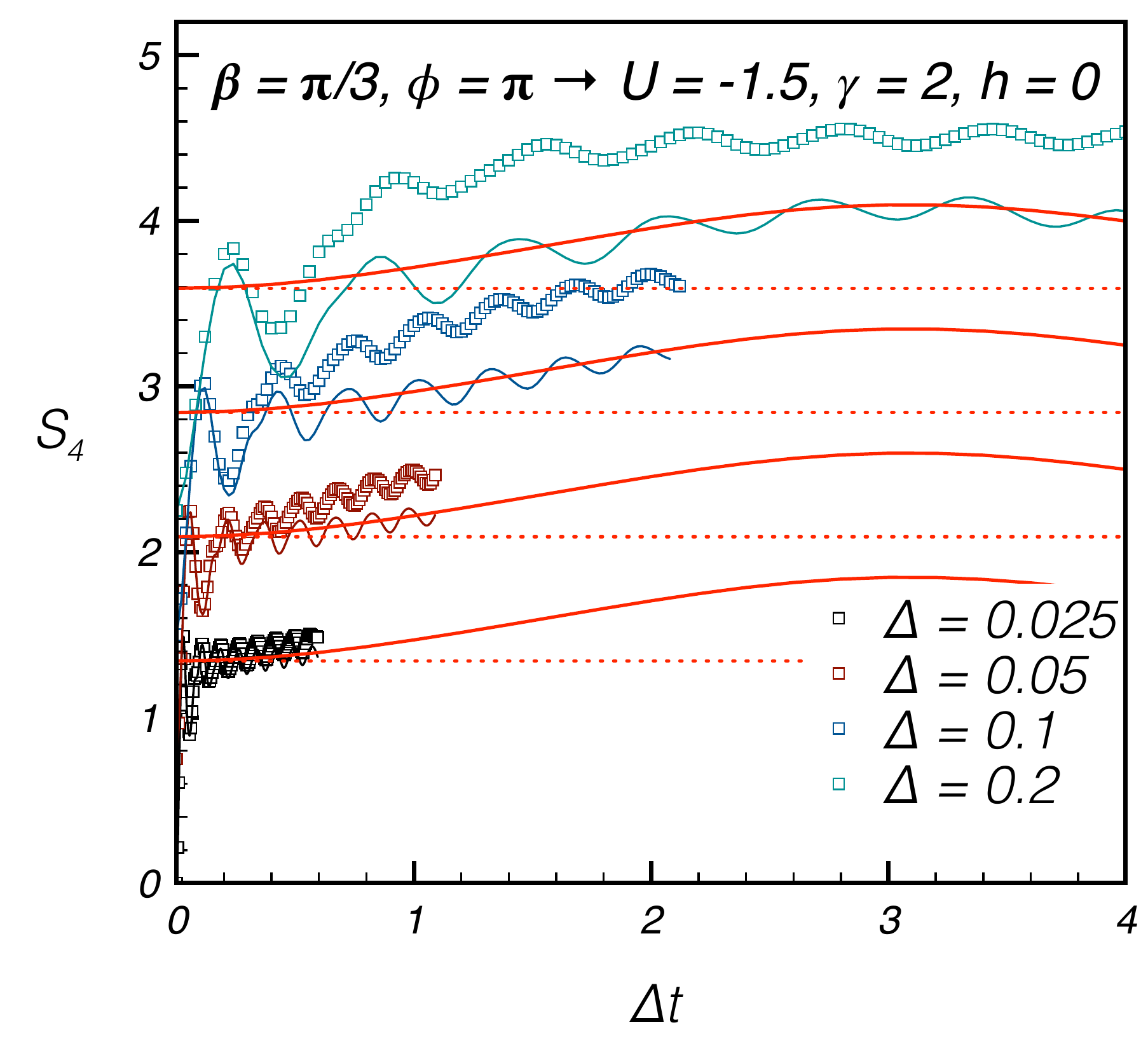}\hspace{\Hspace}\includegraphics[width=\ratioF\textwidth]{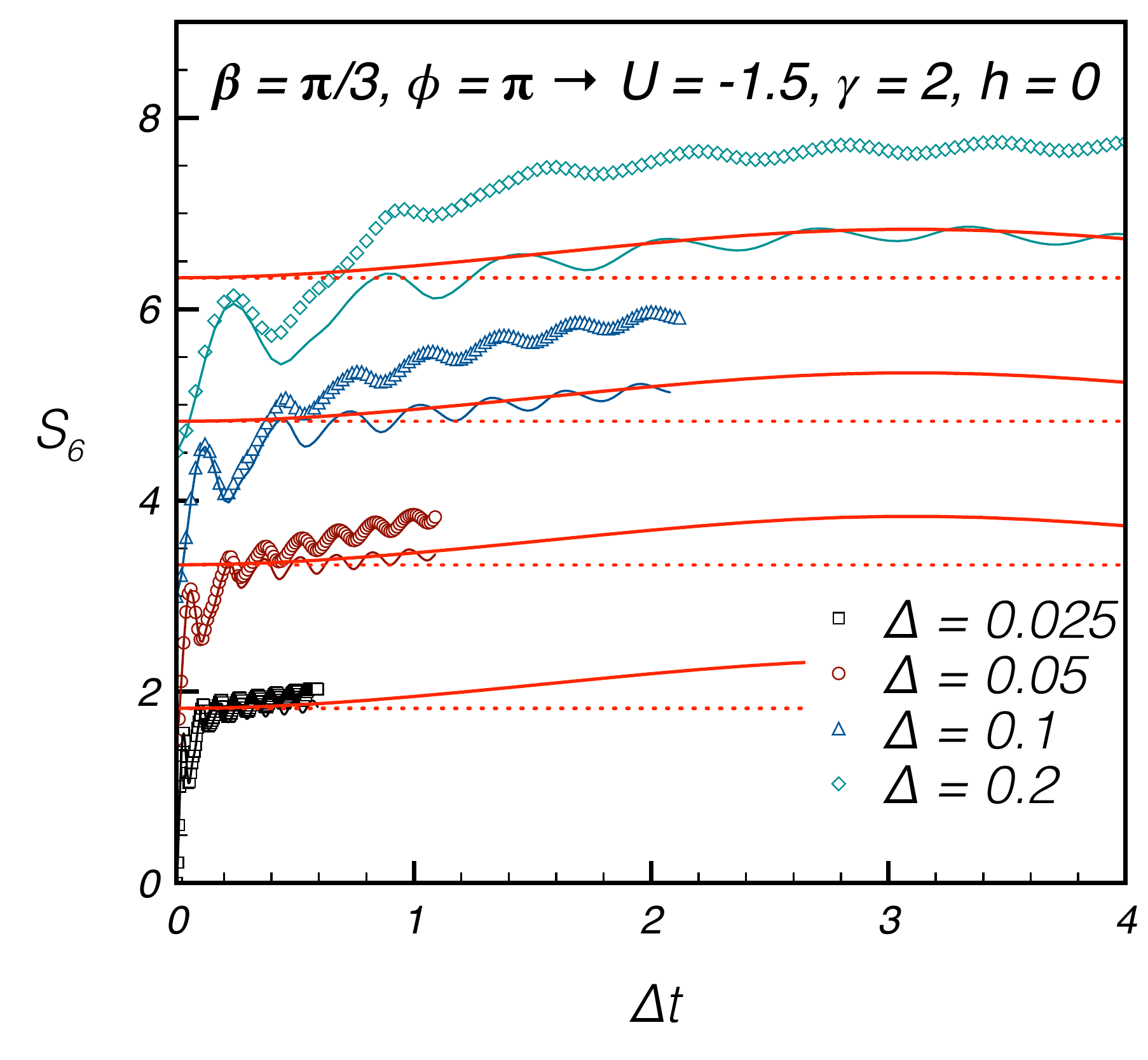}
\end{center}\caption{The same as in Figure \ref{f:S_MG} for the quench \eqref{e:stated} displayed in the panels.
}\label{f:S_betaphi}
\end{figure*}

\begin{figure*}[t!]
\begin{center}
\includegraphics[width=\ratioFthree\textwidth]{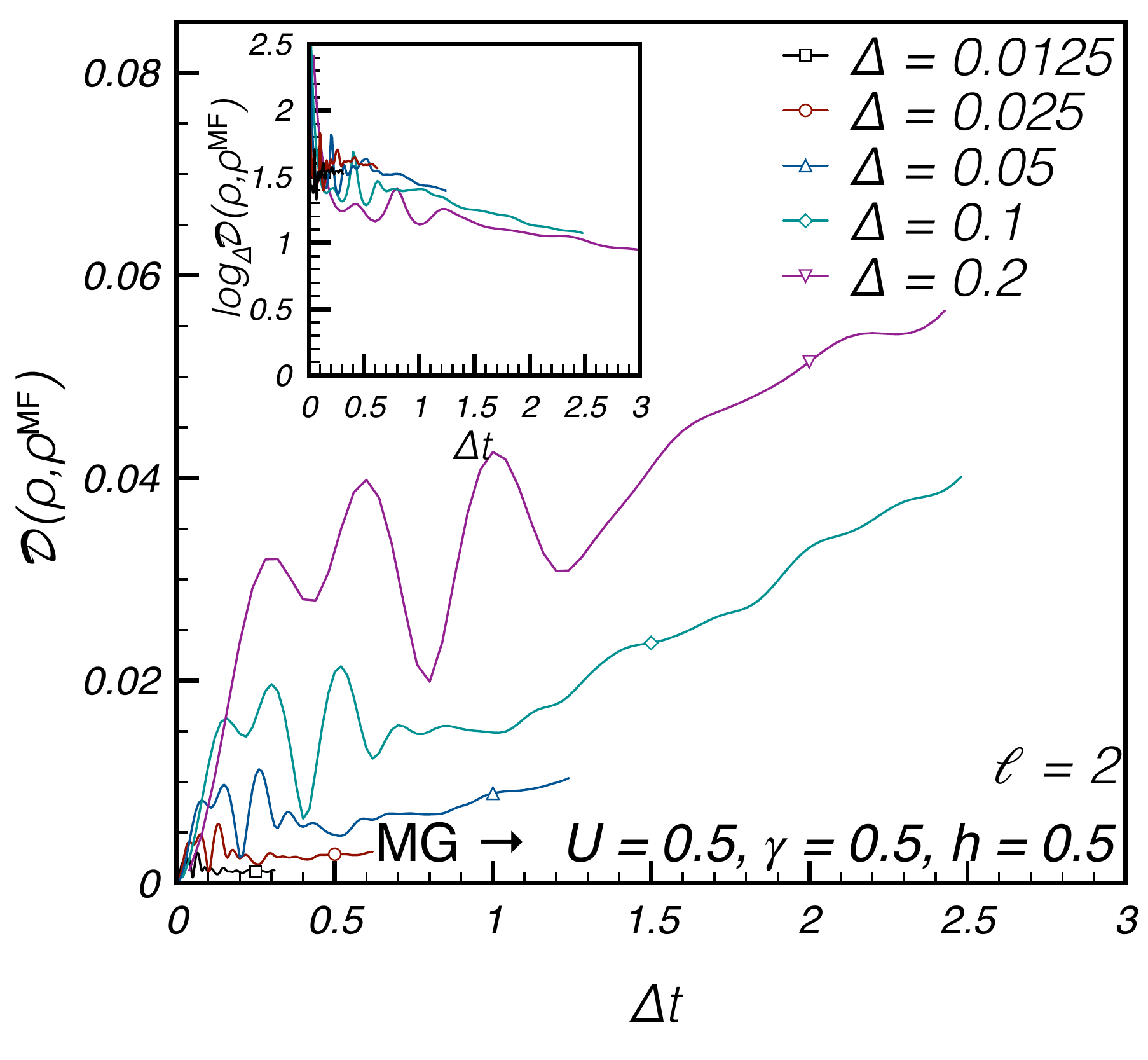}\includegraphics[width=\ratioFthree\textwidth]{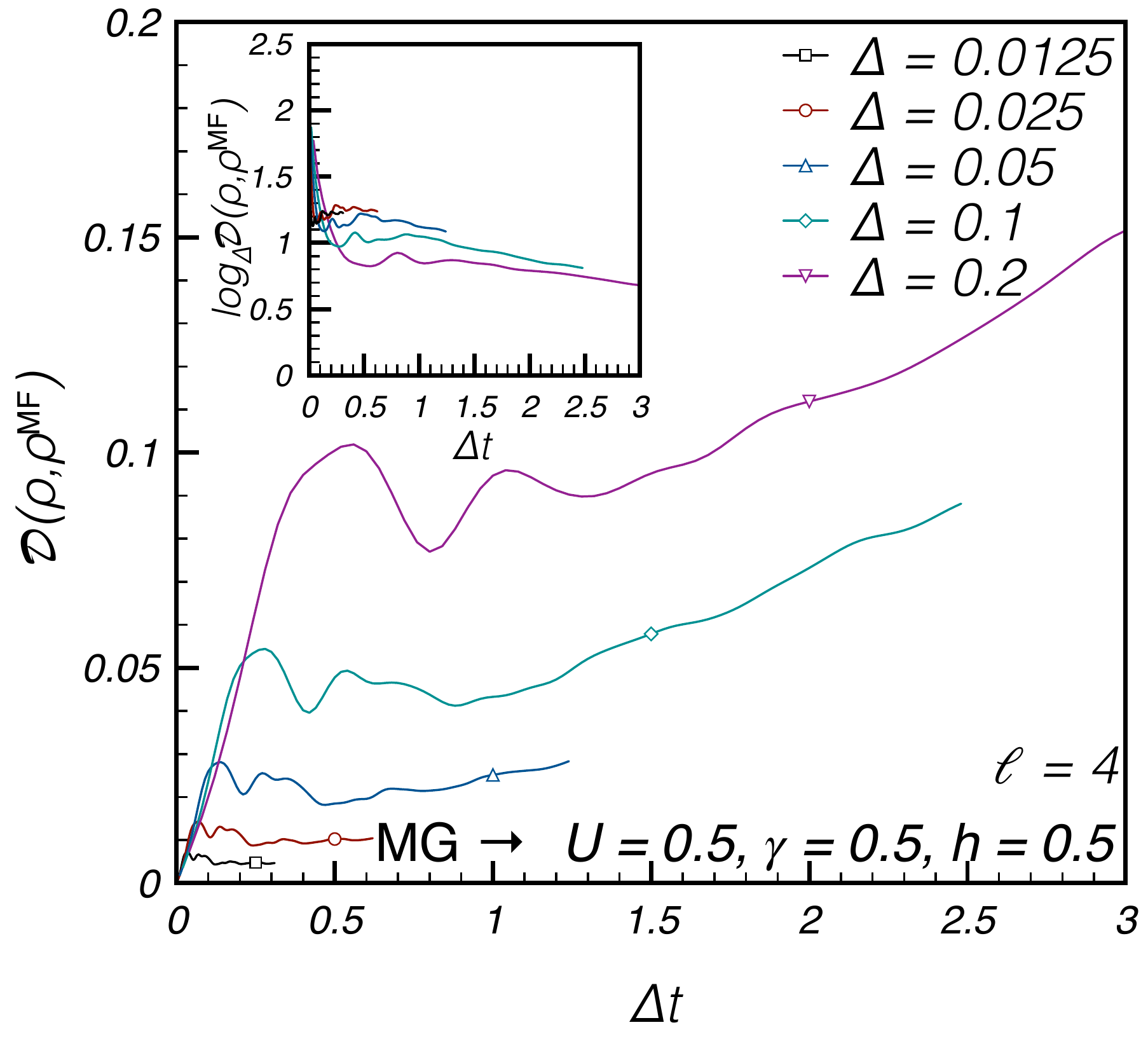}\includegraphics[width=\ratioFthree\textwidth]{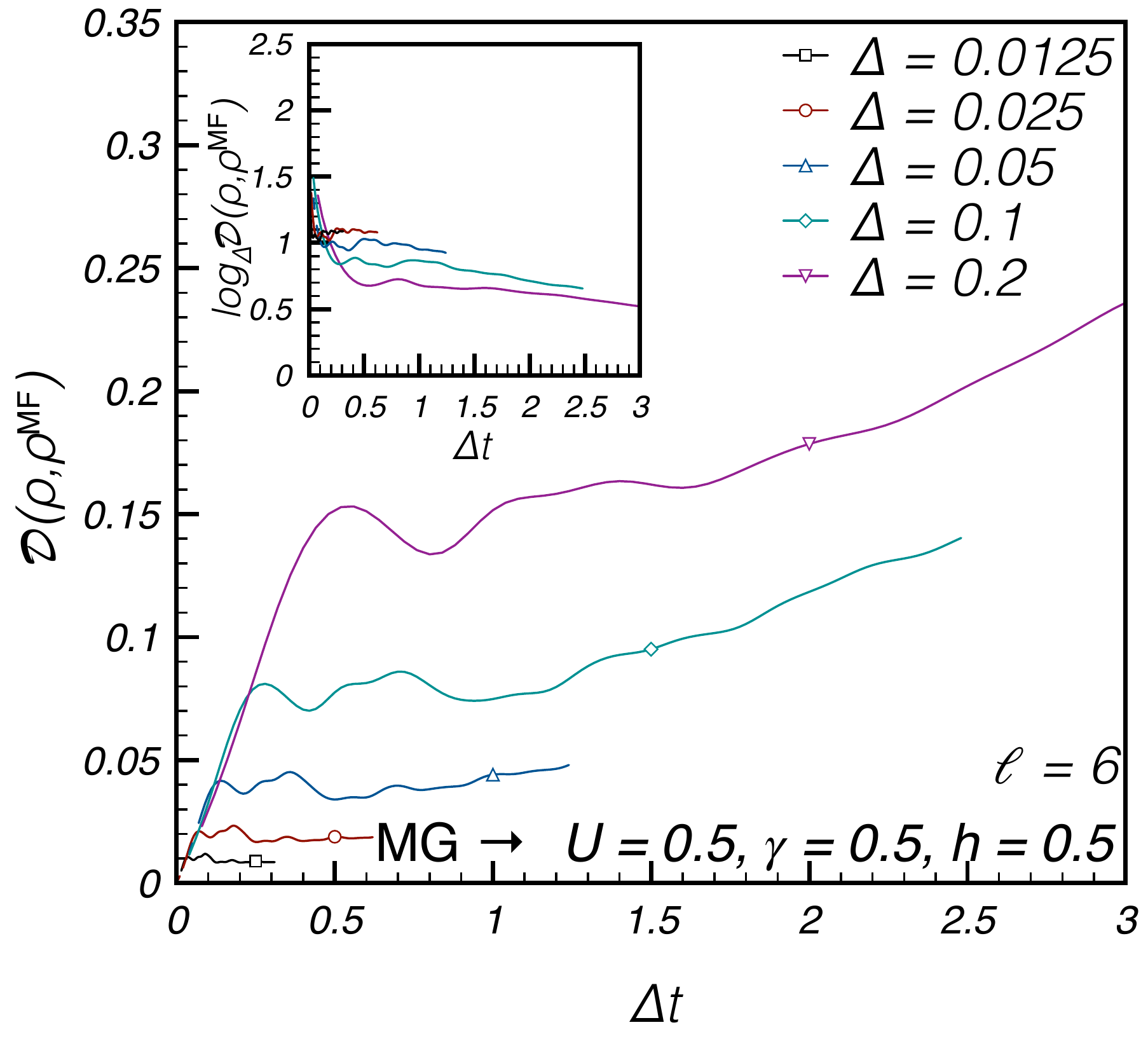}
\includegraphics[width=\ratioFthree\textwidth]{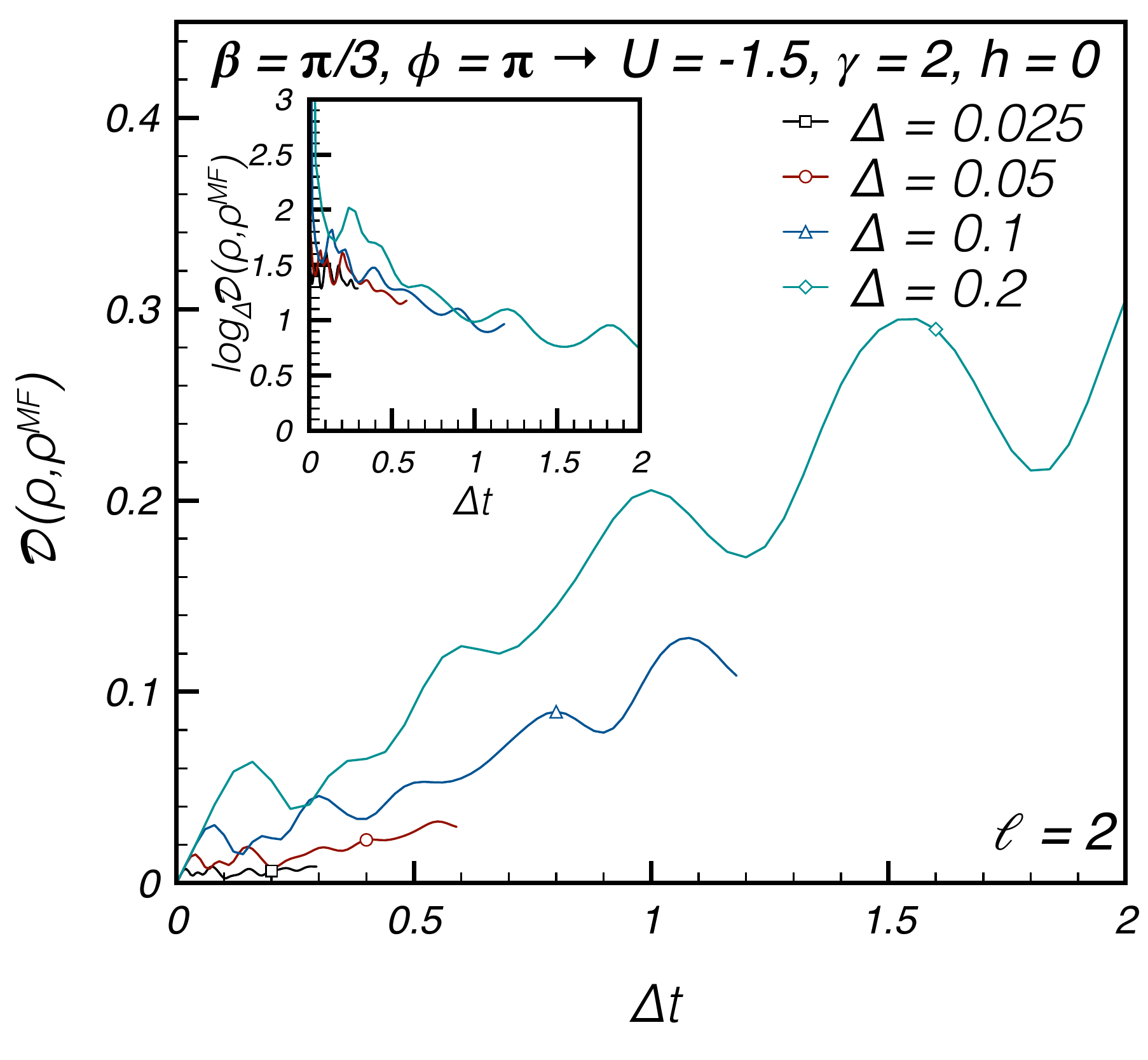}\includegraphics[width=\ratioFthree\textwidth]{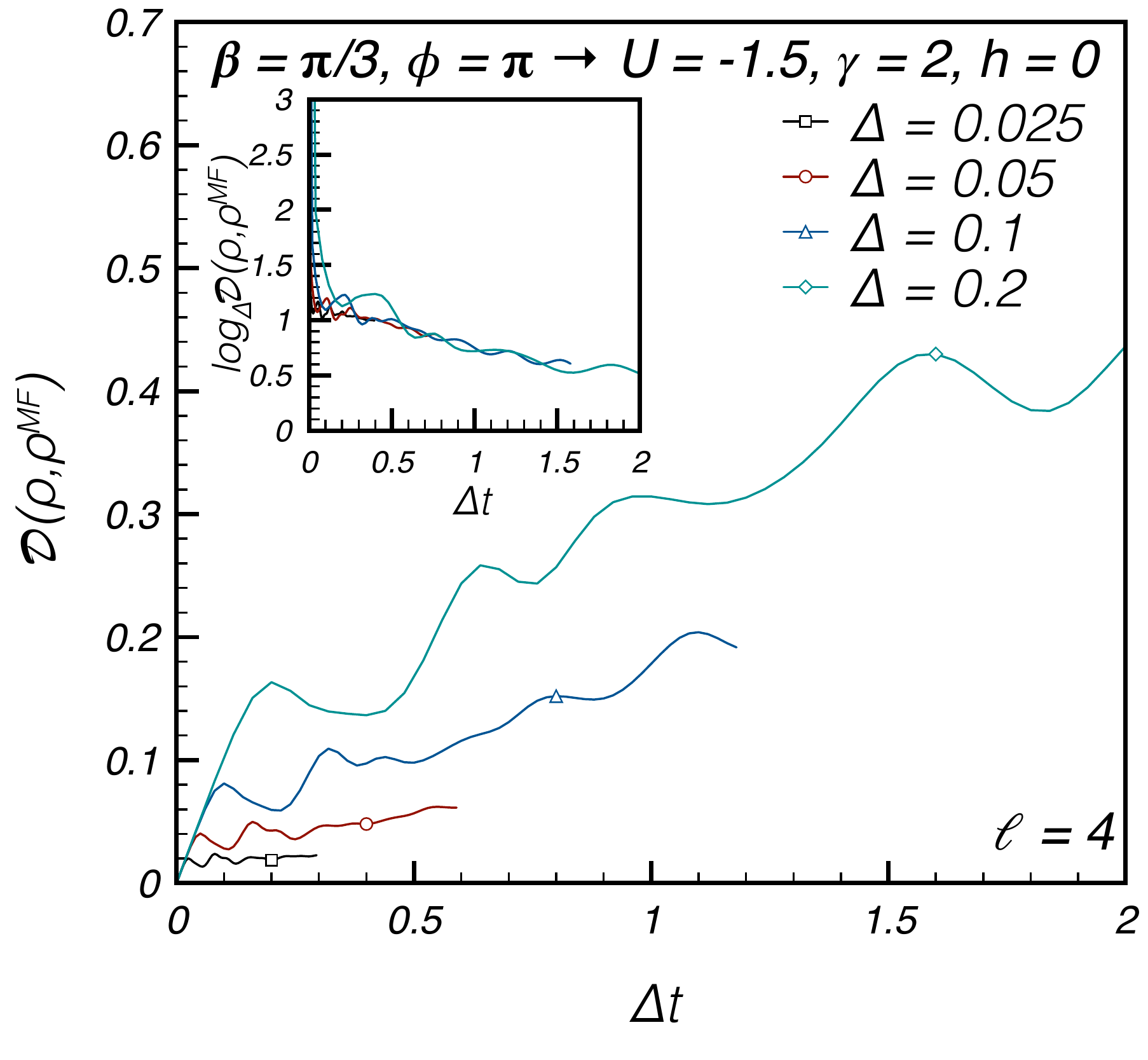}\includegraphics[width=\ratioFthree\textwidth]{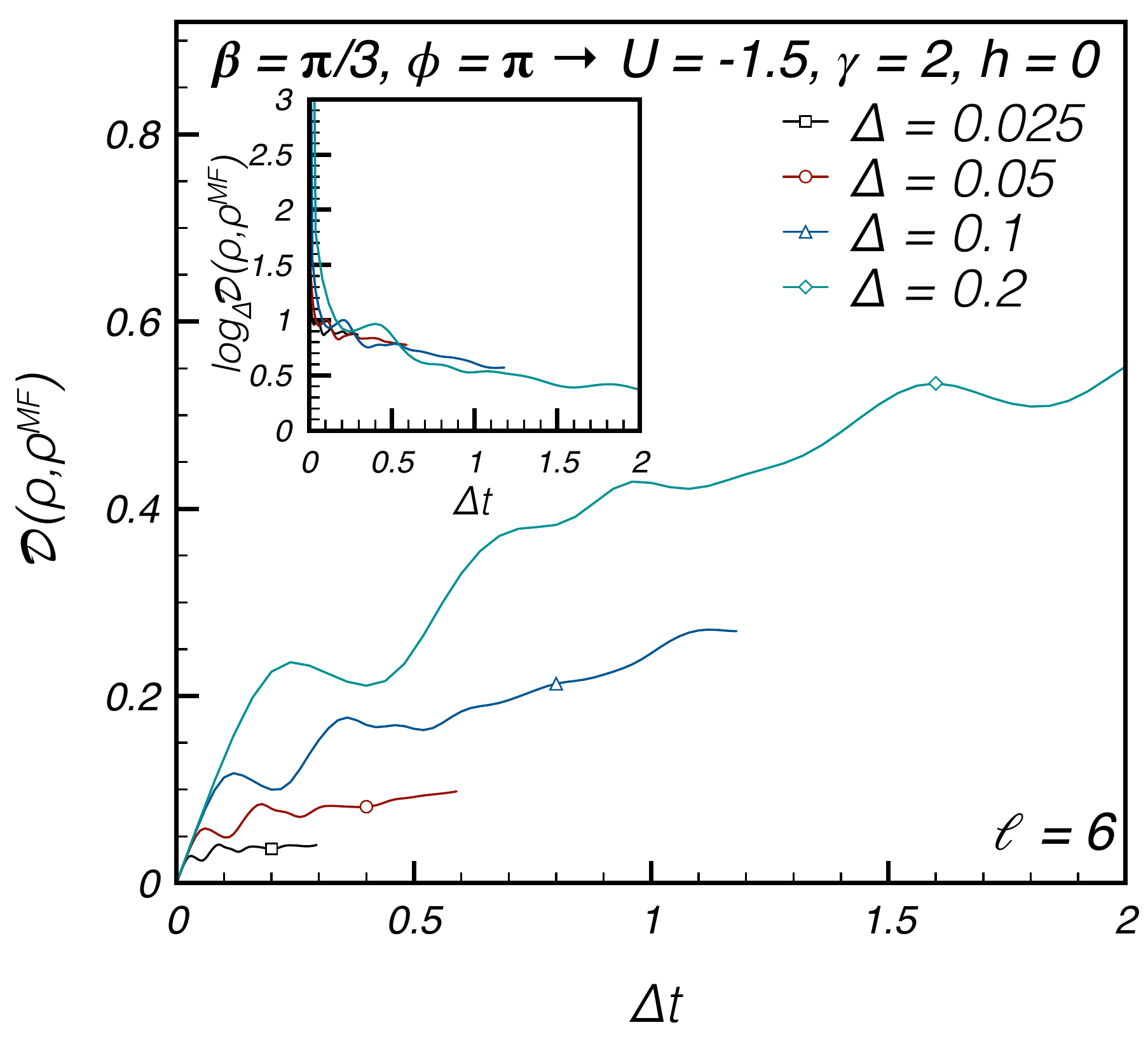}
\end{center}\caption{
The normalized Frobenius distance (see the supplemental material~[\onlinecite{SM}]) 
between the exact RDM and the one obtained via the intermediate mean-field~\eqref{eq:mf0}
for the quenches \eqref{e:statec} and \eqref{e:stated} displayed in the panels. The notations are the same as in Fig. \ref{f:Dist_Neel}.
}\label{f:Dist_MG}\label{f:Dist_betaphi}
\end{figure*}

In this section we show the results of the  iTEBD simulations for the quenches \eqref{e:statec} and \eqref{e:stated}. 
In both cases we focus our attention on the entanglement entropies $S_{\ell}$
for $\ell=1,\, 2,\, 4,$ and $6$ and on the distances between the reduced density matrices of the exact dynamics
and the reduced density matrices obtained through the intermediate mean-field approach.
Below we list the results of such comparison together with some considerations:

\begin{enumerate}[(a)]
\setcounter{enumi}{2}

\item The Majumdar-Ghosh Hamiltonian ground-state
$\ket{\Psi_0} = \ket{{\rm MG}} \equiv \bigotimes(\ket{\uparrow\downarrow}- \ket{\downarrow\uparrow})/\sqrt{2}$ 
breaks translational invariance. Notwithstanding,
the local magnetization $\braket{\sigma^{z}_{\ell}} /2$ is translational invariant at any time after the quench. This is a consequence of reflection symmetry about a bond.
Therefore, we do not show 
the time evolution of the magnetization, as it gives no more information than  the single-site entanglement entropy.
 
In Fig \ref{f:S_MG} we show the evolution of the von Neumann entropies. The agreement with the mean-field prediction is pretty good and improves as  the system length and the interaction strength get smaller.
The discrepancy is more evident at larger rescaled times and could signal the emergence of another regime for $\Delta t\sim \Delta^{-1}$. 
Essentially, the same considerations apply to the distance between reduced density matrices (see Fig \ref{f:Dist_MG}).
This remains rather small (especially for $\ell=2$) even for the largest
value of the interactions we considered ($\Delta=0.2$). 
Finally, also in this case the hypothesis of a power-law scaling 
for the distances seems to be confirmed by the plots of the logarithm of the distance in base $\Delta$.

\item The initial state $\ket{\Psi_0} = \bigotimes(\sqrt{3}\ket{\uparrow\downarrow}- \ket{\downarrow\uparrow})/2$
breaks both one-site shift invariance and reflection symmetry.
We do not show the average magnetization $m_z$, which for $h=0$ is zero at any time. This is a consequence of the combined symmetry under reflection about a bond followed by spin flip. 
 
The evolution of the von Neumann entropy is depicted in Fig \ref{f:S_betaphi}. It is evident that the corrections to the mean-field approximation are stronger than in the previous cases; nevertheless,
they still reduce as $\Delta$ gets closer to zero. 
Interestingly, due to the absence of a magnetic field, the  approximation of neglecting the interacting terms
for this specific quench is trivial, reducing to no evolution at all. Therefore, the nontrivial dynamics of the iTEBD
entropies, which undoubtedly move away from their initial values, 
show unequivocally that the pre-relaxation limit does exist and is not trivial even for a genuine four-fermion perturbation.

The overall behavior and the scaling 
with respect to $\ell$ and $\Delta$ is confirmed by the analysis of the distances between 
the reduced density matrices (see Fig \ref{f:Dist_betaphi}), which basically follow the same qualitative behavior of the difference between the iTEBD entropies and the mean-field ones.  

\end{enumerate}

\section{Conclusions}\label{s:conclusions} %

In this work we have considered the evolution of the entanglement entropies after a quantum quench from initial states that break one-site shift invariance in a perturbed quantum XY model. Our main result is to have identified a long-time behavior that is almost independent of the quench details. We showed that, in the limit $t\sim\Delta^{-1}$, where $\Delta$ is the strength of the perturbation, the entropy, which already reached a plateau of prethermalization/pre-relaxation, grows linearly in $\Delta t$ until a characteristic time 
$
t_{M}\sim \frac{\ell}{\Delta}
$, 
after which it starts saturating to another quasi-stationary value. 
We proposed a semi-classical explanation, based on very general arguments, which  manifests the universality of the phenomenon.
The nontrivial behavior disclosed in the timescale $\Delta t\sim O(\Delta^0)$ is driven by the breaking of what we called ``non-abelian integrability'', namely the existence of an infinite set of local conservation laws that do not commute with one another. 
When the perturbation removes the accidental degeneracies responsible for the non-commuting local conservation laws, at the leading order in $\Delta$ for $t\sim \Delta^{-1}$ local observables become sensible to the slow unitary time evolution that occurs in each \mbox{(quasi-)degenerate} subspace. As one would expect from general reasonings, the entanglement entropy reacts to the symmetry breaking by increasing its value. 

We have studied in details a quench from the N\'eel state under the Hamiltonian of the XYZ spin-$\frac{1}{2}$ chain with an external magnetic field that breaks integrability. In particular, we obtained an analytic expression for the time evolution of the entanglement entropies per unit length in the scaling limit $1\ll \ell\sim\Delta t$. 
Considering other quenches, we showed that the pre-relaxation of the entropy is almost independent of the quench parameters and, even when local observables exhibit persistent oscillations, in the entanglement entropies per unit length the oscillations turn out to be subleading.  

Our analytic findings are based on the method developed in Ref.~[\onlinecite{BF:mf}]:  the dynamics are reduced to mean-field time evolution and the entropy is computed using free-fermion techniques.

We checked our results against iTEBD simulations. We had to face the technical problems behind the pre-relaxation limit, which presupposes very long times. The present-day algorithms are reliable on timescales that, in the cases considered, do not depend much on the strength of the perturbation, so the smaller $\Delta$ is and the shorter the maximal rescaled time $\Delta t$ that is reached. In order to make the comparison more transparent, we analyzed a distance between the reduced density matrices that time evolve with the exact Hamiltonian and those of the mean-field description.  Notwithstanding the complications of the limit, we provided some numerical evidence that the mean-field time evolution is exact in the pre-relaxation limit, confirming that the technical assumptions made in Ref.~[\onlinecite{BF:mf}] are reasonably satisfied for the time evolution under Hamiltonian \eqref{eq:H}. 

Our work raises a number of issues. First and foremost is the perceptible inadequacy of DMRG-like algorithms to study time evolution in the pre-relaxation limit or in other limits in which the time is supposed to be very large. Alternative methods based on a perturbative solution of the equations of motion have been recently employed to study similar problems\cite{BEGR:preT}. Apparently these techniques allow one to reach larger times, although the errors are much less under control than in DMRG algorithms. 
It could be very interesting to  investigate the pre-relaxation limit with these methods,  checking the consistency with mean-field. 

Another relevant question is whether the time-dependent GGE \eqref{eq:GGEt} that emerges in the pre-relaxation limit survives the next order of perturbation theory, like a (deformed) GGE description does at intermediate times in the presence of small integrability breaking perturbations\cite{E:preT}. 

Finally, our analysis suggests that different local conservation laws can respond to perturbations in different timescales. The pre-relaxation time $t_{M}$ seems to characterize the time at which some non-commuting conservation laws of the unperturbed model with range $\ell$ start feeling the effects of the perturbation.
Other local charges are instead expected to remain quasi-conserved for much longer times. A systematic study of these effects could be useful to identify the correct set of conservation laws that contribute to the description of pre-relaxation/prethermalization plateaux.  

We conclude with a speculation. Our findings rely on the non-abelian integrability of the unperturbed model, which in our case is noninteracting. A natural question is if there are also interacting quantum many-body systems with infinitely many non-commuting (quasi-)local conservation laws. Although there is not yet firm evidence, models with a loop algebra symmetry, like the XXZ model with anisotropy root of unity~\cite{DFM:sl2}, are good candidates for possessing similar charges. This is clearly an imperative question to be addressed in the next future.

\acknowledgments  %

We thank Bruno Bertini and Spyros Sotiriadis for helpful discussions. This work was supported by the LabEx ENS-ICFP: ANR-10-LABX-0010/ANR- 10-IDEX-0001-02 PSL* (M.F.) and by the ERC under Starting Grant No. 279391 EDEQS (M.C.).


\onecolumngrid
\appendix
\newpage
\section*{Supplemental Material}

\section{On the iTEBD simulations}\label{a:iTEBD}%
In this supplementary section we discuss the numerical method used to access the pre-relaxation limit of the exact quench dynamics under Hamiltonian
\be\label{eq:H}
H=H_\Delta+\Delta (U S^{z\cdot z}+ h S^z)\, ,
\ee
with
\be\label{eq:HD}
H_\Delta=\sum_\ell\frac{1+\gamma}{4}\sigma_\ell^x\sigma_{\ell+1}^x+\frac{1-\gamma}{4}\sigma_\ell^y\sigma_{\ell+1}^y+\frac{\Delta}{4}\sigma_\ell^z\sigma_{\ell+1}^z\, ,
\ee
\be
S^{z\cdot z}=\frac{1}{4} \sum_\ell \sigma_\ell^z\sigma_{\ell+2}^z\quad\quad S^z=\frac{1}{2} \sum_\ell \sigma_\ell^z\, ,
\ee
and initial state
\be\label{eq:state}
\ket{\Psi_0}=\bigotimes
\Bigl[\cos\frac{\beta}{2}\ket{\uparrow\downarrow}+e^{i\phi}\sin\frac{\beta}{2} \ket{\downarrow\uparrow}\Bigr]\, .
\ee
In particular,
we make use of intensive numerical simulations via infinite time-evolving block-decimation (iTEBD) algorithm\cite{iTEBD}.

The algorithm is based on the matrix product state (MPS) description
 of one-dimensional lattice models  and works directly in the thermodynamic limit.
Since the Hamiltonian (\ref{eq:H}) includes interactions between next-nearest neighbor spins,
 we implemented the algorithm by considering a unit cell of two sites\cite{f:2}, 
 with local dimension $d=4$.
The initial state has then a translational invariant MPS representation.
Nevertheless we need to partially break translational symmetry in order to simulate the action 
of the local evolution operators on even and odd bonds. 
A generic many-body state takes the following MPS canonical representation
\be\label{eq:MPS}
\ket{\Psi} = 
\sum_{\{s\}} 
{\rm Tr}[\cdots {\bf\Gamma}_{o}^{s_{j}} {\bf\Lambda}_{o} 
{\bf \Gamma}_{e}^{s_{j+1}} {\bf\Lambda}_{e} \!\cdots ]
\ket{\cdots s_{j}s_{j+1}\!\cdots}.
\ee
Here ${\bf \Gamma}_{o/e}^{s_{j}}$ are $\chi\times\chi$ matrices associated with odd/even unit cells,
with $s_{j}$ spanning the $j^{\rm th}$ unit-cell space with canonical basis 
$\{\ket{\uparrow\uparrow}, \ket{\uparrow\downarrow}, \ket{\downarrow\uparrow},\ket{\downarrow\downarrow}\}$;
similarly, ${\bf \Lambda}_{o/e}$ are diagonal matrices with entries the singular values associated with the rank-$\chi$ 
Schmidt decomposition for the bipartition of the system on the odd/even bonds. The trace
is taken over the auxiliary space. We point out that \eqref{eq:MPS} is well-defined  thanks to the right/left orthonormalization conditions
\be\label{eq:normalization}
\sum_{s} ({\bf\Gamma}_{o/e}^{s} {\bf\Lambda}_{o/e}) ({\bf\Gamma}_{o/e}^{s} {\bf\Lambda}_{o/e})^{\dag} = {\bf I},\qquad\qquad \sum_{s}  ({\bf\Lambda}_{e/o}{\bf\Gamma}_{o/e}^{s} )^{\dag} ({\bf\Lambda}_{e/o}{\bf\Gamma}_{o/e}^{s}) = {\bf I},
\ee 
which essentially state that  the leading eigenvalue of the right/left transfer matrices is unitary and the corresponding right/left eigenvector is the identity  matrix (reshuffled as a vector). 
These conditions guarantee the correct normalization of the vector $|\Psi\rangle$
as well as the possibility to perform well-defined operations in the  MPS representation.

The initial state $\ket{\Psi_0}$ \eqref{eq:state} admits a trivial rank-$1$ MPS representation with 
a unique singular values ${\bf \Lambda}_{o/e} = 1$ and a 1-by-1 tensor 
${\bf\Gamma}_{o/e}^{s_{j}} =\cos\frac{\beta}{2}\delta_{s_{j},\uparrow\downarrow}
+ e^{i\phi}\sin\frac{\beta}{2}\delta_{s_{j},\downarrow\uparrow}$.

The  time evolution of the MPS in Eq. (\ref{eq:MPS}) is easily accomplished by using
the second order Suzuki-Trotter decomposition of the evolution operator with time step $dt$, namely
\be\label{eq:ST}
e^{-i H dt } \simeq \bigotimes_{j\,odd}e^{-i \mathfrak{h}  dt/2 }
\bigotimes_{j\, even}e^{-i \mathfrak{h} dt }
\bigotimes_{j\,odd}e^{-i  \mathfrak{h}  dt/2 },
\ee
where the index $j$ runs over the unit cells and $\mathfrak{h} $ represents the local interaction between
nearest neighbor cells. Since the original Hamiltonian is translational invariant, 
the local interaction $\mathfrak{h} $ does not depend on the particular bond between cells. 
In practice, the MPS representation of the state at time $t+dt$ is obtained by updating the MPS at time $t$
so as to take account of the repeated action of the local evolution operators in Eq. (\ref{eq:ST}). 
Interestingly, under the action of the unitary operator restricted to the odd (even) bonds, 
only the matrices $\{{\bf\Gamma}_{o}^{s}, {\bf\Gamma}_{e}^{s}, {\bf\Lambda}_{o}\}$ 
($\{{\bf\Gamma}_{o}^{s}, {\bf\Gamma}_{e}^{s}, {\bf\Lambda}_{e}\}$) need to be updated and the  invariance 
of the evolved state under a shift by two unit cells (\emph{i.e.} four chain sites) is preserved.
 
In our numerical simulations we fix $dt=0.05$ (we verified that the data are not affected by the time discretization).
The auxiliary dimension $\chi$ used to describe the reduced Hilbert space is dynamically updated in such a way that, at each local step, 
all the Schmidt vectors corresponding to singular values larger than \mbox{$\lambda_{min} = 10^{-16}$} are retained. For purely practical reasons, this condition is relaxed when $\chi$ reaches a maximal value \mbox{$\chi_{max} \in [400,800]$} dependent on the particular quench.
Because of the (unavoidable) upper bound $\chi_{max}$, the truncation procedure is the main source of errors of the algorithm.
We stop the simulations so as to have a Schmidt error coming from the iterative truncation of the Hilbert space always smaller than $10^{-2}$.
We point out that the bipartite entanglement entropy grows linearly for all explored times, 
as it should do after a global quantum quench. 

Based on the normalization condition \eqref{eq:normalization}, the evaluation of expectation values 
of local quantities with spacial range $\ell$ turns out to be very efficient and scales as $\ell\chi^3$.
 However, for our purpose, we need to reconstruct the full reduced density matrix 
 $\rho_{\ell}$ for a subsystem with $\ell$ adjacent sites, which is actually computationally more demanding.
In particular, for $\ell$ even one has
\be\label{eq:RDMiTEBD}
\rho_{\ell}=
{\rm Tr} 
[ 
({\bf\Lambda}_{e} {\bf\Gamma}_{o}^{s_1}\cdots{\bf\Gamma}_{o/e}^{s_{\ell/2}}{\bf\Lambda}_{o/e})^{\dag}
({\bf\Lambda}_{e} {\bf\Gamma}_{o}^{s_1'}\cdots{\bf\Gamma}_{o/e}^{s_{\ell/2}'}{\bf\Lambda}_{o/e})
]
\ee
where the last two matrices appearing in the strings can be odd or even depending on the parity of $\ell/2$ and
the trace is over the auxiliary space. Clearly, the reduced density matrix
of an odd number of sites is obtained by  tracing out the last spin, 
which corresponds to trace \eqref{eq:RDMiTEBD} over the second site in the last unit cell. 
Finally, from the spectrum of the reduced density matrix one can easily evaluate the von Neumann entropy
$S_{\ell} = -{\rm Tr}( \rho_{\ell} \log \rho_{\ell})$ or, more generally, the R\'enyi
entropies $S^{(\alpha)}_{\ell} =\log[{\rm Tr}(\rho^{\alpha}_{\ell})]/ (1-\alpha)$.

\subsection{Distance between reduced density matrices}%

The main difficulty in comparing the theoretical predictions with
the numerical data is that, in order to describe the
scaling limit $\Delta\to 0$ with $\Delta t$ fixed, the simulations should reach very large times $t\gg 1$. 
In practice, all algorithms based on MPS representations suffer from the growth of the bipartite entanglement entropy,
making it extremely challenging the possibility to investigate the pre-relaxation limit. 
This forces us to choose not too small values of the parameter $\Delta$, introducing unavoidable corrections $o(\Delta^0)$ 
which are beyond our control. The agreement with the mean-field approximation is therefore made evident
via a systematic scaling analysis. 
As $\Delta$ becomes smaller and smaller 
we aim at showing that, for fixed value of the rescaled time $\Delta t$, the expectation values of local quantities converge to the mean field predictions;
more generally, the full reduced density matrix $\rho_{\ell}$ should 
get closer and closer to the corresponding reduced density matrix $\rho^{\rm MF}_{\ell}$ of the mean-field state $\rho^{\rm MF}$ defined as follows
\be\label{eq:rho0}
\ba
& \braket{\mathcal O}_{{\rm GGE}_{0}}^{\rm MF}(T)=\mathrm{Tr}[\rho^{\rm MF}(T) \mathcal O]\\
&i \partial_{T} \rho^{\rm MF}(T)= [H^{\rm MF}(T),\rho^{\rm MF}(T)]\\
&\rho^{\rm MF}(0)=\lim_{|A|\rightarrow\infty}\lim_{t\rightarrow\infty}\mathrm{Tr}_{\bar A}[e^{i H_0 t}\ket{\Psi_0}\bra{\Psi_0}e^{-i H_0 t}]\, .
\ea
\ee  

The approach to the mean-field solution strongly 
depends on the subsystem size $\ell$. Indeed, in the mean-field approximation the condition 
$1/\Delta \gg \ell$ is always verified (since $\Delta$ has been already sent to zero).
On the other hand, in the iTEBD numerical simulations, for a finite small value of $\Delta$, 
one needs to consider small enough subsystems to be in the expected regime of validity of mean-field.

In order to give a quantitative estimation of the deviations between the numerical simulations 
and the mean-field results it is convenient to compute a distance between the corresponding 
reduced density matrices. Ref.~[\onlinecite{FE:13}] analyzed the physical meaning of different 
distances and found a good compromise between practicality and meaningfulness in the following normalized distance:
\be\label{eq:distance}
\mathcal{D}\left(\rho,\tilde \rho \right) = \frac{||\rho-\tilde \rho||}{\sqrt{||\rho||^2+||\tilde \rho||^2}} = \sqrt{1-\frac{2\,{\rm Tr} [\rho\tilde\rho]}{{\rm Tr} [\rho^2]+{\rm Tr} [\tilde\rho^2]}}
\ee
where $||\mathcal O||\equiv\sqrt{{\rm Tr} (\mathcal O^{\dag} \mathcal O)}$ is the Frobenius norm of an operator.
The distance \eqref{eq:distance} has several useful properties~\cite{FE:13}: 

\begin{enumerate}[(i)]

\item It vanishes if and only if the reduced density matrices are identical.

\item It takes values in the interval $[0,1]$.

\item \label{e:lb} It has an interesting lower bound $\mathcal{D}(\rho,\rho') \geq |e^{-S^{(2)}/2}-e^{-S'^{(2)}/2}|/\sqrt{e^{-S^{(2)}}+e^{-S'^{(2)}}}$.

\item  
Even in the limit of large subsystem, if the density matrices are physically different, the distance does not approach zero (this is a problem that affects other distances, like the operator norm~\cite{FE:13}).

\end{enumerate}

Notice that Property \eqref{e:lb} implies that it is sufficient that the R\'enyi entropies $S^{(2)}$ of the two RDMs differ by a constant for $\mathcal{D}(\rho,\rho')$ to have a lower bound independent of $\ell$.

Since the RDMs of the exact dynamics are not gaussian, one can not use free-fermion techniques to compute \eqref{eq:distance} but instead the RDMs within mean-field must be explicitly reconstructed. 
In particular we have
\be\label{eq:gaussianrho}
\rho^{\rm MF}_{\ell} = \frac{\exp ( a_{j} K^{\rm MF}_{ij} a_{i} )}{{\rm Tr}[\exp ( a_{j} K^{\rm MF}_{ij} a_{i} )]},
\ee
where $a_{j}$ are the Majorana fermions
\be\label{eq:J-W}
a_{2\ell-1}=\prod_{j<\ell}\sigma_j^z \sigma_\ell^{x}\qquad a_{2\ell}=\prod_{j<\ell}\sigma_j^z \sigma_\ell^{y}
\ee
and the (implicit) sums are restricted to 
$\ell$ adjacent sites; the matrix elements $K^{\rm MF}_{ij}$ are related to the elements of 
the correlation matrix $\Gamma^{\rm MF}_{ij} = \delta_{ij} - \langle a_{i} a_{j} \rangle_{\rm MF} $
via the matrix relation ${K}^{\rm MF} = \arctanh({\Gamma}^{\rm MF}_\ell)/2$, where  ${ \Gamma}^{\rm MF}_\ell$ is the correlation matrix
restricted to $\ell$ sites. These properties are a consequence of the gaussianity 
of the many-body state in the mean-field description.

\section{Pre-relaxation in quantum XY models}\label{a:free} %

In this section we integrate the main results reported in the paper with the pre-relaxation limit of the entanglement entropies in the noninteracting case described by the XY Hamiltonian $H_0$ with a small transverse field $\Delta h$
\be\label{eq:Hnonint}
H=\sum_\ell\frac{1+\gamma}{4}\sigma_\ell^x\sigma_{\ell+1}^x+\frac{1-\gamma}{4}\sigma_\ell^y\sigma_{\ell+1}^y+\frac{\Delta h}{2}\sigma_\ell^z\, .
\ee
The time evolution under \eqref{eq:Hnonint} of a two-site shift invariant Slater determinant was worked out in Ref.~[\onlinecite{F:super}]. 
We reconsider the same problem here.

It is known\cite{F:super} that in noninteracting models the symbol $\Gamma(k)$ of the correlation matrix time evolves with the symbol $\mathcal H(k)$ of the Hamiltonian as follows
\be
\Gamma(k,t)=e^{-i\mathcal H(k)t}\Gamma(k,0)e^{i\mathcal H(k)t}\, .
\ee
In our specific case $\mathcal H(k)=\mathcal H_0(k)+\Delta h \mathrm I\otimes\sigma^y$ so we have
\be
\Gamma(k,t)=V^\dag(k,t)e^{-i\mathcal H_0(k)t}\Gamma(k,0)e^{i\mathcal H_0(k)t}V(k,t)
\ee
with
\be
i\partial_t V(k,t)=-\Delta h e^{-i\mathcal H_0(k) t}\mathrm I\otimes \sigma^y e^{i\mathcal H_0(k) t} \ V(k,t)\, .
\ee
The mean-field approximation $V_{\rm MF}(k,ht)\sim V(k,t)$ corresponds to replacing $e^{-i\mathcal H_0(k) t}\mathrm I\otimes \sigma^y e^{i\mathcal H_0(k) t}$ by its time average, that is to say (for $\gamma\neq 0$, otherwise the perturbation commutes with $H_0$)
\be
V_{\rm MF}(k,ht)=e^{i h\frac{\mathcal Q_2(k)}{\varepsilon_k^2}(\Delta t)}
\ee
with $\mathcal Q_2(k)$ the symbol associated with the second family of local conservation laws of $H_0$, namely
\be
\ba
 \mathcal{Q}_1(k)&=\varepsilon_k\,\bigl[\sigma^xe^{i\frac{k}{2}\sigma^z}\bigr]\otimes\bigr[\sigma^y e^{i\theta(k)\sigma^z}\bigr]\\
\mathcal{Q}_2(k)&=\cos(k/2) \varepsilon_k \,\1\otimes\bigl[\sigma^y e^{i\theta(k)\sigma^z}\bigr]\\
\mathcal{Q}_3(k)&=\sin( k)\, \1\otimes\1 \\
\mathcal{Q}_4(k)&=\sin(k/2)\,\bigl[\sigma^xe^{i\frac{k}{2}\sigma^z}\bigr]\otimes\1\\
\mathcal{Q}_5(k)&=\varepsilon_k\, \bigl[\sigma^ye^{i\frac{k}{2}\sigma^z}\bigr]\otimes\bigl[\sigma^x e^{i\theta(k)\sigma^z}\bigl]\\
\mathcal{Q}_6(k)&=\cos(k/2)\,\bigl[\sigma^ye^{i\frac{k}{2}\sigma^z}\bigr]\otimes\sigma^z\\
\mathcal{Q}_7(k)&=\sin(k)\,\sigma^z\otimes\sigma^z \\
\mathcal{Q}_8(k)&=\sin(k/2)\varepsilon_k\,\sigma^z\otimes\bigr[\sigma^x e^{i\theta(k)\sigma^z}\bigl]\label{eq:Q}\, ,
\ea
\ee
where $\varepsilon_k=\sqrt{\cos^2\frac{k}{2}+\gamma^2\sin^2\frac{k}{2}}$ and $\tan\theta_k=\gamma \tan\frac{k}{2}$.  
In order to keep the corrections under control, let us isolate the mean-field part
\be
V(k,t)=V_{\rm MF}(k,ht) \delta V(k,t)\, .
\ee
We obtain
\be\label{eq:deltaV}
\ba
&i\partial_{\Delta t} \delta V(k,t)=h \Omega(k,t) \delta V(k,t)\\
&\Omega(k,t)=\frac{\mathcal Q_2(k)}{\varepsilon_k^2}-e^{-i(\mathcal H_0(k)+\Delta h\frac{\mathcal Q_2(k)}{\varepsilon_k^2})t}(\mathrm I\otimes \sigma^y)e^{i(\mathcal H_0(k)+\Delta h\frac{\mathcal Q_2(k)}{\varepsilon_k^2})t}
\ea
\ee
and more explicitly 
\be
\Omega(k,t)=-\sin\theta_k\Bigl[\mathrm I\otimes \sigma^x e^{i\theta_k \sigma^z}e^{\frac{2i h \cos\frac{k}{2}}{\varepsilon_k}\sigma^y e^{i\theta_k \sigma^z}(\Delta t)}\Bigr] e^{-2i\varepsilon_k \sigma^x e^{i\frac{k}{2}\sigma^z}\otimes \sigma^y e^{i\theta_k \sigma^z}t}\, ,
\ee
where $e^{i\theta_k}=\frac{\cos\frac{k}{2}+i\gamma\sin\frac{k}{2}}{\sqrt{\cos^2\frac{k}{2}+\gamma^2\sin^2\frac{k}{2}}}$. 
The rapidly oscillating term on the second line is responsible for the irrelevance of $\delta V$ in the pre-relaxation limit. Roughly speaking, one must go to second order of a perturbation theory for $\delta V$ in $\Delta$ in order to find a term proportional to the time, which is however also proportional to $\Delta^2$, and hence negligible in the pre-relaxation limit (there are no issues related to the re-exponentiation of the perturbative series). On the other hand, the $O(\Delta)$ contributions in \eqref{eq:deltaV} are bounded in time, thus, at fixed $\Delta t$, they simply give $O(\Delta)$ corrections to the time evolution of the observable. 

Finally, we point out that in non-interacting models the second plateau reached at large times in the pre-relaxation limit corresponds to the infinite time limit.  

\begin{figure}[!t]
\begin{center}
\includegraphics[width=0.66\textwidth]{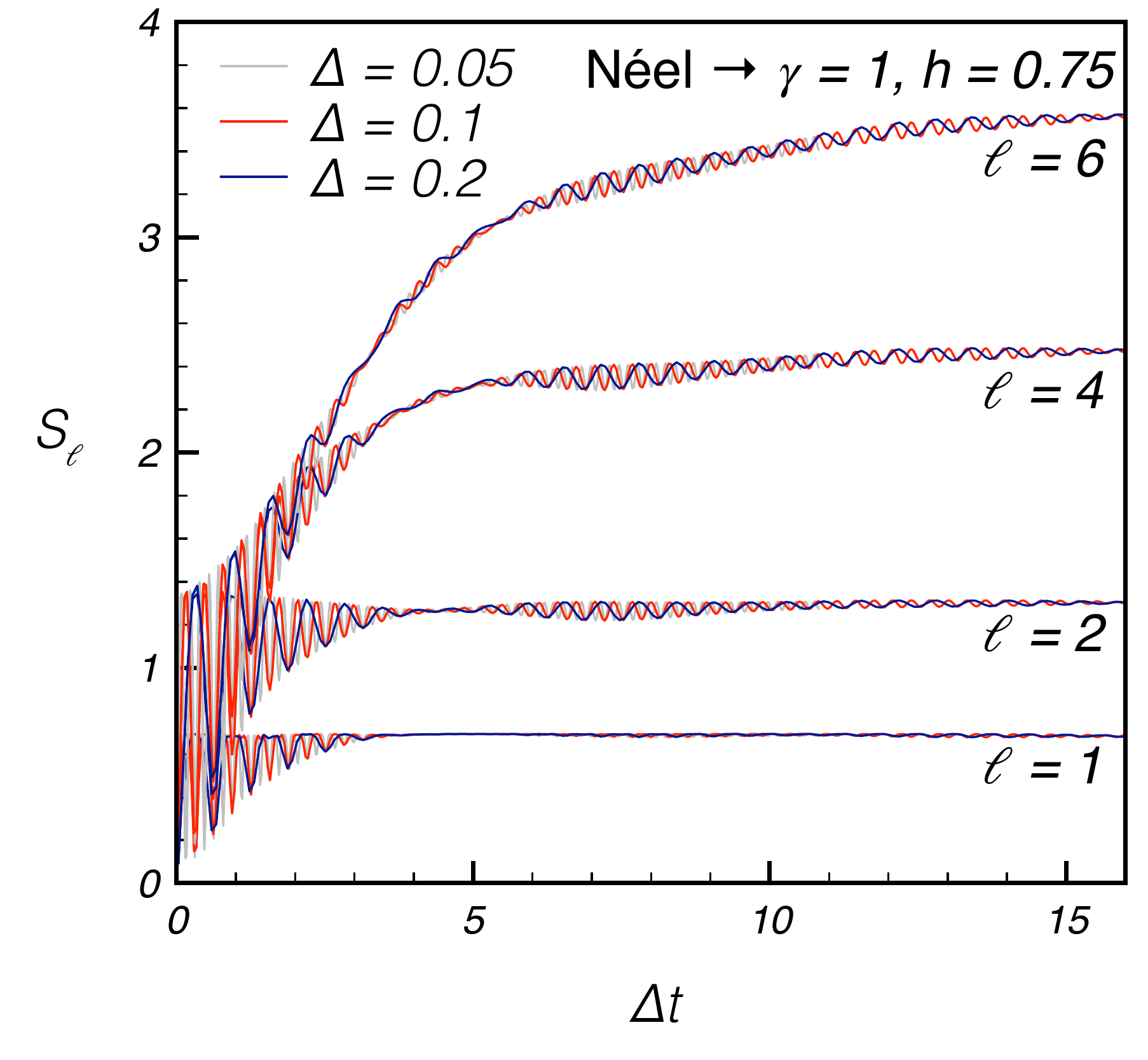}
\end{center}\caption{Evolution of the entanglement entropies for a subsystem with a number of sites 
$\ell = 1,\, 2,\, 4,\,6$ for the quench displayed in the panel and various ``small'' values of $\Delta$.
There is a perfect scaling behavior in terms of $\Delta t$.
}\label{f:S_Neel_g1h075}
\end{figure}

\section{Exceptional case $\gamma=1$}\label{a:g1} %

The mean-field description considered in the main text and in Ref.~[\onlinecite{BF:mf}] breaks down for $\gamma=1$. 
The reason is very simple: the unperturbed Hamiltonian reduces to
$
H_0=\frac{1}{2}\sum_\ell\sigma_\ell^x\sigma_{\ell+1}^x\, ,
$ 
which has infinitely many more local conservation laws than the XY Hamiltonian with $\gamma\neq 1$. 
In addition, the dynamics under $H_0$ never end up with a GGE, being just classical time evolution. 
Therefore, first of all, the mean-field description \eqref{eq:rho0} must be replaced with \be\label{eq:mf0}
\ba
& \braket{\mathcal O}_{\rm MF}(t)=\braket{\Psi_{\rm MF}(t)| \mathcal O|\Psi_{\rm MF}(t)}\\
&i \partial_{t}\ket{\Psi_{\rm MF}(t)}= [H_0+\Delta H^{\rm MF}(\Delta t)]\ket{\Psi_{\rm MF}(t)}\\
&\ket{\Psi_{\rm MF}(0)}=\ket{\Psi_0} ,
\ea
\ee
and, 
second, the additional local conservation laws must be taken into account.
We don't work out this problem analytically but show some numerical 
results that testify to the emergence of a scaling regime for $t\sim \Delta^{-1}$. 
In Figure \ref{f:S_Neel_g1h075} we plot the von Neumann entanglement entropy
for subsystems with length $\ell = 1, 2, 4, 6$ after a quench from the N\'eel state to 
the Hamiltonian (\ref{eq:H}) with $\gamma=1$, $h=0.75$ and $U=0$. 
For all subsystem sizes, the numerical data show a perfect scaling in terms of $\Delta t$ 
at least for the interaction strength we considered, i.e. $\Delta = 0.05,\, 0.1,\, 0.2$.
Interestingly, in terms of the rescaled time and for fixed $\ell$, the different $S_{\ell}$
not only follow the same ``average'' scaling function,
but even the amplitude of the high-frequency 
oscillations is modulated in such a way as to perfectly overlap: 
it seems they construct a sort of  ``universal'' envelop.
It is worth mentioning that, in this particular quench, 
the bipartite entanglement entropy (i.e. the von Neumann entropy of the semi-infinite half-chain)
exhibits the same scaling properties. 
Moreover, it turns out to grow very slowly and the numerical simulations have easily reached large times $O(\Delta^{-1})$.


\begin{thebibliography}{99}


\bibitem{CC:05} P. Calabrese and J. Cardy, J. Stat. Mech. (2005) \href{http://dx.doi.org/10.1088/1742-5468/2005/04/P04010}{P04010}.

\bibitem{CCcorr:05} P. Calabrese and J. Cardy, Phys. Rev. Lett. \href{http://dx.doi.org/10.1103/PhysRevLett.96.136801}{\bf 96}, 136801 (2006).

\bibitem{C-06}
M. A. Cazalilla, Phys. Rev. Lett. \href{http://dx.doi.org/10.1103/PhysRevLett.97.156403} {\bf 97}, 156403 (2006).

\bibitem{rigol} 
M. Rigol, V. Dunjko \emph{et al}
, Phys. Rev. Lett. \href{http://dx.doi.org/10.1103/PhysRevLett.98.050405} {\bf 98}, 050405 (2007).

\bibitem{BS-08}  
T. Barthel and U. Schollw\"ock, Phys. Rev. Lett. \href{http://dx.doi.org/10.1103/PhysRevLett.100.100601} {\bf 100}, 100601 (2008).

\bibitem{C-08}
M. Cramer, C. M. Dawson \emph{et al}
, Phys. Rev. Lett. \href{http://dx.doi.org/10.1103/PhysRevLett.100.030602} {\bf 100}, 030602 (2008).

\bibitem{S:08} A. Silva, Phys. Rev. Lett. \href{http://dx.doi.org/10.1103/PhysRevLett.101.120603}{\bf 101}, 120603 (2008).

\bibitem{CEF} 
P. Calabrese, F.H.L. Essler, and M. Fagotti,  Phys. Rev. Lett. \href{http://dx.doi.org/10.1103/PhysRevLett.106.227203}{\bf 106}, 227203 (2011).

\bibitem{CEF1} 
P. Calabrese, F.H.L. Essler, and M. Fagotti, J. Stat. Mech. (2012) \href{http://dx.doi.org/10.1088/1742-5468/2012/07/P07016}{P07016}.

\bibitem{EEF:dyn} F.H.L. Essler, S. Evangelisti, and M. Fagotti, Phys. Rev. Lett. \href{http://dx.doi.org/10.1103/PhysRevLett.109.247206} {\bf 109}, 247206 (2012).

\bibitem{FE:13}
M. Fagotti and F. H.L. Essler, Phys. Rev. B \href{http://dx.doi.org/10.1103/PhysRevB.87.245107}{\bf 87}, 245107 (2013).

\bibitem{HPK:13} M. Heyl, A. Polkovnikov, and S. Kehrein, Phys. Rev. Lett.  \href{http://dx.doi.org/10.1103/PhysRevLett.110.135704}{\bf 110}, 135704 (2013).

\bibitem{CE-13}
J.-S. Caux and F. H. L. Essler, Phys. Rev. Lett. \href{http://dx.doi.org/10.1103/PhysRevLett.110.257203} {\bf 110}, 257203 (2013).

\bibitem{M-13}
G. Mussardo, Phys. Rev. Lett. \href{http://dx.doi.org/10.1103/PhysRevLett.111.100401} {\bf 111}, 100401 (2013).

\bibitem{CSC-13}
M. Collura, S. Sotiriadis, and P. Calabrese, Phys. Rev. Lett. \href{http://dx.doi.org/10.1103/PhysRevLett.110.245301} {\bf 110}, 245301 (2013).

\bibitem{SC:cluster} S. Sotiriadis and P. Calabrese, J. Stat. Mech. (2014) \href{http://dx.doi.org/10.1088/1742-5468/2014/07/P07024}{P07024}.

\bibitem{E:preT} F.H.L. Essler, S. Kehrein \emph{et al}
, Phys. Rev. B \href{http://dx.doi.org/10.1103/PhysRevB.89.165104}{\bf 89}, 165104 (2014).


\bibitem{QAXXZ-14}
B. Wouters, M. Brockmann \emph{et al}
, Phys. Rev. Lett. \href{http://dx.doi.org/10.1103/PhysRevLett.113.117202} {\bf 113}, 117202 (2014).

\bibitem{PMWKZT-14}
B. Pozsgay, M. Mesty\'{a}n \emph{et al}
,  Phys. Rev. Lett. \href{http://dx.doi.org/10.1103/PhysRevLett.113.117203} {\bf 113}, 117203 (2014).

\bibitem{B:sG} 
B. Bertini, D. Schuricht, and F.H.L. Essler,  J. Stat. Mech. (2014) \href{http://dx.doi.org/10.1088/1742-5468/2014/10/P10035}{P10035}.

\bibitem{D-14}
G. Delfino, J. Phys. A  \href{http://dx.doi.org/10.1088/1751-8113/47/40/402001}{\bf 47} (2014) 402001.

\bibitem{F:super} M. Fagotti, J. Stat. Mech. (2014) \href{http://dx.doi.org/10.1088/1742-5468/2014/03/P03016}{P03016}.

\bibitem{CBSF:glassy} G. Carleo, F. Becca \emph{et al}
, Scientific Reports \href{http://dx.doi.org/10.1038/srep00243}{\bf 2}, 243 (2011).

\bibitem{BEL:14} L. Bonnes, F.H.L. Essler, and A.M. L\"auchli, Phys. Rev. Lett. \href{http://dx.doi.org/10.1103/PhysRevLett.113.187203}{\bf 113}, 187203 (2014).


\bibitem{RSMS:08} D. Rossini, A. Silva \emph{et al}
, Phys. Rev. Lett. \href{http://dx.doi.org/10.1103/PhysRevLett.102.127204}{\bf 102}, 127204 (2008). 

\bibitem{ban-11}
M. C. Ba\~{n}uls, J. I. Cirac, and M. B. Hastings,
Phys. Rev. Lett. \href{http://dx.doi.org/10.1103/PhysRevLett.106.050405} {\bf 106}, 050405 (2011).

\bibitem{CK}
J.-S. Caux and R. M. Konik, Phys. Rev. Lett. \href{http://dx.doi.org/10.1103/PhysRevLett.109.175301} {\bf 109}, 175301 (2012).

\bibitem{KRSM:12} C. Karrasch, J. Rentrop \emph{et al}
, Phys. Rev. Lett. \href{http://dx.doi.org/10.1103/PhysRevLett.109.126406}{\bf 109}, 126406 (2012).

\bibitem{FCEC-14}
M. Fagotti, M. Collura \emph{et al}
, Phys. Rev. B \href{http://dx.doi.org/10.1103/PhysRevB.89.125101} {\bf 89}, 125101 (2014).

\bibitem{rigol:num} M. Rigol, Phys. Rev. Lett. \href{http://dx.doi.org/10.1103/PhysRevLett.112.170601}{\bf 112}, 170601 (2014). 

\bibitem{CCS-P:15} L. Cevolani, G. Carleo, and L. Sanchez-Palencia, arXiv:\href{http://arxiv.org/abs/1503.01786}{1503.01786}.

\bibitem{gm-02}
M. Greiner, O. Mandel \emph{et al}
, Nature \href{http://dx.doi.org/10.1038/415039a} {\bf 415}, 39 (2002).

\bibitem{kww-06}
T. Kinoshita, T. Wenger, and D. S. Weiss, Nature \href{http://dx.doi.org/10.1038/nature04693} {\bf 440}, 900 (2006).

\bibitem{Getal1}
M. Gring, M. Kuhnert \emph{et al}
, Science \href{http://dx.doi.org/10.1126/science.1224953} {\bf 337}, 1318 (2012).

\bibitem{Getal2} T. Langen, M. Gring \emph{et al}
, Eur. Phys. J. Special Topics \href{http://dx.doi.org/10.1140/epjst/e2013-01752-0} {\bf 217}, 43 (2013).

\bibitem{Tetal-12}
S. Trotzky, Y.-A. Chen \emph{et al}
, Nature Phys. \href{http://dx.doi.org/10.1038/nphys2232} {\bf 8}, 325 (2012).


\bibitem{chetal-12}
M. Cheneau, P. Barmettler \emph{et al}
,  Nature \href{http://dx.doi.org/10.1038/nature10748}  {\bf 481}, 484 (2012).

\bibitem{schetal-12}
U. Schneider, L. Hackerm\"uller \emph{et al}
,  Nature Physics \href{http://dx.doi.org/10.1038/nphys2205} {\bf 8}, 213 (2012).
    

\bibitem{Metal-13} 
F. Meinert, M. J. Mark \emph{et al}
, Phys. Rev. Lett. \href{http://dx.doi.org/10.1103/PhysRevLett.111.053003} {\bf 111}, 053003 (2013).

\bibitem{Fateal-13}
T. Fukuhara, A. Kantian \emph{et al}
, Nature Phys. \href{http://dx.doi.org/10.1038/nphys2561} {\bf 9}, 235 (2013).

\bibitem{FSetal-13}
T. Fukuhara, P. Schau{\ss} \emph{et al}
, Nature \href{http://dx.doi.org/10.1038/nature12541} {\bf 502}, 76 (2013).

\bibitem{Metal-14}
F. Meinert, M. J. Mark \emph{et al}
, Phys. Rev. Lett. \href{http://dx.doi.org/10.1103/PhysRevLett.112.193003} {\bf 112}, 193003 (2014).


\bibitem{FC:08} M. Fagotti and P. Calabrese, Phys. Rev. A \href{http://dx.doi.org/10.1103/PhysRevA.78.010306}{\bf 78}, 010306(R) (2008).

\bibitem{lk2008}
A. M. L\"auchli and C. Kollath, J. Stat. Mech. (2008) \href{http://dx.doi.org/10.1088/1742-5468/2008/05/P05018}{P05018}.

\bibitem{dmcf2008}
G. De Chiara, S. Montangero \emph{et al}
, J. Stat. Mech. (2008) \href{http://dx.doi.org/10.1088/1742-5468/2006/03/P03001}{P03001}.

\bibitem{LR:velocity} E.H. Lieb and D.W. Robinson, Commun. Math. Phys. \href{http://dx.doi.org/10.1007/BF01645779}{\bf 28}, 251 (1972). 

 
\bibitem{isl2012}
F. Igl\'oi, Z. Szatm\'ari, and Y.-C. Lin, Phys. Rev. B \href{http://dx.doi.org/10.1103/PhysRevB.85.094417}{\bf 85}, 094417 (2012).

\bibitem{ttd2014}
G. Torlai, L. Tagliacozzo, and G. De Chiara, J. Stat. Mech. (2014) \href{http://dx.doi.org/10.1088/1742-5468/2014/06/P06001}{P06001}.

\bibitem{KBC:exc} M. Kormos, L. Bucciantini, and P. Calabrese, EPL, \href{http://dx.doi.org/10.1209/0295-5075/107/40002}{\bf 107} (2014) 40002. 

\bibitem{CTC:14} A. Coser, E. Tonni, and P. Calabrese, J. Stat. Mech. (2014) \href{http://dx.doi.org/10.1088/1742-5468/2014/12/P12017}{P12017}.

\bibitem{SLRD:13} J. Schachenmayer \emph{et al}
, Phys. Rev. X \href{http://dx.doi.org/10.1103/PhysRevX.3.031015}{\bf 3}, 031015 (2013). 

\bibitem{gr2014}
M. Ghasemi Nezhadhaghighi and M. A. Rajabpour, Phys. Rev. B \href{http://dx.doi.org/10.1103/PhysRevB.90.205438}{\bf 90}, 205438 (2014).




\bibitem{sd2011} 
J.-M. St\'ephan and J. Dubail, J. Stat. Mech. (2011) \href{http://dx.doi.org/10.1088/1742-5468/2011/08/P08019}{P08019}.

\bibitem{ep2012}
V. Eisler and I. Peschel, (2012) EPL \href{http://dx.doi.org/10.1209/0295-5075/99/20001}{99} 20001.

\bibitem{CCo:13} M. Collura and P. Calabrese, J. Phys. A: Math. Theor. \href{http://dx.doi.org/10.1088/1751-8113/46/17/175001}{\bf 46}, 175001 (2013).

\bibitem{ah2014}
V. Alba and F. Heidrich-Meisner, Phys. Rev. B \href{http://dx.doi.org/10.1103/PhysRevB.90.075144}{\bf 90}, 075144 (2014).




\bibitem{cardy2011}
J. Cardy, Phys. Rev. Lett. \href{http://dx.doi.org/10.1103/PhysRevLett.106.150404}{\bf 106}, 150404 (2011).

\bibitem{ad2012}
D. A. Abanin and E. Demlerm, Phys. Rev. Lett. \href{http://dx.doi.org/10.1103/PhysRevLett.109.020504}{\bf 109}, 020504 (2012).


\bibitem{deu-91}
J. M. Deutsch, Phys. Rev.  A \href{http://dx.doi.org/10.1103/PhysRevA.43.2046} {\bf 43} , 2046 (1991).

\bibitem{sred-94}
M. Srednicki, Phys. Rev. E \href{http://dx.doi.org/10.1103/PhysRevE.50.888} {\bf 50}, 888 (1994).

\bibitem{rignat-08}
M. Rigol, V. Dunjko, and M. Olshanii
, Nature  \href{http://dx.doi.org/10.1038/nature06838} {\bf 452}, 854 (2008).

\bibitem{bir-10}
G. Biroli, C. Kollath, and A. M. L\"auchli
, Phys. Rev. Lett. \href{http://dx.doi.org/10.1103/PhysRevLett.105.250401} {\bf 105}, 250401 (2010).

\bibitem{sir-14}
J. Sirker, N. P. Konstantinidis \emph{et al}
, Phys. Rev. A \href{http://dx.doi.org/10.1103/PhysRevA.89.042104} {\bf 89}, 042104 (2014).

\bibitem{A:ETH} V. Alba, Phys. Rev. B \href{http://dx.doi.org/10.1103/PhysRevB.91.155123}{\bf 91}, 155123 (2015).


\bibitem{SUPER} W. Miller, S. Post, and P. Winternitz, J. Phys. A: Math. Theor. \href{http://dx.doi.org/10.1088/1751-8113/46/42/423001}{\bf 46} 423001 (2013).
\bibitem{Baxter} R.J. Baxter, Phys. Rev. Lett. \href{http://dx.doi.org/10.1103/PhysRevLett.26.834}{\bf 26}, 834 (1971).
\bibitem{LSM:XY} E. Lieb, T. Schultz, and D. Mattis, Ann. of Phys. \href{http://dx.doi.org/10.1016/0003-4916(61)90115-4}{\bf 13}, 407 (1961).
\bibitem{BEGR:preT} B. Bertini, F.H.L. Essler \emph{et al}
, arXiv:\href{http://arxiv.org/abs/1506.02994}{1506.02994}.
\bibitem{BF:mf} B. Bertini and M. Fagotti, J. Stat. Mech. (2015) \href{http://dx.doi.org/10.1088/1742-5468/2015/07/P07012}{P07012}.
\bibitem{SM} Supplemental material can be found at \dots.
\bibitem{KIB:ABA} V.E. Korepin, A.G. Izergin, and N.M Bogoliubov, 1993, \emph{Quantum Inverse Scattering Method, Correlation Functions and Algebraic Bethe Ansatz} (Cambridge: Cambridge University Press).
\bibitem{f:0} It is customary to qualify it as ``local'' a translation invariant  operator with a local density. 
\bibitem{P:XXZbc} T. Prosen, Nucl. Phys. B \href{http://dx.doi.org/10.1016/j.nuclphysb.2014.07.024}{\bf 886} (2014) 1177.
\bibitem{PPSA:XXZ} R.G. Pereira, V. Pasquier \emph{et al}
, J. Stat. Mech. (2014) \href{http://dx.doi.org/10.1088/1742-5468/2014/09/P09037}{P09037}.
\bibitem{IMP:XXX} E. Ilievski, M. Medenjak, and T. Prosen, arXiv:\href{http://arxiv.org/abs/1506.05049}{1506.05049}.

\bibitem{CT:QM} C. Cohen-Tannoudji, B. Diu, and F. Laloe, 1977, \emph{Quantum Mechanics, Volume 2} (Wiley). 

\bibitem{QHubbard1} M. Moeckel and S. Kehrein, Phys. Rev. Lett. \href{http://dx.doi.org/10.1103/PhysRevLett.100.175702}{\bf 100}, 175702 (2008).
\bibitem{QHubbard2} M. Moeckel and S. Kehrein, Ann. Phys. \href{http://dx.doi.org/10.1016/j.aop.2009.03.009}{\bf 324}, 2146 (2009).

\bibitem{metaOpt} A. Rosch \emph{et al}
, Phys. Rev. Lett. \href{http://dx.doi.org/10.1103/PhysRevLett.101.265301}{\bf 101}, 265301 (2008). 

\bibitem{Kollar} M. Kollar, F.A. Wolf, and M. Eckstein, Phys. Rev. B \href{http://dx.doi.org/10.1103/PhysRevB.84.054304}{\bf 84}, 054304 (2011).

\bibitem{IsingNI} M. Marcuzzi, J. Marino \emph{et al}
, Phys. Rev. Lett. \href{http://dx.doi.org/10.1103/PhysRevLett.111.197203}{\bf 111}, 197203 (2013).

\bibitem{NI:15} N. Nessi and A. Iucci, arXiv:\href{http://arxiv.org/abs/1503.02507}{1503.02507}.

\bibitem{BDK:15} M. Babadi, E. Demler, and M. Knap, arXiv:\href{http://arxiv.org/abs/1504.05956}{1504.05956}.

\bibitem{f:m1} The mean-field Hamiltonian defined in \eqref{eq:HMFform} differs from the one defined in Ref.~[\onlinecite{BF:mf}] for having removed an additional term $H_0$, which is in fact irrelevant. 


\bibitem{CKC:2014} M. Collura, M. Kormos, and P. Calabrese, J. Stat. Mech. (2014) \href{http://dx.doi.org/10.1088/1742-5468/2014/01/P01009}{P01009}.

\bibitem{iTEBD}
G. Vidal, Phys. Rev. Lett. \href{http://dx.doi.org/10.1103/PhysRevLett.98.070201}{\bf 98}, 070201 (2007).

\bibitem{f:2}
 When $U=0$ the Hamiltonian has only nearest neighbor interactions and one  can also use the usual iTEBD implementation with unit cell containing only one site.
 
 \bibitem{f:3} 
 In the two-site representation, the local space is doubled and the Hamiltonian is diagonalized by two species of fermions with the same dispersion relation.
 
 \bibitem{GM:charges} M.P. Grabowski and P. Mathieu, Ann. of Phys. \href{http://dx.doi.org/10.1006/aphy.1995.1101}{\bf 243}, 299 (1995).
 
 \bibitem{f:4} 
 From \eqref{eq:ssymb} it follows that the decay length of the operator density's tail coincides with $s$ times the characteristic length at which the Fourier transform of the  $s$-site representation of the symbol decays.

\bibitem{BHV:06} S. Bravyi, M. B. Hastings, and F. Verstraete, Phys. Rev. Lett. \href{http://dx.doi.org/10.1103/PhysRevLett.97.050401}{\bf 97}, 050401 (2006).
\bibitem{W:trToeplitz} H. Widom, Advances in Math. \href{http://dx.doi.org/10.1016/0001-8708(76)90113-4}{\bf 21}, 1 (1976).

\bibitem{f:6}
This can be immediately inferred from the initial conditions in \eqref{eq:initialcond} being zero for the variables associated with one-site shift invariant charges.

\bibitem{f:5}
Notice that $\bar \Sigma_T(k)$ is not the eigenvalue of the correlation matrix at the time $T$, but shares with it the infinite time limit $\lim_{T\rightarrow\infty}\bar \Sigma_T(k)=\Sigma_\infty(k)$.

\bibitem{DFM:sl2} T. Deguchi, K. Fabricius, and B.M. McCoy, J. Stat. Phys. \href{http://dx.doi.org/10.1023/A:1004894701900}{\bf 102}, 701.








\end{thebibliography}
\end{document}